\documentclass[final]{svjour3}
\pdfoutput=1
\usepackage{etex} 
\usepackage{amsmath,amssymb}
\usepackage{mathtools}
\usepackage{floatrow}
\usepackage{tikz} \usetikzlibrary{arrows.meta,shapes,arrows,bending,automata,calc}
\usepackage{booktabs}
\usepackage{caption}
\usepackage{geometry}
\usepackage[plainpages=false,pdfpagelabels,colorlinks,linkcolor=blue,citecolor=blue,urlcolor=blue,linktoc=all]{hyperref}
\usepackage{subcaption}
\usepackage[T1]{fontenc}  
\usepackage{IEEEtrantools}
\usepackage{longtable}
\usepackage[normalem]{ulem} 
\usepackage{enumerate}
\usepackage[toc,page]{appendix}
\usepackage{cases}
\usepackage{pifont}
\usepackage{pgfplots,wrapfig}

\newcommand{\defref}[1]{Definition~\ref{#1}}
\newcommand{\propref}[1]{Property~\ref{#1}}
\newcommand{\exref}[1]{Example~\ref{#1}}
\newcommand{\figref}[1]{Figure~\ref{#1}}
\newcommand{\secref}[1]{Section~\ref{#1}}
\newcommand{\tabref}[1]{Table~\ref{#1}}
\newcommand{\thref}[1]{Theorem~\ref{#1}}
\newcommand{\equalref}[1]{Equality~(\ref{#1})}
\newcommand{\ineqref}[1]{Inequality~(\ref{#1})}
\newcommand{\lemref}[1]{Lemma~\ref{#1}}
\newcommand{\proposref}[1]{Proposition~\ref{#1}}

\newcommand{\condref}[1]{Condition~\ref{#1}}

\newcommand{\Integers}{\mathbb{Z}}
\newcommand{\IntegersStar}{\mathbb{Z}^*}

\newcommand{\Naturals}{\mathbb{N}}
\newcommand{\AllWords}{\Sigma^*}

\newcommand{\SetRegExpressions}{\mathcal{R}_{\Sigma}}
\newcommand{\Undefined}{\textnormal{undefined}}
\newcommand{\PowerSet}[1]{\mathcal{P}(#1)}

\newcommand{\Set}[1]{\{#1\}}
\newcommand{\Curly}[1]{\{#1\}}
\newcommand{\Tuple}[1]{\left \langle#1 \right \rangle} 

\newcommand{\Constraint}[1]{\textsc{#1}}

\newcommand{\Result}{N}

\newcommand{\MaxAggr}{\mathtt{Max}}
\newcommand{\MinAggr}{\mathtt{Min}}
\newcommand{\SumAggr}{\mathtt{Sum}}

\newcommand{\highlight}[2][yellow!15]{\mathchoice%
  {\colorbox{#1}{$\displaystyle#2$}}%
  {\colorbox{#1}{$\textstyle#2$}}%
  {\colorbox{#1}{$\scriptstyle#2$}}%
  {\colorbox{#1}{$\scriptscriptstyle#2$}}}%

\newcommand{\MaxFeature}{\mathtt{max}}
\newcommand{\MinFeature}{\mathtt{min}}
\newcommand{\One}{\mathtt{one}}
\newcommand{\Surf}{\mathtt{surf}}
\newcommand{\Width}{\mathtt{width}}

\newcommand{\BumpOnDecreasingSequencePatternName}{\mathtt{BumpOnDecreasingSequence}}
\newcommand{\BumpOnDecreasingSequencePattern}{\textnormal{`}>><>>\textnormal{'}}
\newcommand{\BumpOnDecreasingSequencePatternInduced}{\textnormal{`}\highlight{>><>>}\textnormal{'}}
\newcommand{\BumpOnDecreasingSequenceInduced}{\textnormal{`}>><>>\textnormal{'}}
\newcommand{\BumpOnDecreasingSequencePatternWidth}{\textnormal{`}\highlight{>><>>}\textnormal{'}}

\newcommand{\DecreasingPatternName}{\mathtt{Decreasing}}
\newcommand{\DecreasingPattern}{\textnormal{`>'}}
\newcommand{\DecreasingPatternInduced}{\textnormal{`}\highlight{>}\textnormal{'}}
\newcommand{\DecreasingInduced}{\textnormal{`}>\textnormal{'}}
\newcommand{\DecreasingPatternWidth}{\textnormal{`}\highlight{>}\textnormal{'}}

\newcommand{\DecreasingSequencePatternName}{\mathtt{DecreasingSequence}}
\newcommand{\DecreasingSequencePattern}{\textnormal{`(>(>|=)*)*>'}}
\newcommand{\DecreasingSequencePatternInduced}{\textnormal{`}(>(>|=)^*)^*\highlight{>}\textnormal{'}}
\newcommand{\DecreasingSequenceInduced}{\textnormal{`}>\textnormal{'}}
\newcommand{\DecreasingSequencePatternWidth}{\textnormal{`}(>(>|=)^*)^*\highlight{>}\textnormal{'}}

\newcommand{\DecreasingTerracePatternName}{\mathtt{DecreasingTerrace}}
\newcommand{\DecreasingTerracePattern}{\textnormal{`}>=^+>\textnormal{'}}
\newcommand{\DecreasingTerracePatternInduced}{\textnormal{`}\highlight{>=}=^*\highlight{>}\textnormal{'}}
\newcommand{\DecreasingTerraceInduced}{\textnormal{`}>=>\textnormal{'}}
\newcommand{\DecreasingTerracePatternWidth}{\textnormal{`}\highlight{>=}=^*\highlight{>}\textnormal{'}}

\newcommand{\DipOnIncreasingSequencePatternName}{\mathtt{DipOnIncreasingSequence}}
\newcommand{\DipOnIncreasingSequencePattern}{\textnormal{`}<<><<\textnormal{'}}
\newcommand{\DipOnIncreasingSequencePatternInduced}{\textnormal{`}\highlight{<<><<}\textnormal{'}}
\newcommand{\DipOnIncreasingSequenceInduced}{\textnormal{`}<<><<\textnormal{'}}
\newcommand{\DipOnIncreasingSequencePatternWidth}{\textnormal{`}\highlight{<<><<}\textnormal{'}}

\newcommand{\GorgePatternName}{\mathtt{Gorge}}
\newcommand{\GorgePattern}{\textnormal{`(>(>|=)*)*><((<|=)*<)*'}}
\newcommand{\GorgePatternInduced}{\textnormal{`}(>(>|=)^*)^*\highlight{><}((<|=)^*<)^*\textnormal{'}}
\newcommand{\GorgeInduced}{\textnormal{`}><\textnormal{'}}
\newcommand{\GorgePatternWidth}{\textnormal{`}(\highlight{>}|>(>|=)^*>)(\highlight{<}|<(<|=)^*<)\textnormal{'}}

\newcommand{\IncreasingPatternName}{\mathtt{Increasing}}
\newcommand{\IncreasingPattern}{\textnormal{`<'}}
\newcommand{\IncreasingPatternInduced}{\textnormal{`}\highlight{<}\textnormal{'}}
\newcommand{\IncreasingInduced}{\textnormal{`}<\textnormal{'}}
\newcommand{\IncreasingPatternWidth}{\textnormal{`}\highlight{<}\textnormal{'}}

\newcommand{\IncreasingSequencePatternName}{\mathtt{IncreasingSequence}}
\newcommand{\IncreasingSequencePattern}{\textnormal{`(<(<|=)*)*<'}}
\newcommand{\IncreasingSequencePatternInduced}{\textnormal{`}(<(<|=)^*)^*\highlight{<}\textnormal{'}}
\newcommand{\IncreasingSequenceInduced}{\textnormal{`}<\textnormal{'}}
\newcommand{\IncreasingSequencePatternWidth}{\textnormal{`}(<(<|=)^*)^*\highlight{<}\textnormal{'}}

\newcommand{\IncreasingTerracePatternName}{\mathtt{IncreasingTerrace}}
\newcommand{\IncreasingTerracePattern}{\textnormal{`}<=^+<\textnormal{'}}
\newcommand{\IncreasingTerracePatternInduced}{\textnormal{`}\highlight{<=}=^*\highlight{<}\textnormal{'}}
\newcommand{\IncreasingTerraceInduced}{\textnormal{`}<=<\textnormal{'}}
\newcommand{\IncreasingTerracePatternWidth}{\textnormal{`}\highlight{<=}=^*\highlight{<}\textnormal{'}}

\newcommand{\InflexionPattern}{\textnormal{`<(<|=)*>~|~>(>|=)*<'}}
\newcommand{\InflexionPatternName}{\mathtt{Inflexion}}
\newcommand{\InflexionPatternInduced}{\textnormal{`}\highlight{<}(<|=)^*\highlight{>}~|~\highlight{>}(>|=)^*\highlight{<}\textnormal{'}}
\newcommand{\InflexionInduced}{\textnormal{`}<>\textnormal{'},\textnormal{`}><\textnormal{'}}
\newcommand{\InflexionPatternWidth}{\textnormal{`}\highlight{<}(<|=)^*\highlight{>}~|~>(>|=)^*<\textnormal{'}}

\newcommand{\PeakPatternName}{\mathtt{Peak}}
\newcommand{\PeakPattern}{\textnormal{`<(<|=)* (>|=)*>'}}
\newcommand{\PeakPatternInduced}{\textnormal{`}\highlight{<}(<|=)^*(>|=)^*\highlight{>}\textnormal{'}}
\newcommand{\PeakInduced}{\textnormal{`}<>\textnormal{'}}
\newcommand{\PeakPatternWidth}{\textnormal{`}\highlight{<}(<|=)^*(>|=)^*\highlight{>}\textnormal{'}}

\newcommand{\PlainPatternName}{\mathtt{Plain}}
\newcommand{\PlainPattern}{\textnormal{`>=*<'}}
\newcommand{\PlainPatternInduced}{\textnormal{`}\highlight{>}=^*\highlight{<}\textnormal{'}}
\newcommand{\PlainInduced}{\textnormal{`}><\textnormal{'}}
\newcommand{\PlainPatternWidth}{\textnormal{`}\highlight{>}=^*\highlight{<}\textnormal{'}}

\newcommand{\PlateauPatternName}{\mathtt{Plateau}}
\newcommand{\PlateauPattern}{\textnormal{`<=*>'}}
\newcommand{\PlateauPatternInduced}{\textnormal{`}\highlight{<}=^*\highlight{>}\textnormal{'}}
\newcommand{\PlateauInduced}{\textnormal{`}<>\textnormal{'}}
\newcommand{\PlateauPatternWidth}{\textnormal{`}\highlight{<}=^*\highlight{>}\textnormal{'}}

\newcommand{\ProperPlainPatternName}{\mathtt{ProperPlain}}
\newcommand{\ProperPlainPattern}{\textnormal{`}>=^+<\textnormal{'}}
\newcommand{\ProperPlainPatternInduced}{\textnormal{`}\highlight{>=}=^*\highlight{<}\textnormal{'}}
\newcommand{\ProperPlainInduced}{\textnormal{`}>=<\textnormal{'}}
\newcommand{\ProperPlainPatternWidth}{\textnormal{`}\highlight{>=}=^*\highlight{<}\textnormal{'}}

\newcommand{\ProperPlateauPatternName}{\mathtt{ProperPlateau}}
\newcommand{\ProperPlateauPattern}{\textnormal{`}<=^+>\textnormal{'}}
\newcommand{\ProperPlateauPatternInduced}{\textnormal{`}\highlight{<=}=^*\highlight{>}\textnormal{'}}
\newcommand{\ProperPlateauInduced}{\textnormal{`}<=>\textnormal{'}}
\newcommand{\ProperPlateauPatternWidth}{\textnormal{`}\highlight{<=}=^*\highlight{>}\textnormal{'}}

\newcommand{\SteadyPatternName}{\mathtt{Steady}}
\newcommand{\SteadyPattern}{\textnormal{`='}}
\newcommand{\SteadyPatternInduced}{\textnormal{`}\highlight{=}\textnormal{'}}
\newcommand{\SteadyInduced}{\textnormal{`}=\textnormal{'}}
\newcommand{\SteadyPatternWidth}{\textnormal{`}\highlight{=}\textnormal{'}}

\newcommand{\SteadySequencePatternName}{\mathtt{SteadySequence}}
\newcommand{\SteadySequencePattern}{\textnormal{`}=^+\textnormal{'}}
\newcommand{\SteadySequencePatternInduced}{\textnormal{`}\highlight{=}=^*\textnormal{'}}
\newcommand{\SteadySequenceInduced}{\textnormal{`}=\textnormal{'}}
\newcommand{\SteadySequencePatternWidth}{\textnormal{`}\highlight{=}=^*\textnormal{'}}

\newcommand{\StrictlyDecreasingSequencePatternName}{\mathtt{StrictlyDecreasingSequence}}
\newcommand{\StrictlyDecreasingSequencePattern}{\textnormal{`}>^+\textnormal{'}}
\newcommand{\StrictlyDecreasingSequencePatternInduced}{\textnormal{`}\highlight{>}>^*\textnormal{'}}
\newcommand{\StrictlyDecreasingSequenceInduced}{\textnormal{`}>\textnormal{'}}
\newcommand{\StrictlyDecreasingSequencePatternWidth}{\textnormal{`}\highlight{>}>^*\textnormal{'}}

\newcommand{\StrictlyIncreasingSequencePatternName}{\mathtt{StrictlyIncreasingSequence}}
\newcommand{\StrictlyIncreasingSequencePattern}{\textnormal{`}<^+\textnormal{'}}
\newcommand{\StrictlyIncreasingSequencePatternInduced}{\textnormal{`}\highlight{<}<^*\textnormal{'}}
\newcommand{\StrictlyIncreasingSequenceInduced}{\textnormal{`}<\textnormal{'}}
\newcommand{\StrictlyIncreasingSequencePatternWidth}{\textnormal{`}\highlight{<}<^*\textnormal{'}}

\newcommand{\SummitPatternName}{\mathtt{Summit}}
\newcommand{\SummitPattern}{\textnormal{`(<(<|=)*)*<>((>|=)*>)*'}}
\newcommand{\SummitPatternInduced}{\textnormal{`}(<(<|=)^*)^*\highlight{<>}((>|=)^*>)^*\textnormal{'}}
\newcommand{\SummitInduced}{\textnormal{`}<>\textnormal{'}}
\newcommand{\SummitPatternWidth}{\textnormal{`}(\highlight{<}|<(<|=)^*<)(\highlight{>}|>(>|=)^*>)\textnormal{'}}

\newcommand{\ValleyPatternName}{\mathtt{Valley}}
\newcommand{\ValleyPattern}{\textnormal{`>(>|=)* (<|=)*<'}}
\newcommand{\ValleyPatternInduced}{\textnormal{`}\highlight{>}(>|=)^*(<|=)^*\highlight{<}\textnormal{'}}
\newcommand{\ValleyInduced}{\textnormal{`}><\textnormal{'}}
\newcommand{\ValleyPatternWidth}{\textnormal{`}\highlight{>}(>|=)^*(<|=)^*\highlight{<}\textnormal{'}}

\newcommand{\ZigzagPatternName}{\mathtt{Zigzag}}
\newcommand{\ZigzagPattern}{\textnormal{`}(<>)^+<(>|\varepsilon)~|~(><)^+>(<|\varepsilon)\textnormal{'}}
\newcommand{\ZigzagPatternInduced}{\textnormal{`}(<>)^*\highlight{<><}(>|\varepsilon)~|~(><)^*\highlight{><>}(<|\varepsilon)\textnormal{'}}
\newcommand{\ZigzagInduced}{\textnormal{`}<><\textnormal{'},\textnormal{`}><>\textnormal{'}}
\newcommand{\ZigzagPatternWidth}{\textnormal{`}(<>)^*\highlight{<><}(>|\varepsilon)~|~(><)^*\highlight{><>}(<|\varepsilon)\textnormal{'}}

\newcommand{\After}{a}
\newcommand{\Before}{b}
\newcommand{\pattern}{\sigma} 
\newcommand{\reg}[1]{\text{\mbox{`$#1$'}}} 
\newcommand{\RegVerb}[1]{\verb/{#1}/}
\newcommand{\RegExp}{r}

\newcommand{\FullExtLinCoeff}{e}
\newcommand{\FullExtFreeCoeff}{c}

\newcommand{\NbOcc}{\tau}
\newcommand{\nbocc}{\mathit{P}}

\newcommand{\plength}{m}
\newcommand{\superposition}{superposition}

\newcommand{\SeqRest}{\seqlength_r}

\newcommand{\Char}[2]{#1_{#2}}
\newcommand{\CharPar}[3]{#1_{#2}^{\Tuple{#3}}}
\newcommand{\CharArg}[3]{#1_{#2}(#3)}
\newcommand{\CharAll}[4]{#1_{#2}^{\Tuple{#3}}(#4)}

\newcommand{\CharArgUp}[3]{\overline{#1}_{#2}(#3)}
\newcommand{\CharAllUp}[4]{\overline{#1}_{#2}^{\Tuple{#3}}(#4)}

\newcommand{\Intro}[1]{Consider a regular expression~$#1$}
\newcommand{\IntroDom}[2]{\Intro{#1} and an
  integer interval domain~$#2$}
\newcommand{\IntroLength}[2]{\Intro{#1} and a time series length~$#2$}

\newcommand{\DefinitionArg}[7]{The~\emph{#1} of~#2 wrt~$\Tuple{#3}$,
  denoted by~$#4$, is a function that maps an element of~$#5$ to~$#6$#7.}
\newcommand{\Definition}[6]{The~\emph{#1} of~#2,
  denoted by~$#3$, is a function that maps an element of~$#4$
  to~$#5$#6.}

\newcommand{\width}{\omega}

\newcommand{\Height}{\eta}

\newcommand{\Overlap}{o}

\newcommand{\VariationOfMax}{\delta} 

\newcommand{\SetSuperposition}{\Gamma}

\newcommand{\IndWordSet}{\Theta}
\newcommand{\Range}{\phi}
\newcommand{\CompleteExtension}{\phi} 
\newcommand{\SupportSet}{\Omega}
\newcommand{\Shift}{\nu}
\newcommand{\Rest}{\rho}

\newcommand{\Language}[1]{\ensuremath{\mathcal{L}_{#1}}}
\newcommand{\word}{w}

\newcommand{\wordz}{z}
\newcommand{\wordq}{q_\mathit{opt}}
\newcommand{\wordb}{\tilde{q}_\mathit{opt}}

\newcommand{\wordu}{\upsilon}
\newcommand{\wordv}{v}

\newcommand{\WordLength}{k}

\newcommand{\alphabet}{\mathcal{A}}
\newcommand{\Alphabet}{\Sigma}

\newcommand{\Domain}{[\DomainMin, \DomainMax]}
\newcommand{\DomTuple}{\Tuple{\DomainMin, \DomainMax}}
\newcommand{\DomainMin}{\ell}
\newcommand{\DomainMax}{u}
\newcommand{\TimeSeries}{t}
\newcommand{\seqlength}{n}

\newcommand{\XSeq}{\Tuple{X_1, X_2, \dots, X_\seqlength}}
\newcommand{\NbPat}{p}
\newcommand{\MinDiff}{d}

\newcommand{\sequence}[1]{\langle #1 \rangle}
\newcommand{\abs}[1]{|#1|}
\newcommand{\Parentheses}[1]{ \left( #1 \right )}
\newcommand{\Frac}[2]{\left \lfloor \frac{#1}{#2} \right \rfloor}
\newcommand{\ekaterina}[1]{#1}

\newcommand{\NbWidthLowerSimple}{$\underline{\textsc{nb}}$\nobreakdash-simple}
\newcommand{\NbUpperOverlap}{$\overline{\textsc{nb}}$\nobreakdash-overlap}
\newcommand{\NbUpperNoOverlap}{$\overline{\textsc{nb}}$\nobreakdash-no\nobreakdash-overlap}
\newcommand{\WidthUpper}{$\overline{\textsc{width}}$\nobreakdash-max}
\newcommand{\WidthUpperSum}{$\overline{\textsc{width}}$\nobreakdash-sum}
\newcommand{\NecessaryCondition}{necessary-and-sufficient condition}
\newcommand{\MinWidth}{$\underline{\textsc{width}}$\nobreakdash-occurrence}
\newcommand{\MaxMin}{Condition of~Theorem~1 in~\cite{CP16}}

\newcommand\xqed[1]{%
  \leavevmode\unskip\penalty9999 \hbox{}\nobreak\hfill
  \quad\hbox{#1}}
\newcommand\qedexample{\xqed{\scriptsize$\triangle$}}
\newcommand{\Wlog}{w.l.o.g.}
\newcommand{\Resp}{respectively}
\newcommand{\If}{\textnormal{~if~}}
\newcommand{\Otherwise}{\textnormal{~otherwise}}

\newcommand{\AndText}{~\textnormal{and}~}

\newcommand{\Property}{Prop.}

\newcommand{\Disj}{\hspace*{2pt}|\hspace*{2pt}}

\newcounter{ItemCounter}

\newcommand{\shiftedqed}{\smartqed \qed}

\definecolor{MySmokenGrey}{cmyk}{0, 0, 0, 0.17}

\newcommand\PlanStyle{%
   \renewcommand{\labelitemi}{$\diamond$}
  \renewcommand{\labelitemii}{$\ast$}}

\newcommand\ListStyle{%
   \renewcommand{\labelitemi}{$\circ$}}

\newcommand\ExFullStyle{%
   \renewcommand{\labelitemi}{$\bullet$}
  \renewcommand{\labelitemii}{$\ast$}
\renewcommand{\labelitemiii}{$\circ$}}

\newcommand\ExStyle{%
  \renewcommand{\labelitemi}{$\ast$}
\renewcommand{\labelitemii}{$\circ$}}

\makeatletter
\newcommand{\setword}[2]{%
  \phantomsection
  #1\def\@currentlabel{\unexpanded{#1}}\label{#2}%
}
\makeatother

\newif\ifpaper
\papertrue

\newif\ifthesis
\thesisfalse

\begin{document}

\author{
   Ekaterina~Arafailova 
  \and Nicolas~Beldiceanu 
  \and Helmut~Simonis
  }

\title{Deriving Generic Bounds for Time-Series Constraints \\
       Based on Regular Expressions Characteristics\thanks{This is an extended version of parts
       of the CP~2016 paper~\cite{CP16} which involves a subset of the original authors.
       This paper introduces $6$ new characteristics of regular expressions
       to express generic bounds on time-series constraints, which
       were not discussed in the original paper~\cite{CP16}.
       Ekaterina Arafailova is supported by the EU H2020 programme under grant 640954 for project GRACeFUL.
       Nicolas~Beldiceanu is partially supported by the GRACeFUL project and
       by the Gaspard Monge Program for Optimization and Operations Research (PGMO). Helmut Simonis is supported by Science Foundation Ireland (SFI) under
grant SFI/10/IN.1/I3032; the Insight Centre for Data Analytics is supported by
SFI under grant SFI/12/RC/2289. \\
}}
   \institute{E. Arafailova \and N. Beldiceanu \at TASC (LS2N) IMT
     Atlantique, FR -- 44307 Nantes, France \\
     \email{\{Ekaterina.Arafailova,Nicolas.Beldiceanu\}@imt-atlantique.fr
     }
\and H. Simonis \at Insight Centre for Data Analytics,
     University College Cork, Cork, Ireland  \\
     \email{Helmut.Simonis@insight-centre.org}
   }

\maketitle

\begin{abstract}
We introduce the concept of regular expression characteristics as a unified way to concisely express bounds on time-series constraints. This allows us not only to define time-series constraints in a compositional way, but also to deal with their combinatorial aspect in a compositional way, without developing ad-hoc bounds for each time-series constraint separately.
\end{abstract}

\section{Introduction}
\label{sec:intro}

A \emph{time series} is here a sequence of integers, corresponding to
measurements taken over a time interval.
Time series are common in many application areas,
such as the output of electric power stations over multiple
days~\cite{beldiceanu:EDF:2013}, or the manpower required in a call-centre~\cite{ASTRA:CPAIOR16},
or the daily capacity of a hospital clinic over a period of years.
Time series are constrained by physical or organisational
limits, which restrict the evolution of the series.

We showed in~\cite{Beldiceanu:synthesis} that many constraints
$\gamma(\sequence{X_1,X_2,\dots,X_n},\Result)$ on an unknown time series
$X = \sequence{X_1,X_2,\dots,X_n}$ can be specified in a \emph{compositional way}
by a triple $\Tuple{\pattern,f,g}$, where
$\pattern$ is a regular expression over the alphabet $ \Sigma=\Set{\reg{<},\reg{=},\reg{>}}$
(we assume the reader is familiar with regular expressions~\cite{Hopcroft:automata}),
while $f \in \Set{\MaxFeature, \MinFeature, \One, \Surf, \Width}$ is called a \emph{feature function}, and
$g \in \Set{\MaxAggr, \MinAggr, \SumAggr}$ is called an \emph{aggregator function}.
Volume~II of the global constraint catalogue~\cite{ArafailovaBeldiceanuDouenceCarlssonFlenerRodriguezPearsonSimonis16}
contains $266$ such functional time-series constraints.

It is currently unknown in general, how to maintain efficiently domain consistency for such time\nobreakdash-series constraints.
Computing bounds on the result variable $\Result$ of a time-series constraint
is a way to handle the combinatorial aspect and thus improve propagation.
Since we have too many time-series constraints deriving such bounds needs to be done in a systematic way.
Motivated by this, we sketched in~\cite{CP16} a methodology to obtain such bounds
and illustrated it only for time-series constraints when $g=\MaxAggr$ and $f=\MinFeature$.

The contribution of this paper, which makes explicit the approach sketched in~\cite{CP16},
is to introduce the notion of \emph{regular expression characteristics} that provides a unified way
to concisely express bounds on the result variable~$\Result$ of a time-series constraint. Six regular expression characteristics
are introduced, which allows coming up in a compositional way with bounds when
$\Tuple{g,f}\in\Set{\Tuple{\SumAggr,\One},\Tuple{\MaxAggr,\Width},$ $\Tuple{\MinAggr,\Width},\Tuple{\SumAggr,\Width}}$:
five main theorems (see~Theorems~\ref{th:nb-patterns-lower-bound}, \ref{th:nb-patterns},
\ref{th:max-width}, \ref{th:sum-width}, and \ref{th:min-width}) allow
obtaining $95$ bounds implemented in Volume~II of the global constraint catalogue~\cite{ArafailovaBeldiceanuDouenceCarlssonFlenerRodriguezPearsonSimonis16}.
When the time-series variables $\sequence{X_1,\dots,X_n}$ have the same interval integer domain,
these bounds are sharp for all the $22$ regular expression of~\cite{ArafailovaBeldiceanuDouenceCarlssonFlenerRodriguezPearsonSimonis16}.
We now put in perspective with existing work the contribution of this paper.

Going back to the work of Sch\"utzenberger~\cite{Schutzenberger61},
\emph{regular cost functions} are quantitative extensions of regular languages
that correspond to a function mapping a word to an integer value or infinity (QRE for short).
Recently there was a rise of interest in this area, both from a theoretical perspective~\cite{ColcombetDaviaud16}
with max\nobreakdash-plus automata, and from a practical point of view with the synthesis of cost register
automata~\cite{AlurAntoniDeshmukhRaghothamanYuan13} for data streams~\cite{AlurFismanRaghothaman16}.
Within constraint programming automata constraints were introduced in~\cite{Pesant:seqs}
and in~\cite{Beldiceanu:automata,CostRegular}, the later also computing an integer value from a word.
More recently, the work on synthesising automata from transducers~\cite{Beldiceanu:synthesis} for identifying
all maximal occurrences of a pattern in a time series is part of the QRE line of research.
While most previous mentioned works give quantitative extensions of regular languages as their general motivation,
to the best of our knowledge none of them introduced the concept of regular expression characteristics, which is the key abstraction proposed here.
The paper is structured in the following way:
\begin{itemize}
\PlanStyle
\item
  In~\secref{sec:background}, we recall the background both on regular expressions~\cite{Hopcroft:automata},
  and on the way of describing time-series constraints in a compositional way~\cite{Beldiceanu:synthesis}.
\item
  In~\secref{sec:characteristics}, we first introduce a notation system for denoting regular expression characteristics.
  Then we present six regular expressions characteristics, which characterise different quantitative aspects
  of regular expressions useful for capturing some of their combinatorial flavour.
  Finally, based on two of these characteristics, we provide a necessary condition for the occurrence of a regular
  expression in a time-series.
\item
  In~\secref{sec:nb-patterns}, we show how to obtain generic bounds
  for time-series constraints 
  whose result variable is constrained to be the number of occurrences of a regular expression in a time-series,
  i.e.,~time-series constraints where $g=\SumAggr$ and $f=\One$.
\item
  In~\secref{sec:width-constraints}, we show how to obtain generic bounds
  for the result variables of time-series constraints for which the feature $f$ is~$\Width$,
  and the aggregator $g$ is in~$\Curly{\MaxAggr, \MinAggr, \SumAggr}$.
\item
  In~\secref{sec:synthesis}, we synthesise all the results on bounds we have so far from the
  CP~paper~\cite{CP16}, and from Sections~\ref{sec:nb-patterns} and \ref{sec:width-constraints}
  of this paper: for each bound we recall
  (1)~the regular expression characteristics it uses,
  (2)~the generic theorem it comes from, and
  (3)~the properties under which the bound is sharp.
 \item
  We evaluate in \secref{sec:evaluation} the beneficial propagation impact of the derived bounds.
\end{itemize}

\section{Background}
\label{sec:background}

First we give the necessary background on word and regular languages.
Then we recall the time\nobreakdash-series constraints introduced in~\cite{Beldiceanu:synthesis}.

\subsection{Regular Languages \label{sec:reg-definitions}}

An \emph{alphabet} $\alphabet$ is a finite set of symbols and
a symbol of~$\alphabet$ is called a~\emph{letter}.
A \emph{word} on $\alphabet$ is a finite sequence of symbols
belonging to $\alphabet$. 
The empty word is denoted by $\varepsilon$. 
The \emph{length} of a word $\word$ is the number of letters in
$\word$  and is denoted by $|\word|$.
For $i \in [1,|\word|]$, $\word[i]$ denotes the letter~$i$ of a word~$\word$.
The concatenation of two words is denoted by putting them side by side,
with an implicit infix operator between them.
A word $\word$ is a \emph{factor} of a word $x$ if there exist two words
$\wordv$ and $\wordz$ such that $x = \wordv \word \wordz$;
when $\wordv = \varepsilon$, $\word$ is a \emph{prefix} of $x$, 
when $\wordz = \varepsilon$, $\word$ is a \emph{suffix} of $x$.  
If both~$\word$ is not empty and different from~$x$, then it is
a~\emph{proper} factor of~$x$.
Given a word $\word$ and a positive integer $k>0$, $\word^k$ denotes
 the \emph{concatenation of $k$ occurrences} of $\word$.
 Given an integer $k$ and a language $\mathcal{L}$, $\mathcal{L}^k$ is defined by
 $\mathcal{L}^0=\{\epsilon\}$, $\mathcal{L}^1=\mathcal{L}$ and $\mathcal{L}^k=\mathcal{L}\cdot\mathcal{L}^{k-1}$ where $\reg{\cdot}$ is the concatenation operator.
 Then the Kleene closure of $\mathcal{L}$ is defined by $\cup_{n\geq 0}\mathcal{L}^n$ and denoted by $\mathcal{L}^*$.

\begin{definition}
  A regular expression~\cite{Crochemore} $r$ on an alphabet~$\alphabet$ and the language $\Language{\RegExp}$
  it describes, the regular language, are recursively defined as follows:
  \begin{enumerate}[(1)]
  \item
    $0$ and $1$ are regular expressions that respectively describe
    $\emptyset$ (the empty set) and~$\{\varepsilon\}$.
  \item
    For every letter~$\ell$ of~$\alphabet$,
    $\ell$ is a regular expression that
    describes the singleton $\Curly{\ell}$.
  \item
   If $r_1$ and $r_2$ are regular expressions,
    respectively describing the regular languages $\Language{r_1}$ and $\Language{r_2}$,
    then $r_1 + r_2$, $r_1*r_2$ and $r_1^*$ are regular expressions
    that respectively describe the regular languages $\Language{r_1}
    \cup \Language{r_2}$, $\Language{r_1} \cap \Language{r_2}$,
    and~$\Language{r_1}^*$.
  \end{enumerate}
\end{definition}

\begin{example}
  \ExFullStyle
  Consider the alphabet~$\Sigma = \Curly{\reg{<}, \reg{=}, \reg{>}}$.
  \begin{itemize}
    \item
      $\DecreasingPatternName = \DecreasingPattern$ is a regular
      expression on~$\Sigma$.
      The word~$\wordv = \reg{>}$ is a word of length~$1$ on~$\Sigma$
      that belongs to~$\Language{\DecreasingPatternName}$, and it does
      not have any proper factors.
      The word~$\reg{>>}$ is a word of length~$2$ on~$\Sigma$,
      which does not belong to~$\Language{\DecreasingPatternName}$.
    \item
      $\InflexionPatternName = \InflexionPattern$ is a regular
      expression on~$\Sigma$.
      The word~$\wordv = \reg{>=<}$ is a word of length~$3$ on~$\Sigma$
      that belongs to~$\Language{\InflexionPatternName}$.
      The word~$\wordu$ has multiple proper factors,
      e.g.,~$\reg{>}$,~$\reg{<}$.
      The word~$\reg{>=<<}$ does not belong to~$\Language{\InflexionPatternName}$. 
\qedexample
  \end{itemize}
\end{example}

\begin{definition}
\label{def:non-fixed-length} 
A regular expression~$\RegExp$ is a~\emph{non-fixed length regular expression}
if not all words of~$\Language{\RegExp}$ have the same length.
\end{definition}

\begin{example}
  \ExFullStyle
\label{ex:non-fixed-length}
  We give two examples of regular expressions,
  a first one with a fixed length and a second one with a non-fixed length.
  \begin{itemize}
    \item
      The~$\DecreasingPatternName = \DecreasingPattern$ regular expression has a fixed length
      since~$\Language{\DecreasingPatternName}$ contains a single word.
    \item
      The~$\InflexionPatternName = \InflexionPattern$ regular expression
      does not have a fixed length since $\Language{\InflexionPatternName}$ contains words of different length.
      \qedexample
  \end{itemize}
\end{example}

\begin{definition}
  \label{def:disjunction-capsuled}
  A regular expression over an alphabet~$\alphabet$ is~\emph{disjunction-capsuled}
  if it is in the form of~`$r_1r_2\dots r_p$', where every~$r_i$ (with $i \in [1,p]$) is,
  either a letter of the alphabet~$\alphabet$, or a regular expression whose regular language contains
  the empty word.
\end{definition}

Note that~\defref{def:disjunction-capsuled} is a slight extension of
a similar notion introduced in~\cite{DBPL}.

\begin{example}
\label{ex:disjunction-capsuled}
\tabref{tab:patterns} recalls the~$22$ regular expressions used for
describing time-series
constraints in~\cite{ArafailovaBeldiceanuDouenceCarlssonFlenerRodriguezPearsonSimonis16,Beldiceanu:synthesis}.
Every regular expression~$\pattern$ in column~2 of~\tabref{tab:patterns} is
in the form of~$\pattern =\pattern_1|\pattern_2|\dots|\pattern_t$ with~$t\geq 1$,
and every~$\pattern_i$ (with $i \in [1,t]$) is a disjunction-capsuled regular expression.
Then~$\Language{\pattern}$ is the union of the~$\Language{\pattern_i}$
(with $i\in[1,t]$).

The~$\reg{(>|>(>|=)^*>) (<|<(<|=)^*<)}$ regular expression
has the same regular language as $\GorgePatternName$, but is not
disjunction-capsuled.
\qedexample
\end{example}

\subsection{Time-Series Constraints}
\label{sec:time-series-constraints}

A time series here is a sequence of integers corresponding to measurements taken over the time.
We showed in~\cite{Beldiceanu:synthesis} that many constraints
$\gamma(\Tuple{X_1,X_2,\dots,X_\seqlength},\Result)$ on an unknown time series
$\Tuple{X_1,X_2,\dots,X_\seqlength}$ of given length $\seqlength$,
where every~$X_i$ is an \emph{integer variable}, can be specified by
a triple $\Tuple{\pattern,f,g}$, where
$\pattern$ is a regular expression on the alphabet
$\Sigma=\Set{\reg{<},\reg{=},\reg{>}}$ that is characterised by
two integer constants~$\Char{\After}{\pattern}$ and~$\Char{\Before}{\pattern}$,
whose role is to trim the left and right borders of the regular expression,
while $f \in \Set{\MaxFeature, \MinFeature, \One, \Surf, \Width}$ is called a \emph{feature}, and
$g \in \Set{\MaxAggr, \MinAggr, \SumAggr}$ is called an
\emph{aggregator}.
Let the sequence $S = \Tuple{S_1,S_2,\dots,S_{\seqlength-1}}$,
called the \emph{signature} and containing \emph{signature variables},
be linked to the sequence $X$ via the \emph{signature constraints}
$(X_i<X_{i+1}\Leftrightarrow S_i=\reg{<})$
$\land~(X_i=X_{i+1}\Leftrightarrow S_i=\reg{=})$
$\land~(X_i>X_{i+1}\Leftrightarrow S_i=\reg{>})$
for all $i \in [1,\seqlength-1]$.
If a sub-signature~$\Tuple{S_i,S_{i+1},\dots,S_j}$ is a maximal word matching~$\pattern$ in the signature of~$X$,
then~the subseries~$\Tuple{X_{i+\Char{\Before}{\pattern}},X_{i+\Char{\Before}{\pattern}+1},\dots,X_{j+1-\Char{\After}{\pattern}}}$ is called a
$\pattern$-\emph{pattern} and the subseries~$\Tuple{X_{i},X_{i+1},\dots,X_{j+1}}$ is called
an~\emph{extended~$\pattern$-pattern}. 
Integer variable~$\Result$ is constrained to be the aggregation,
computed using~$g$, of the list of values of feature~$f$ for all
$\pattern$-patterns in~$\XSeq$.
We name a time-series constraint specified by $\Tuple{\pattern,f,g}$
as $g\_f\_\pattern$. 
\ekaterina{A time series is \emph{maximal} for~$g\_f\_\pattern(\Tuple{X_1,X_2,\dots,X_\seqlength},\Result)$ if
it yields the maximum value of~$\Result$
among all time series of length~$\seqlength$ that have the same initial
domains for the time-series variables.} \\

\begin{table*}[!h]\centering

\setlength{\tabcolsep}{6pt}
\begin{tabular}{@{}llll@{}} \toprule
  name~$\pattern$                 		
  & regular expression 
  & $a_\pattern$ 
  & $b_\pattern$ \\\midrule
  $\BumpOnDecreasingSequencePatternName$   & $\BumpOnDecreasingSequencePattern$   & 1 & 2 \\
  $\DecreasingPatternName$                 & $\DecreasingPattern$                 & 0 & 0 \\
  $\DecreasingSequencePatternName$         & $\DecreasingSequencePattern$         & 0 & 0 \\
  $\DecreasingTerracePatternName$          & $\DecreasingTerracePattern$          & 1 & 1 \\
  $\DipOnIncreasingSequencePatternName$    & $\DipOnIncreasingSequencePattern$    & 1 & 2 \\
  $\GorgePatternName$                      & $\GorgePattern$                      & 1 & 1 \\
  $\IncreasingPatternName$                 & $\IncreasingPattern$                 & 0 & 0 \\
  $\IncreasingSequencePatternName$         & $\IncreasingSequencePattern$         & 0 & 0 \\
  $\IncreasingTerracePatternName$          & $\IncreasingTerracePattern$          & 1 & 1 \\
  $\InflexionPatternName$                  & $\InflexionPattern$                  & 1 & 1 \\
  $\PeakPatternName$                       & $\PeakPattern$                       & 1 & 1 \\
  $\PlainPatternName$                      & $\PlainPattern$                      & 1 & 1 \\
  $\PlateauPatternName$                    & $\PlateauPattern$                    & 1 & 1 \\
  $\ProperPlainPatternName$                & $\ProperPlainPattern$                & 1 & 1 \\
  $\ProperPlateauPatternName$              & $\ProperPlateauPattern$              & 1 & 1 \\
  $\SteadyPatternName$                     & $\SteadyPattern$                     & 0 & 0 \\
  $\SteadySequencePatternName$             & $\SteadySequencePattern$             & 0 & 0 \\
  $\StrictlyDecreasingSequencePatternName$ & $\StrictlyDecreasingSequencePattern$ & 0 & 0 \\
  $\StrictlyIncreasingSequencePatternName$ & $\StrictlyIncreasingSequencePattern$ & 0 & 0 \\
  $\SummitPatternName$                     & $\SummitPattern$                     & 1 & 1 \\
  $\ValleyPatternName$                     & $\ValleyPattern$                     & 1 & 1 \\
  $\ZigzagPatternName$                     & $\ZigzagPattern$                     & 1 & 1 \\
\bottomrule
\end{tabular}
\caption{\label{tab:patterns} Regular expression names $\pattern$,
  corresponding regular expressions, values of the parameters~$a_\pattern$ and~$b_\pattern$}
\end{table*}

We consider the set of $22$ regular expressions
used in~\cite{ArafailovaBeldiceanuDouenceCarlssonFlenerRodriguezPearsonSimonis16},
which is given in~\tabref{tab:patterns}.
Most of these regular expressions capture topological patterns that one wants to control when generating time-series,
while some of them, like $\ZigzagPatternName$, correspond to abnormal situations one wants to catch from existing time-series.
Within a $\pattern$-\emph{pattern} the two integer constants~$\Char{\Before}{\pattern}$ and~$\Char{\After}{\pattern}$
trim respectively the left and right borders of the interval $[i,j+1]$ to the leftmost and rightmost variable of $\XSeq$ used to compute the corresponding feature:
for example for $\IncreasingTerracePatternName=\IncreasingTerracePattern$,
since $\Before_\IncreasingTerracePatternName=\After_\IncreasingTerracePatternName=1$,
we only consider the $X_i$ that are involved in an equality,
i.e.,~the $X_i$ of the flat part of the terrace.
\begin{figure}[t]
\centering
\begin{tikzpicture}
\begin{scope}[scale=0.5,yshift=-2.8cm]
\draw[fill=black!6,draw=black!6,rounded corners=2pt] (4,-0.2) -- (4,3/2.5) -- (5,3/2.5) -- (5,5/2.5) -- (7,5/2.5) -- (7,8/2.5) -- (8,8/2.5) -- (8,5/2.5) -- (9,5/2.5) -- (9,-0.2) -- cycle;
\draw[fill=black!6,draw=black!6,rounded corners=2pt] (11,-0.2) -- (11,3/2.5) -- (17,3/2.5) -- (17,-0.2) -- cycle;
\draw[draw=black,line width=0.75pt,rounded corners=2pt] (0,5/2.5) -- (2,5/2.5) -- (2,1/2.5) -- (4,1/2.5) -- (4,3/2.5) -- (5,3/2.5) -- (5,5/2.5) -- (7,5/2.5) -- (7,8/2.5) -- (8,8/2.5) -- (8,5/2.5) -- (9,5/2.5) -- (9,1/2.5) -- (11,1/2.5) -- (11,3/2.5) -- (17,3/2.5) -- (17,1/2.5) -- (18,1/2.5);
\node at (0.5,5/2.5) [below=0] {\scriptsize{$4$}};
\node at (1.5,5/2.5) [below=0] {\scriptsize{$4$}};
\node at (2.5,1/2.5) [below=0] {\scriptsize{$0$}};
\node at (3.5,1/2.5) [below=0] {\scriptsize{$0$}};
\node at (4.5,3/2.5) [below=0] {\scriptsize{$2$}};
\node at (5.5,5/2.5) [below=0] {\scriptsize{$4$}};
\node at (6.5,5/2.5) [below=0] {\scriptsize{$4$}};
\node at (7.5,8/2.5) [below=0] {\scriptsize{$7$}};
\node at (8.5,5/2.5) [below=0] {\scriptsize{$4$}};
\node at (9.5,1/2.5) [below=0] {\scriptsize{$0$}};
\node at (10.5,1/2.5) [below=0] {\scriptsize{$0$}};
\node at (11.5,3/2.5) [below=0] {\scriptsize{$2$}};
\node at (12.5,3/2.5) [below=0] {\scriptsize{$2$}};
\node at (13.5,3/2.5) [below=0] {\scriptsize{$2$}};
\node at (14.5,3/2.5) [below=0] {\scriptsize{$2$}};
\node at (15.5,3/2.5) [below=0] {\scriptsize{$2$}};
\node at (16.5,3/2.5) [below=0] {\scriptsize{$2$}};
\node at (17.5,1/2.5) [below=0] {\scriptsize{$0$}};
\draw[<->,color=black] (4,-0.2) -- (9,-0.2);
\draw[<->,color=black] (11,-0.2) -- (17,-0.2);
\node at (6.5,0.05) {\scriptsize$\mathbf{5}$};
\node at (14,0.05) {\scriptsize{$\mathbf{6}$}};
\end{scope}
\end{tikzpicture}
\caption{\label{fig:peak-time-series}
  Illustrating the $\Constraint{min\_width\_peak}(5,
  \sequence{4,4,0,0,2,4,4,7,4,0,0,2,2,2,2,2,2,0})$ time-series constraint with its two peaks of respective width $5$ and $6$} 
\end{figure}

\begin{example}
  \label{ex:intro_peak}
\sloppypar{
  Consider the time series $X = \Tuple{4,4,0,0,2,4,4,7,4,0,0,2,2,2,2,2,2,0}$
  with its signature $\reg{=>=<<=<>>=<=====>}$ and the
  regular expression $\PeakPatternName = \PeakPattern$
  with~$\Char{\Before}{\PeakPatternName}$ and~$\Char{\After}{\PeakPatternName}$ being both
  equal to~$1$:
  a $\PeakPatternName$-pattern, called a \emph{peak}, within a time
  series corresponds, except for its first and last elements, to a
  maximal occurrence of $\PeakPatternName$ in the signature, and the
  $\Width$ feature value of a peak is its number of elements.
  The time series~$X$ contains two peaks, namely
  $\Tuple{2,4,4,7,4}$ and $\Tuple{2,2,2,2,2,2}$, visible the way
  $X$ is plotted in \figref{fig:peak-time-series}, of widths~$5$
  and~$6$ respectively, hence the minimal-width peak, obtained by
  using the aggregator~$\MinAggr$, has width~$\Result = 5$:
  the corresponding constraint is named
  $\Constraint{min\_width\_peak}$.}
\qedexample
\end{example}

\section{Regular Expressions Characteristics}
\label{sec:characteristics}

To get parametrised bounds, this section introduces
regular expressions characteristics used for deriving
sharp lower and upper bounds on the result variable of a time series
constraint when the feature is in~$\Curly{\One,\Width}$.
For all characteristics we use a notation system, which is
described in~\secref{sec:notation}.
We introduce the following characteristics:
\begin{itemize}
\ListStyle
\item
The \emph{width} of a regular expression in \secref{sec:width}.
\item
The \emph{height} of a regular expression in \secref{sec:height}.
\item
The \emph{range} of a regular expression wrt a time series length in \secref{sec:range}.
\item
The \emph{set of inducing words} of a regular expression in \secref{sec:set-of-inducing-words}.
\item
The \emph{overlap} of a regular expression wrt an integer interval domain in \secref{sec:overlap}.
\item
The \emph{smallest variation of maxima} of a regular expression wrt an integer interval domain in \secref{sec:variation-of-maxima}.
\end{itemize}

\secref{sec:summary-example} presents a summary example combining
all the introduced regular expressions characteristics
for the~$\DecreasingTerracePatternName$ regular expression.
\secref{sec:necessary-and-sufficient} introduces a necessary and sufficient
condition for the existence of at least one occurrence of a regular expression
within the signature of a time series under some hypothesis on the
domain of time-series variables.
%
\tabref{tab:characteristics} provides for each of the $22$ regular expressions in~\tabref{tab:patterns}
the corresponding value of each regular expression characteristics.

\subsection{A Notation System for Regular Expression Characteristics}
\label{sec:notation}

We introduce a notation system for naming the characteristics of regular expressions.
A regular expression characteristic~$C$ is a function, denoted
by~$C_R^P(V)$, whose arguments are~$R$,~$P$, and~$V$ explained below:

\begin{itemize}
\ListStyle
\item
  $R$ is a regular expression over~$\Sigma=\Curly{<, =, >}$.
\item
  $P$ is a subset, possible empty, drawn from~$\Curly{\DomainMin, \DomainMax, \seqlength}$,
  where~$\Domain$ is the domain of the variables of a time series,
  and~$\seqlength$ is the length of a given time series.
\item
  $V$ is a vector of additional arguments, which are different from~$R$,
  $\DomainMin$, $\DomainMax$, and $\seqlength$.
  If~$V$ is empty, then we simply write~$C_R^P$.
  Quite often these additional arguments correspond to words in $\Language{R}$
  since a characteristics $C_R^P$ will be defined in terms of an other characteristic $C_R^P(V)$:
  for instance the height of a regular expression $R$
  will be defined in terms of the heights of words in $\Language{R}$.
\end{itemize}

The domain of the function~$C_R^P(V)$ is the cartesian product of the
following elements in the given order:
\begin{itemize}
\ListStyle
\item 
 The domain of~$R$, namely~$\SetRegExpressions$, which is the set of regular
 expressions over~$\Sigma$.
\item The cartesian product of the domains of the elements
  of~$P$, if~$P$ is not empty.
\item
  The cartesian product of the domains of the arguments of~$V$, if~$V$ is not empty.
\end{itemize}

The font used for the symbol~`$C$' depends on the type of values returned by~$C_R^P(V)$:
\begin{itemize}
\ListStyle
\setlength\itemsep{0.1cm}
\item
  If the codomain of~$C_R^P(V)$ is~$\Integers$, then `$C$' is a lower-case Greek letter, e.g.,~$\delta$.
\item
  If the codomain of~$C_R^P(V)$ is the power set of some set, then `$C$' is an upper-case
  Greek letter, e.g.,~$\Delta$.
\end{itemize}

Some characteristics are associated with,
either the lower or the upper bound on the value of the result variable of a time-series constraint.
In this case, the ones associated with the upper (respectively lower)  bound are denoted by~$\overline{C}_R^P(V)$
(respectively $\underline{C}_R^P(V)$).

\subsection{Width}
\label{sec:width}

This section introduces the \emph{width} characteristic of a regular
expression; it will be used in~\thref{th:nb-patterns} for computing 
the sharp upper bound on the number of
occurrences of the regular expression within the signature of a time
series.
This characteristics is also used for defining a necessary and
sufficient condition, see~\propref{prop:necessary-condition}, for
the existence of at least one occurrence of a regular expression
within the signature of a time series over an integer interval domain.

\begin{definition}
  \label{def:width}
  \Intro{\pattern}.
  \Definition{width}{$\pattern$}{\Char{\width}{\pattern}}{\SetRegExpressions}{\Naturals}{}
  It is defined by 
  $\Char{\width}{\pattern} = \min \limits_{\wordv \in \Language{\pattern}} |\wordv|$.
\end{definition}

\begin{example}
\label{ex:width}
  \renewcommand{\labelitemi}{$\bullet$}
    Consider the~$\pattern = \DecreasingTerracePatternName$ regular expression.
    There is a single shortest word
    in~$\Language{\pattern}$, namely~$\reg{>=>}$.
    Thus the width of~$\pattern$ is~$3$.
    Hence, any extended~$\pattern$-pattern has
    at least~$3+1$ time-series variables. \qedexample
\end{example}

\subsection{Height}
\label{sec:height}

We introduce the notion of height of a regular expression,
which is used for defining a necessary and sufficient condition, see~\propref{prop:necessary-condition}, for
the existence of at least one occurrence of a regular expression
within the signature of a time series.
This characteristics is  also used in~\thref{th:nb-patterns} of~\secref{sec:nb-patterns} for computing 
a sharp upper bound on the number of
occurrences of the regular expression within the signature of a time
series.
Definitions~\ref{def:support-set}
and~\ref{def:height-word} are only used
for introducing~\defref{def:height}.

\begin{definition}
\label{def:support-set}
\IntroDom{\pattern}{\Domain}.
\DefinitionArg{set of supporting time series}{\emph{a word} $\wordv$ in~$\Language{\pattern}$}{\DomainMin,
  \DomainMax}{\CharAll{\SupportSet}{\pattern}{\DomainMin,
    \DomainMax}{\wordv}}{\SetRegExpressions \times \Integers \times \Integers
\times \AllWords }{\PowerSet{\IntegersStar}}{,
where~$\PowerSet{\IntegersStar}$ is the power set of~$\IntegersStar$}
Each element of~$\CharAll{\SupportSet}{\pattern}{\DomainMin,
  \DomainMax}{\wordv}$ is a time series over~$\Domain$ whose signature
is~$\wordv$, and is called a \emph{supporting time series of~$\wordv$
  wrt~$\Tuple{\DomainMin, \DomainMax}$}.
\end{definition}

\begin{definition}
\label{def:height-word}
\Intro{\pattern}.
\Definition{height}{a word~$\wordv$
  in~$\Language{\pattern}$}{\CharArg{\Height}{\pattern}{\wordv}}{\SetRegExpressions
\times \AllWords}{\Naturals}{}
It is defined by~$\CharArg{\Height}{\pattern}{\wordv} =
 \min \limits_{\CharAll{\SupportSet}{\pattern}{\DomainMin,
      \DomainMax}{\wordv} \neq \emptyset} (\DomainMax - \DomainMin)$,
  where~$\Domain$ is an integer interval domain.

\end{definition}

\begin{definition}
\label{def:height}
\Intro{\pattern}.
\Definition{height}{$\pattern$}{\Char{\Height}{\pattern}}{\SetRegExpressions}{\Naturals}{}
It is defined by~$\Char{\Height}{\pattern} =
\min \limits_{\wordv \in \Language{\pattern}} \CharArg{\Height}{\pattern}{\wordv}$.
\end{definition}

\begin{example}
  \label{ex:height}
Consider the~$\pattern = \DecreasingTerracePatternName $
    regular expression and an integer interval domain~$\Domain$.
     When~$\DomainMax-\DomainMin \leq 1$, there does not exist a time
    series over~$\Domain$ whose signature is a word
    in~$\Language{\pattern}$;
    but when~$\DomainMax - \DomainMin = 2$, there exists a time series over~$\Domain$
    whose signature is a word, for example~\reg{>=>},
    in~$\Language{\pattern}$.
    Hence, the height of~$\pattern$ equals~$2$.
\qedexample
\end{example}

\subsection{Range}
\label{sec:range}

This section introduces a characteristics needed by
Theorems~\ref{th:max-width},~\ref{th:sum-width},
and~\ref{th:min-width} for characterising 
sharp bounds on the result value of a time-series constraint when the
feature is~$\Width$.
This characteristics, described in~\defref{def:range-regular-expression}, is called the \emph{range} of a regular
expression~$\pattern$, and shows the variation of the minimum height
of words of~$\Language{\pattern}$ for words of increasing length.

\begin{definition}
\label{def:range-regular-expression}
\IntroLength{\pattern}{\seqlength}. 
\DefinitionArg{range}{$\pattern$}{\seqlength}{\CharPar{\Range}{\pattern}{\seqlength}}{\SetRegExpressions
  \times \Naturals}{\Naturals}{}
It is defined by~$\CharPar{\Range}{\pattern}{\seqlength}
  = \min \limits_{\wordv \in \Language{\pattern},~|\wordv| =
    \seqlength-1} \CharArg{\Height}{\pattern}{\wordv}$,
  where~$\CharArg{\Height}{\pattern}{\wordv}$ is the height of the
  word~$\wordv$.
If~$\Language{\pattern}$ does not contain any word of length~$\seqlength
- 1$, then the value of~$\CharPar{\Range}{\pattern}{\seqlength}$ is undefined.

\end{definition}

\begin{example}
  \label{ex:steady-sequence-range}
  Consider the~$\pattern = \SteadySequencePatternName$ regular
  expression.
  For every integer~$\seqlength > \Char{\width}{\pattern}$,
  the language~$\Language{\pattern}$ contains a word
  with~$\seqlength - 1$ equalities.
  Any word of this type has a height of~$0$.
  Hence, the range of~$\pattern$ is a constant
  function of~$\seqlength$, which equals~$0$.
\end{example}

\subsection{Set of Inducing Words}
\label{sec:set-of-inducing-words}

Given a disjunction-capsuled regular expression $\pattern$,
we first introduce the notion of~\emph{inducing word} of~$\Language{\pattern}$,
which is a maximum sequence of letters that appears in every word
of~$\Language{\pattern}$ in a fixed order. 
Then we generalise this notion to any disjunction of
disjunction-capsuled regular expression.

\begin{definition}
  \label{def:inducing-word-disjunction-capsuled}
  Consider a disjunction-capsuled regular expression~$\pattern$.
  The (unique) non-empty shortest word of $\Language{\pattern}$ is the
  \emph{inducing word} of~$\Language{\pattern}$.
\end{definition}

\begin{definition}
  \label{def:set-of-inducing-words}
  Consider a regular expression~$\pattern$ that is in the form of
  $\pattern =\pattern_1 \Disj \pattern_2 \Disj\dots \Disj \pattern_t$ with~$t\geq
    1$, where every~$\pattern_i$~(with $i\in[1,t]$)
      is a disjunction-capsuled regular expression.
 \Definition{set of inducing
      words}{$\pattern$}{\Char{\IndWordSet}{\pattern}}{\SetRegExpressions}{\PowerSet{\AllWords}}{,
      where~$\PowerSet{\AllWords}$ is the power set of~$\AllWords$.
      The value of~$\Char{\IndWordSet}{\pattern}$
      is the union of inducing words of every~$\pattern_i$}
\end{definition}

\begin{example}
  \label{ex:inducing-words}
    Consider the~$\InflexionPatternName = \InflexionPattern$ regular expression. 
    It can be represented as~$\InflexionPatternName_1|\InflexionPatternName_2$,
    where~$\InflexionPatternName_1=\reg{<(<|=)^*>}$,
    $\InflexionPatternName_2=\reg{>(>|=)^*<}$,
    and both $\InflexionPatternName_1$ and $\InflexionPatternName_2$ are disjunction-capsuled.
    The word~$\wordu = \reg{<>}$ is the inducing word
    of~$\Language{\InflexionPatternName_1}$, the word~$\wordv = \reg{><}$ is the
    inducing word of~$\Language{\InflexionPatternName_2}$, and both $\wordu$ and
    $\wordv$ are inducing words of~$\Language{\InflexionPatternName}$.
    Hence,~$\Char{\IndWordSet}{\InflexionPatternName} =
    \Curly{\reg{<>},\reg{><}}$.
\qedexample

\end{example}

\subsection{Overlap}
\label{sec:overlap}

This section introduces the \emph{overlap} characteristic of a regular
expression; it will be used in~\thref{th:nb-patterns} for computing 
the sharp upper bound on the number of
occurrences of the regular expression within the signature of a time
series.
To define the overlap of a regular expression $\pattern$
wrt to an integer interval domain~$\Domain$,~\defref{def:superposition}
first introduces the notion of \emph{set of superpositions} of two
words~$\wordv$ and~$\word$ in~$\Language{\pattern}$
wrt~$\Tuple{\DomainMin, \DomainMax}$.
Intuitively the superposition of $\wordv$ and~$\word$
wrt~$\Tuple{\DomainMin, \DomainMax}$ is the signature~$\wordz$
of some ground time series over~$\Domain$ that contains exactly
two~$\pattern$-patterns,
i.e.,~$\wordv$ as a prefix and $\word$ as a suffix of~$\wordz$,
and whose length does not exceed the length of~$\wordv\word$.

\begin{definition}
\label{def:superposition}
\IntroDom{\pattern}{\Domain}.
\DefinitionArg{set of superpositions}{two words, $\wordv$ and~$\word$ in~$\Language{\pattern}$,}{\DomainMin,
  \DomainMax}{\CharAll{\SetSuperposition}{\pattern}{\DomainMin,
    \DomainMax}{\wordv, \word}}{\SetRegExpressions \times \Integers \times \Integers
\times \AllWords \times  \AllWords }{\PowerSet{\AllWords}}{,
where~$\PowerSet{\AllWords}$ is the power set of~$\AllWords$}
Each element~$\wordz$ in~$\CharAll{\SetSuperposition}{\pattern}{\DomainMin,
  \DomainMax}{\wordv, \word}$ is a word over~$\Alphabet$, called a superposition of~$\wordv$
and~$\word$ wrt~$\Tuple{\DomainMin, \DomainMax}$ and satisfying all the
following conditions:

\setword{(1)}{Cond:i}~$\wordz \notin \Language{\pattern}$,
\hspace*{5pt} \setword{(2)}{Cond:ii}~$\CharAll{\SupportSet}{\pattern}{\DomainMin,
  \DomainMax}{\wordz} \neq \emptyset$,
\hspace*{5pt}\setword{(3)}{Cond:iii}~$\wordv$ is a prefix of $\wordz$,
\hspace*{5pt}\setword{(4)}{Cond:iv}~$\word$ is a suffix of $\wordz$,
\hspace*{5pt}\setword{(5)}{Cond:v}~$|\wordz|\leq|\wordv\word|$.

\end{definition}

\begin{example}
    \ExStyle
  \label{ex:superpositions}

    Consider~$\pattern = \DecreasingTerracePatternName$,
    and an integer interval domain $\Domain$ allowing to have at least
    one occurrence of~$\pattern$ in the signature of a time series
    over~$\Domain$, i.e.,~$\DomainMax-\DomainMin\geq 2$.
    We compute a superposition of the pair~$\Tuple{\wordv, \wordv}$,
    where~$\wordv =
    \reg{>=>}\thinspace\in\thinspace\Language{\pattern}$.
    Let~$\wordz$ denote~$\reg{>=>=>}$.
    \begin{itemize}
    \item
      First, assume that~$\DomainMax-\DomainMin=2$.
      The word $\wordz$ is not a
      \superposition~of~$\wordv$ and~$\wordu$,
      since the number of consecutive increases in the word~$\wordz$ is~$3$,
      which is strictly greater than~$\DomainMax-\DomainMin=2$,
      and thus~$\CharAll{\SupportSet}{\pattern}{\DomainMin,
        \DomainMax}{\wordz} = \emptyset$.
      Indeed, when $\DomainMax-\DomainMin=2$, there is no superposition
      of~$\wordv$ and~$\wordv$,
      because any word different from~$\wordz$ satisfying the first four conditions
      of~\defref{def:superposition} will violate Condition~\ref{Cond:v}
      of~\defref{def:superposition},
      i.e.,~will be strictly longer than~$2 \cdot|\wordv|$, 
      thus~$\CharAll{\SetSuperposition}{\pattern}{\DomainMin,
          \DomainMax}{\wordv, \wordv} = \emptyset$.
    
  \item
    Now assume that~$\DomainMax-\DomainMin=3$.
    Then,~$\CharAll{\SupportSet}{\pattern}{\DomainMin,
        \DomainMax}{\wordz} \neq \emptyset$, and the word~$\wordz$
    is the only superposition of~$\wordv$ and~$\wordv$, 
     thus~$\CharAll{\SetSuperposition}{\pattern}{\DomainMin,
          \DomainMax}{\wordv, \wordv} = \Curly{\reg{>=>=>}}$.
    \item
      Finally, assume that~$\DomainMax-\DomainMin\geq 4$.
      The sets of supporting time series of both words~$\reg{>=>=>}$
      and~$\reg{>=>>=>}$ wrt~$\Tuple{\DomainMin, \DomainMax}$ are not
      empty, and these two words are
      the only superpositions of~$\wordu$ and $\wordu$
      wrt~$\Tuple{\DomainMin, \DomainMax}$, 
      thus~$\CharAll{\SetSuperposition}{\pattern}{\DomainMin,
          \DomainMax}{\wordv, \wordv} =
        \Curly{\reg{>=>=>},\reg{>=>>=>}}$. \qedexample
    \end{itemize}
\end{example}

For a regular expression~$\pattern$ and an integer interval
domain~$\Domain$,  we now introduce the \emph{overlap} characteristic
  of~$\pattern$ wrt~$\Tuple{\DomainMin, \DomainMax}$,
which is a crucial component in the sharp upper bound formula stated
in~\thref{th:nb-patterns}.
It corresponds to the maximum number of time-series variables that can be shared by two
consecutive extended~$\pattern$-patterns:
when maximising the number of~$\pattern$-patterns in a time series, we
need to enforce as many consecutive extended~$\pattern$-patterns as
possible to have as many common time-series variables as possible.

\begin{definition}
  \label{def:overlap-words}
  \IntroDom{\pattern}{\Domain}.
  \DefinitionArg{overlap}{two words,~$\wordv$ and~$\word$ in~$\Language{\pattern}$,}{\DomainMin,
    \DomainMax}{\CharAll{\Overlap}{\pattern}{\DomainMin,
      \DomainMax}{\wordv, \word}}{\SetRegExpressions  \times \Integers \times \Integers
    \times \AllWords \times  \AllWords}{\Naturals}{}
  It is defined by
  \begin{numcases}{
    \CharAll{\Overlap}{\pattern}{\DomainMin,
    \DomainMax}{\wordv, \word} =}
\left(|\wordv\word| - \min \limits_{ \wordz \in\CharAll{\SetSuperposition}{\pattern}{\DomainMin,
      \DomainMax}{\wordv, \word}}
  |\wordz|\right)+1 & 
\text{~if } $\CharAll{\SetSuperposition}{\pattern}{\DomainMin,
  \DomainMax}{\wordv,
  \word}\neq\emptyset$  \label{eq:overlap-1}
\\
0, & \text{~otherwise.}
\label{eq:overlap-2} 
  \end{numcases}

\end{definition}

Case~(\ref{eq:overlap-1}) of~\defref{def:overlap-words} corresponding
to condition~$\CharAll{\SetSuperposition}{\pattern}{\DomainMin,
  \DomainMax}{\wordv,
  \word}\neq\emptyset$ 
states that the overlap is strictly greater than $0$ iff there exists at least one ground time
series over~$\Domain$ that is not strictly longer than~$|\wordv\word|$
and that has exactly two~$\pattern$\nobreakdash-patterns corresponding to the occurrences of~$\wordv$
and~$\word$ in its signature.
The term $|\wordv\word| - \min \limits_{ \wordz \in \CharAll{\SetSuperposition}{\pattern}{\DomainMin,
  \DomainMax}{\wordv, \word}}|\wordz|$
denotes the maximum possible overlap that is actually achieved by the
shortest superposition.
The increment $+1$ denotes that we count the number of
time-series variables rather than the number of signature variables.

We now generalise in~\defref{def:overlap} the notion of overlap wrt~$\Tuple{\DomainMin,
  \DomainMax}$ upon a regular expression.

\begin{definition}
  \label{def:overlap}
  \IntroDom{\pattern}{\Domain}.
  \DefinitionArg{overlap}{$\pattern$}{\DomainMin,
    \DomainMax}{\CharPar{\Overlap}{\pattern}{\DomainMin,
      \DomainMax}}{\SetRegExpressions \times \Integers \times \Integers}{\Naturals}{}
  If there exists a constant~$c$ in~$\Naturals$ such that for any pair of
  words~$\wordv$,~$\word$ in~$\Language{\pattern}$, the value
  of~$\CharAll{\Overlap}{\pattern}{\DomainMin, \DomainMax}{\wordv,
    \word}$ is bounded by~$c$, then the overlap of~$\pattern$
  wrt~$\Tuple{\DomainMin, \DomainMax}$ is defined
  by~$\CharPar{\Overlap}{\pattern}{\DomainMin,
    \DomainMax} = \max \limits_{\wordv,\word \in \Language{\pattern}} 
  \CharAll{\Overlap}{\pattern}{\DomainMin,
    \DomainMax}{\wordv, \word} $.
  Otherwise, $\CharPar{\Overlap}{\pattern}{\DomainMin,
    \DomainMax}$ is undefined.

\end{definition}

By~\defref{def:overlap}, we need to compute the overlap of~$\pattern$
wrt every pair of values~$\Tuple{\DomainMin, \DomainMax}$, i.e.,
every domain~$[\DomainMin, \DomainMax]$.
Note that it is enough to compute the overlap of~$\pattern$
wrt~$\Tuple{\DomainMin, \DomainMax}$
once for every value of the difference~$\DomainMax-\DomainMin$
permitting an occurrence of~$\pattern$ in the signature of a time
series, i.e.,~for a difference that is greater than
or equal to the height of the regular expression~$\pattern$.
While in the general case, we do need to consider every value
of~$\DomainMax -\DomainMin$, this is not required
for all the~$22$ regular expressions in~\tabref{tab:patterns}, where
we only need to consider at most two different values of~$\DomainMax -
\DomainMin$.

\begin{example}
  \label{ex:overlap}
 We successively consider values of the overlap of two regular
  expressions.

\ExFullStyle
  \begin{itemize}
    
    \item
      Consider the~$\pattern = \DecreasingTerracePatternName$ regular expression, whose
    height~$\Char{\Height}{\pattern}$  is~$2$.
      \begin{itemize}
      \item
        If~$\DomainMax - \DomainMin = \Char{\Height}{\pattern} = 2$,
        then~$\CharPar{\Overlap}{\pattern}{\DomainMin,
          \DomainMax} = 0$, because
        as shown in~\exref{ex:superpositions},
        for any pair of words in~$\Language{\pattern}$,
        the set of their superpositions wrt~$\Tuple{\DomainMin,\DomainMax}$ is empty.
      \item
        If~$\DomainMax - \DomainMin \geq \Char{\Height}{\pattern}+1 = 3$,
        then~$\CharPar{\Overlap}{\pattern}{\DomainMin,
          \DomainMax} = 2$ and is
        obtained, for example, for the pair~$\reg{>=>}$
        and~$\reg{>=>}$, and their superposition~$\reg{>=>=>}$.
    \item
      For any other value of~$\DomainMax - \DomainMin \geq 4$, the value of the overlap
      of~$\pattern$ wrt~$\Tuple{\DomainMin,
        \DomainMax}$ equals~$2$ as well.
    \end{itemize}
  
\item Consider the~$\pattern = \reg{<=^* | =^*>}$ regular
  expression and an integer interval domain~$\Domain$ such
  that~$\DomainMax > \DomainMin$.
  The overlap of~$\pattern$ wrt~$\Tuple{\DomainMin, \DomainMax}$ is
  undefined, because for any constant~$c$ in~$\Naturals$, there always
  exists a pair of words of length~$c+1$ whose overlap
  wrt~$\Tuple{\DomainMin, \DomainMax}$ equals~$c+1$. 
\qedexample
  \end{itemize}
\end{example}

\subsection{Smallest Variation of Maxima}
\label{sec:variation-of-maxima}

This section introduces the \emph{smallest variation of maxima}
characteristics of a regular expression, which is used in~\thref{th:nb-patterns} for computing 
the sharp upper bound on the number of
occurrences of the regular expression within the signature of a time
series.
To maximise the number of occurrences of a regular
expression~$\pattern$ in a time series over an integer interval domain~$\Domain$,
we select extended $\pattern$-patterns of minimum length $\Char{\width}{\pattern}+1$ such that
two consecutive extended $\pattern$-patterns maximise the number of shared time-series variables,
i.e.,~share $\CharPar{\Overlap}{\pattern}{\DomainMin,
          \DomainMax} $ variables.
Unfortunately, for a few regular expressions like $\DecreasingTerracePatternName$,
it is not always possible that all~$\pattern$-patterns share~$\CharPar{\Overlap}{\pattern}{\DomainMin,
  \DomainMax} $ time-series variables: since we
decrease by at least one unit between two consecutive overlapping extended~$\pattern$-patterns
we will be blocked at some point by the lower limit~$\DomainMin$, even if we start
from the upper limit~$\DomainMax$.
To maximise the number of~$\pattern$-patterns in a time series, we must decrease as little
as possible on two consecutive overlapping extended
$\pattern$-patterns. \defref{def:variation-of-maxima}
formalises the notion of \emph{smallest variation of the maxima} of a
regular expression wrt~$\Tuple{\DomainMin, \DomainMax}$.
First,~\defref{def:shift-two-words} defines the notion of shift
of a proper factor in a word in the
language of a regular expression wrt some integer interval domain.
Then, using this notion,~\defref{def:variation-of-maxima-two-words}
(\Resp~\defref{def:variation-of-maxima}) introduces
the smallest variation of
the maxima of two words (\Resp~a language~$\Language{\pattern}$) 
wrt~$\Tuple{\DomainMin, \DomainMax}$.

\begin{definition} 
\label{def:shift-two-words}
\IntroDom{\pattern}{\Domain}.
\DefinitionArg{shift}{a proper factor~$\word$ in a word~$\wordv$ in~$\Language{\pattern}$}{\DomainMin,
  \DomainMax}{\CharAllUp{\Shift}{\pattern}{\DomainMin,
    \DomainMax}{\wordv, \word, i}}{\SetRegExpressions \times \Integers \times \Integers
\times \AllWords \times \AllWords \times \Naturals}{\Naturals}{}
It is defined by
$$\CharAllUp{\Shift}{\pattern}{\DomainMin,
    \DomainMax}{\wordv, \word, i} = \min_{\TimeSeries \in
    \CharAll{\SupportSet}{\pattern}{\DomainMin,
  \DomainMax}{\wordv}} \min_{x \in \TimeSeries_{\word_i}} (\max(\TimeSeries) -
x),$$
 where~$\max(\TimeSeries)$ is the maximum value of a
time series~$\TimeSeries$, a supporting time series
of~$\wordv$ wrt~$\Tuple{\DomainMin, \DomainMax}$,
 and~$\TimeSeries_{\word_i}$ is a subseries of~$\TimeSeries$ 
 corresponding to the~$i^\textrm{th}$
extended~$\pattern$-pattern whose signature is~$\word$.
If~$\word$ is not a proper factor of~$\wordv$, or~$i$ is strictly greater
than the number of occurrences of~$\word$ in~$\wordv$,
then~$\CharAllUp{\Shift}{\pattern}{\DomainMin,
    \DomainMax}{\wordv, \word, i} $ is undefined.
\end{definition}

  \IntroDom{\pattern}{\Domain}.
 For any~$\wordv$ in~$\Language{\pattern}$, if~$\DomainMax -
 \DomainMin \geq \CharArg{\Height}{\pattern}{\wordv}$, then
 the value of~$\CharAllUp{\Shift}{\pattern}{\DomainMin,
   \DomainMax}{\wordv, \word, i}$ does not depend on the
 domain~$\Domain$, because there always exists a time series
 in~$\CharAll{\SupportSet}{\pattern}{\DomainMin,
   \DomainMax}{\wordv}$ where each variable has its largest value
 compared to the other time series of~$\CharAll{\SupportSet}{\pattern}{\DomainMin,
   \DomainMax}{\wordv}$.
 Then,~$\CharAllUp{\Shift}{\pattern}{\DomainMin,
    \DomainMax}{\wordv, \word, i}$ does not depend on the values
 in the domain, but only on the structure of the word~$\wordv$. 
 Hence, \Wlog~the notation for~$\CharAllUp{\Shift}{\pattern}{\DomainMin,
   \DomainMax}{\wordv, \word, i}$ can be simplified
 to~$\CharArgUp{\Shift}{\pattern}{\wordv, \word, i}$.

\begin{example}
  Consider $\pattern = \DecreasingTerracePatternName$
  when~$\DomainMax - \DomainMin \geq 3$, and two words~$\wordv =
  \reg{<=<=<}$ and~$\word = \reg{<=<}$.
  The word~$\wordv$ contains two occurrences of~$\word$, thus the value
  of~$\CharArgUp{\Shift}{\pattern}{\wordv, \word, i}$ is defined
  when~$i \in \Curly{1,2}$:

 \ExStyle
 \begin{itemize}
 \item
   When~$i$ is~$1$, the value of~$\CharArgUp{\Shift}{\pattern}{\wordv,
     \word, 1}$ equals~$0$, since the first
   extended~$\pattern$-pattern whose signature
   is~$\word$ necessarily contains the maximum of any time
   series in~$\CharAll{\SupportSet}{\pattern}{\DomainMin,
     \DomainMax}{\wordv}$.
   
 \item
   When~$i$ is~$2$, the value of~$\CharArgUp{\Shift}{\pattern}{\wordv,
     \word, 2}$ equals~$1$, since the maximum of
   the second extended~$\pattern$\nobreakdash-pattern whose signature
   is~$\word$ has a difference of at least one with the maximum of any time
   series in~$\CharAll{\SupportSet}{\pattern}{\DomainMin,
     \DomainMax}{\wordv}$. \qedexample
 \end{itemize}
  
\end{example}

\begin{definition}
\label{def:variation-of-maxima-two-words}
\IntroDom{\pattern}{\Domain}.
\DefinitionArg{smallest variation of maxima}{superpositions of two
  words~$\word$ and~$\wordv$ in~$\Language{\pattern}$}{\DomainMin,
  \DomainMax}{\CharAll{\VariationOfMax}{\pattern}{\DomainMin,
  \DomainMax}{\wordv, \word}}{\SetRegExpressions \times \Integers
\times \Integers \times \AllWords \times \AllWords}{\Naturals}{}
It is defined by 

\begin{equation*}
  \CharAll{\VariationOfMax}{\pattern}{\DomainMin,
    \DomainMax}{\wordv, \word} = 
  \begin{cases}
    \CharArgUp{\Shift}{\pattern}{\wordz_*, \wordv, 1} -
        \CharArgUp{\Shift}{\pattern}{\wordz_*, \word, 1}, &
      \textnormal{~if~} \wordv \neq \word \textnormal{~and~} \CharAll{\SetSuperposition}{\pattern}{\DomainMin,
          \DomainMax}{\wordv, \word} \neq \emptyset \\
         \CharArgUp{\Shift}{\pattern}{\wordz_{**}, \wordv, 1} -
        \CharArgUp{\Shift}{\pattern}{\wordz_{**}, \word, 2}, &
      \textnormal{~if~} \wordv = \word \textnormal{~and~} \CharAll{\SetSuperposition}{\pattern}{\DomainMin,
          \DomainMax}{\wordv, \word} \neq \emptyset\\
       0, &\textnormal{~if~} \CharAll{\SetSuperposition}{\pattern}{\DomainMin,
          \DomainMax}{\wordv, \word} = \emptyset\\
      
 \end{cases}
\end{equation*}

  where the words~$\wordz_*$ and~$\wordz_{**}$ both  belongs to~$\CharAll{\SetSuperposition}{\pattern}{\DomainMin,
          \DomainMax}{\wordv, \word}$, and the value
        $ \min \limits_{\wordz \in \CharAll{\SetSuperposition}{\pattern}{\DomainMin,
    \DomainMax}{\wordv, \word}}\abs{\CharArgUp{\Shift}{\pattern}{\wordz, \wordv, 1} -
        \CharArgUp{\Shift}{\pattern}{\wordz, \word, 1}}$ (respectively $ \min \limits_{\wordz \in
        \CharAll{\SetSuperposition}{\pattern}{\DomainMin,
    \DomainMax}{\wordv, \word}}\abs{\CharArgUp{\Shift}{\pattern}{\wordz, \wordv, 1} -
        \CharArgUp{\Shift}{\pattern}{\wordz, \word, 2}}$) is reached
      when~$\wordz$ is~$\wordz_*$ (respectively~$\wordz_{**}$).

\end{definition}

\begin{sloppypar}
In~\defref{def:variation-of-maxima-two-words},
either~$\CharArgUp{\Shift}{\pattern}{\wordz_*, \wordv, 1}$
(respectively~$\CharArgUp{\Shift}{\pattern}{\wordz_{**}, \wordv, 1}$)
or~$\CharArgUp{\Shift}{\pattern}{\wordz_*, \word, 1}$
(respectively~$\CharArgUp{\Shift}{\pattern}{\wordz_{**}, \word, 2}$)
equals zero, since for any time series~$\TimeSeries$ whose signature is~$\wordz_*$
(respectively~$\wordz_{**}$), at least one of the two
extended~$\pattern$-patterns contains the maximum of~$\TimeSeries$.
\end{sloppypar}

The next lemma introduces a property of words whose
smallest variation of maxima wrt some integer interval domain is not zero.

\begin{lemma}
\label{lemma:variation-of-maxima}
\IntroDom{\pattern}{\Domain}.
If~$\CharAll{\VariationOfMax}{\pattern}{\DomainMin,
  \DomainMax}{\wordv, \word}$, the smallest variation of maxima of two words~$\wordv$ and~$\word$
in~$\Language{\pattern}$ wrt~$\Tuple{\DomainMin, \DomainMax}$, is positive (respectively negative),
then~$\wordv$ (respectively~$\word$) does not contain any~$\reg{>}$ (respectively~$\reg{<}$).
\end{lemma}

\begin{proof}
For brevity, we consider only the case
of~$\CharAll{\VariationOfMax}{\pattern}{\DomainMin,
  \DomainMax}{\wordv, \word}$ being positive, the case of a
negative value of~$\CharAll{\VariationOfMax}{\pattern}{\DomainMin,
  \DomainMax}{\wordv, \word}$ being symmetric, and \Wlog~we assume
that~$\wordv \neq \word$.

Since~$\CharAll{\VariationOfMax}{\pattern}{\DomainMin,
  \DomainMax}{\wordv, \word} > 0$, there exists at least one
superposition~$\wordz$ of~$\wordv$ and~$\word$ wrt~$\Tuple{\DomainMin,
  \DomainMax}$ such that $\CharArgUp{\Shift}{\pattern}{\wordz, \wordv,
  1 } = \CharAll{\VariationOfMax}{\pattern}{\DomainMin,
  \DomainMax}{\wordv, \word} $,
and~$\CharArgUp{\Shift}{\pattern}{\wordz, \word, 1 }  = 0$.
Assume that~$\wordv$  contains at least
one~$\reg{>}$.
Let $i$ denote the position of the first~$\reg{>}$ in~$\wordz$, which
is necessarily within its proper factor~$\wordv$.
Since
there exists a time series
in~$\CharAll{\SupportSet}{\pattern}{\DomainMin, \DomainMax}{\wordz}$
such that the time-series variable at position~$i$
equals~$\DomainMax$, $\CharArgUp{\Shift}{\pattern}{\wordz, \wordv, 1 }$ equals~$0$.
This contradicts the fact
that~$\CharArgUp{\Shift}{\pattern}{\wordz, \wordv, 1 } =
\CharAll{\VariationOfMax}{\pattern}{\DomainMin,
  \DomainMax}{\wordv, \word} > 0$, thus the word~$\wordv$ does not
contain any~$\reg{>}$. \shiftedqed
\end{proof}

\begin{definition}
\label{def:variation-of-maxima}
\IntroDom{\pattern}{\Domain}.
\DefinitionArg{smallest variation of maxima}{$\pattern$}{\DomainMin,
  \DomainMax}{\CharPar{\VariationOfMax}{\pattern}{\DomainMin,
  \DomainMax}}{\SetRegExpressions \times \Integers \times \Integers}{\Naturals}{}
It is defined by
\begin{equation*}
  \CharPar{\VariationOfMax}{\pattern}{\DomainMin,
    \DomainMax}  = 
  \begin{cases}
    \Undefined, & \If \exists~\wordv_1, \wordv_2, \word_1, \word_2 \in \Language{\pattern}
  \textnormal{~s.t.~} \CharAll{\VariationOfMax}{\pattern}{\DomainMin,
    \DomainMax}{\wordv_1, \word_1} > 0 \AndText \CharAll{\VariationOfMax}{\pattern}{\DomainMin,
    \DomainMax}{\wordv_2, \word_2} < 0 \\
     0, & \If \CharPar{\Overlap}{\pattern}{\DomainMin,
    \DomainMax}= 0 \\
     \CharAll{\VariationOfMax}{\pattern}{\DomainMin,
    \DomainMax}{\wordv_*, \word_*}, & \Otherwise\\
 \end{cases}
\end{equation*}

where the words~$\wordv_*$ and~$\word_*$ both belong
to~$\Language{\pattern}$ and the value~$\min
\limits_{\substack{\wordv, \word \in \Language{\pattern} \\ 
  \CharAll{\Overlap}{\pattern}{\DomainMin,
    \DomainMax}{\wordv, \word} \neq 0
} } \abs{\CharAll{\VariationOfMax}{\pattern}{\DomainMin,
    \DomainMax}{\wordv, \word}}$ is reached when~$\wordv$ is~$\wordv_*$
  and~$\word$ is~$\word_*$.

\end{definition}

\begin{example}
\label{ex:variation-of-maxima}
  Consider the~$\pattern = \DecreasingTerracePatternName$ regular expression, an integer
  interval domain~$\Domain$ such that~$\DomainMax - \DomainMin \geq 3$,
  and the superposition~$\wordz = \reg{>=>=>}$ of the
  words~$\wordv = \reg{>=>}$ and~$\wordv = \reg{>=>}$ in~$\Language{\pattern}$.
  The value of~$\CharArgUp{\Shift}{\pattern}{\wordz, \wordv, 1} -
    \CharArgUp{\Shift}{\pattern}{\wordz,
      \wordv, 2}  $ is equal to $0 - 1 =- 1$.
  For any other pair of words
  of~$\Language{\pattern}$ whose set of
  superpositions wrt~$\Tuple{\DomainMin, \DomainMax}$ is not empty, we
  obtain a same or a smaller negative value.
  Hence, if two extended~$\pattern$-patterns
  share some time-series variables, then the maximum of a second
  extended~$\pattern$-pattern
  is at least one unit below, i.e., $\CharPar{\VariationOfMax}{\pattern}{\DomainMin,
  \DomainMax} = -1$,  from the maximum of the
  first extended~$\pattern$-pattern. \qedexample
\end{example}

If~$\CharPar{\VariationOfMax}{\pattern}{\DomainMin,
    \DomainMax}$ is positive (respectively negative),
  then for any two extended~$\pattern$-patterns that have at least one
  common time-series variable, the maximum of the first
  extended~$\pattern$-pattern is strictly
  less (respectively greater) than the maximum of the second
  extended~$\pattern$-pattern, e.g., 
  for~$\DecreasingTerracePatternName$,
  $\CharPar{\VariationOfMax}{\pattern}{\DomainMin,
    \DomainMax}$ equals~$-1$, but
  for~$\IncreasingTerracePatternName$,
  $\CharPar{\VariationOfMax}{\pattern}{\DomainMin,
    \DomainMax}$ equals~$+1$.

\subsection{Summary Example Illustrating All Characteristics }
\label{sec:summary-example}

This section illustrates the various regular expression characteristics
introduced in the previous sections.

\begin{example}
\label{ex:summary-example}
Consider the~$\pattern = \DecreasingTerracePatternName$ regular
expression and a time series~$X$ of length~$6$ over an integer
interval domain~$[0,3]$. 
Let us recall the characteristics mentioned in
Examples~\ref{ex:height}, \ref{ex:overlap}, \ref{ex:width}, and
\ref{ex:variation-of-maxima}, which are illustrated
in~\figref{fig:characteristics}.

\ListStyle
\begin{itemize}
  \item
  The \emph{width} of~$\pattern$, denoted by~$\Char{\width}{\pattern}$, equals~$3$.
\item
  The \emph{height} of~$\pattern$, denoted by~$\Char{\Height}{\pattern}$, equals~$2$.
  This is the difference between the~$y$-coordinates of the points~$L_1$ and~$S$
  in~\figref{fig:characteristics}, which are respectively the maximum and the
  minimum points of the first extended~$\pattern$-pattern of~$X$.
\item
  The \emph{range} of~$\pattern$ wrt~$\Tuple{\seqlength}$, denoted by
  $\CharPar{\Range}{\pattern}{\seqlength}$, equals~$2$,
  with~$\seqlength \in \Naturals$ being
  greater than or equal to~$\Char{\width}{\pattern} = 3$.
\item
  The \emph{overlap} of~$\pattern$ wrt~$\Tuple{0,3}$, denoted by
  $\CharPar{\Overlap}{\pattern}{0, 3}$, equals~$2$.
  It is the number of common points of the first and the second
  extended~$\pattern$-patterns in~\figref{fig:characteristics}, i.e.,
  the number points coloured in violet.
\item
  The \emph{smallest variation of maxima} of~$\pattern$ wrt~$\Tuple{0,3}$,
  denoted by~$\CharPar{\VariationOfMax}{\pattern}{0, 3}$,
  equals~$1$.
  It is the difference
  between the~$y$-coordinates of the~$L_1$ and the~$L_2$ points
  in~\figref{fig:characteristics}, which are the maxima points of the
  first, respectively the second, extended~$\pattern$-pattern of~$X$.
\qedexample
\end{itemize}
\end{example}

\begin{figure}[!h]
\floatbox[{\capbeside\thisfloatsetup{capbesideposition={left,top}, 
capbesidewidth=0.65\textwidth}}]{figure}[\FBwidth]
{\caption{A time series of length~$\seqlength = 6$ over the integer interval domain~$[0,3]$
    containing two extended~$\pattern$-patterns, where~$\pattern$
    is~$\DecreasingTerracePatternName$.
    The $x$-axis is for time-series variables,
    the~$y$-axis is for domain values.
    The first (respectively~second)
    extended~$\pattern$-pattern is shown
    in red (respectively~blue).
    %
    %
    The common time-series variables of the two extended~$\pattern$-patterns
    are coloured in violet.
    $L_1$ (respectively~$L_2$) is the point whose~$y$\nobreakdash-coordinate is maximum
    among all points of the first (respectively second) extended~$\pattern$\nobreakdash-pattern.
    $S$ is the point whose~$y$\nobreakdash-coordinate is minimum
    among all points of the first 
    extended~$\pattern$\nobreakdash-pattern.
    %
    %
    %
\label{fig:characteristics}}}
{\begin{tikzpicture}
\begin{scope}[scale=0.38,yshift=-4cm]
\def\xmin{0}
\def\xmax{5}
\def\ymin{0}
\def\ymax{3}
\filldraw[color=violet!10] (2,1) -- (2,2) -- (3,1);
\filldraw[color=violet!10] (2,-4) rectangle (3,1);
\draw[] (\xmin,\ymin) -- (\xmax,\ymin) node[right] {};
\draw[] (\xmin,\ymin) -- (\xmin,\ymax) node[above] {};
\foreach \x in {1,...,6} \node at (\x-1, \ymin) [below ]  {\tiny $X_{\x}$};
\foreach \y in {0,...,3} \node at (\xmin,\y) [left] {\tiny \y};
\draw[style=help lines, ystep=1, xstep=1, color=MySmokenGrey] (\xmin,\ymin) grid (\xmax,\ymax);
\draw (0,3) -- (1,2) [line width=0.25mm, color = red]  node {};
\draw (1,2) -- (2,2) [line width=0.25mm, color = red]  node {};
\draw (2,2) -- (3,1) [line width=0.25mm, color = red]  node {};
\draw (3,1) -- (4,1) [line width=0.25mm, color = blue] node {};
\draw (4,1) -- (5,0) [line width=0.25mm, color = blue] node {};
\draw (2,2) -- (3,1) [color = blue]                    node {};
\draw [line width=0.8pt,densely dotted] (0,3) -- (3,3) [color = red]  node {};
\draw [line width=0.8pt,densely dotted] (2,2) -- (3,2) [color = blue] node {};
\foreach \x/\y/\r in {0/3/3.8,1/2/3.2}
\fill[color=red] (\x,\y) circle [radius=\r pt, color = red];
\foreach \x/\y/\r in {2/2/3.8,3/1/3.2}
\fill[color =violet] (\x,\y) circle [radius=\r pt, color=violet];
\foreach \x/\y/\r in {4/1/3.2,5/0/3.2}
\fill[color=blue] (\x,\y) circle [radius=\r pt, color=blue];
\draw [<->] (3,-1-0.2) -- (0,-1-0.2) node [pos=0.5,below] {\scriptsize$\Char{\width}{\pattern}=3$};
\draw [<->] (2,-2-0.5) -- (5,-2-0.5) node [pos=0.5,below] {\scriptsize$\Char{\width}{\pattern} = 3$};
\draw [<->] (2,-3-1.0) -- (3,-3-1.0) node [pos=0.3,below] {\scriptsize$\CharPar{\Overlap}{\pattern}{0,3} = 2$};
\draw [<->] (2,2) -- (2,3) node [pos=0.5,right] { \scriptsize
  $\CharPar{\VariationOfMax}{\pattern}{0,3}=1$};
\draw [line width=0.8pt,densely dotted] (0,3) -- (7,3) [color = red]  node {};
\draw [line width=0.8pt,densely dotted] (3,1) -- (7,1) [color = blue]
node {};
\draw [<->] (5.7,3) -- (5.7,1) node [pos=0.55,right] { \scriptsize
  $\begin{array}{ll}\Char{\Height}{\pattern}&
     =\CharPar{\Range}{\pattern}{\seqlength}  \\ &= 2 \end{array}$};
\draw (0,3)  [above]  node {\tiny $L_1$};
\draw (2,2)  [below]  node {\tiny $L_2$};
\draw (4,1)  [below]  node {\tiny $S$};
\end{scope}
\end{tikzpicture}
}
\end{figure}


\subsection{Necessary and Sufficient Condition for the Existence of
  an Occurrence of a Regular Expression}
\label{sec:necessary-and-sufficient}

Consider a regular expression~$\pattern$  and a time series~$X  =
\XSeq$ with every~$X_i$ ranging over the same integer interval domain.
There exists a necessary and sufficient condition, based on the
domains and the number of time-series variables, for~$\pattern$ to occur at least once
within the signature of~$X$. 
In order to define this condition we use the~\emph{width} of a regular expression, introduced
in~\defref{def:width}, and the \emph{height} of a regular expression,
introduced in~\defref{def:height}.

\begin{property}
\label{prop:necessary-condition}
\Intro{\pattern} and a time series~$\XSeq$ with every~$X_i$ ranging
over the same integer interval domain~$\Domain$.
The \emph{\NecessaryCondition} is satisfied if the two following
conditions hold:
\begin{enumerate}[(i)]
\item
  The value of~$\seqlength$ is strictly greater
  than~$\Char{\width}{\pattern}$, the width of~$\pattern$.
\item
  The difference between~$\DomainMax$ and~$\DomainMin$ is greater than
  or equal to~$\Char{\Height}{\pattern}$, the height of~$\pattern$.
\end{enumerate}
\end{property}

\begin{example}
  \label{ex:nec:cond:inflexion}
 Consider the~$\pattern = \DecreasingTerracePatternName$ regular
 expression and a time series of length~$\seqlength$ over an integer
 interval domain~$\Domain$.
  We recall the values computed in~Examples~\ref{ex:height}
  and~\ref{ex:width}, namely the height of~$\pattern$ is~$2$, and the
  width of~$\pattern$ is~$3$.
  Hence, the~\NecessaryCondition~is satisfied
  if~$\seqlength > 3$ and~$\DomainMax - \DomainMin \geq 2$.
  \qedexample
\end{example} 

All formulae presented in all the next sections assume that
\propref{prop:necessary-condition}~holds.

\section{Constraints that Restrict the Number of Occurrences of a
Regular Expression}
\label{sec:nb-patterns}

The first family of time-series constraints we consider is the
$\Constraint{nb}\_\pattern(\XSeq,\Result)$ family.
Given a sequence $X=\XSeq$, 
where all~$X_i$ are integer variables, it enforces the
number of occurrences
of pattern~$\pattern$ in $X$ to be equal to $\Result$.
Within this constraint family 
the aggregator is~$\SumAggr$, and
the feature is~$\One$.
The main results of~\secref{sec:nb-patterns} are described by
Theorems~\ref{th:nb-patterns-lower-bound} and \ref{th:nb-patterns}, which respectively
provide a sharp lower bound and a sharp upper bound on the number of
occurrences of a regular expression~$\pattern$ in the signature of a
time series provided all~$X_i$ (with $i\in[1,\seqlength]$)
have the same integer interval domain~$\Domain$.
\secref{sec:nb-patterns} is structured in the following way:
\PlanStyle

\begin{itemize}
\item
  First, \secref{sec:nb-patterns-lower-bound} introduces
  \propref{prop:nb-patterns-lb}, and 
  gives a sharp lower bound on~$\Result$ provided \propref{prop:nb-patterns-lb}
  holds.
\item
  Second, \secref{sec:nb-patterns-upper-bound-not-sharp} provides
  an upper bound, not necessarily sharp, on~$\Result$.
  This bound is valid for any regular expression~$\pattern$ for which the
  overlap characteristics is defined and does not exceed the width of
  $\pattern$.
\item
  Third, \secref{sec:nb-patterns-upper-bound} extends the 
   upper bound on~$\Result$
   of~\secref{sec:nb-patterns-upper-bound-not-sharp}, and shows that the
   extended formula is sharp under some hypothesis
   on the regular expression characteristics:
  \begin{itemize}
    \item 
      \secref{sec:nb-patterns-properties}
      defines~Properties~\ref{prop:nb-patterns-first}
      and~\ref{prop:nb-patterns-second} of regular expressions that
      must hold to obtain a sharp upper bound.
    \item
      \secref{sec:nb-patterns-upper-bound-optimal-structure}
      describes the structure of a \ekaterina{maximal time series for $\Constraint{nb}\_\pattern(\XSeq,\Result)$}
      provided either~\propref{prop:nb-patterns-first}
      or~\propref{prop:nb-patterns-second} holds.
    \item
      Based on the structure of a \ekaterina{maximal time series for $\Constraint{nb}\_\pattern(\XSeq,\Result)$}, \secref{sec:nb-patterns-sharp-upper-bound}
      provides a sharp upper bound on~$\Result$,
      provided either~\propref{prop:nb-patterns-first}
      or~\propref{prop:nb-patterns-second} holds.
  \end{itemize}
\item
  Finally, \secref{sec:nb-patterns-exceptions} gives a sharp upper
  bound on~$\Result$ in a special case of~$\pattern$ being
  $\SteadySequencePatternName$, where
  neither~\propref{prop:nb-patterns-first} nor 
  \propref{prop:nb-patterns-second} is satisfied.
\end{itemize}

\subsection{A Sharp Lower Bound on the Number of Pattern Occurrences} 
\label{sec:nb-patterns-lower-bound}

Consider a~$\Constraint{nb}\_\pattern(X,\Result)$ time-series constraint
with $X=\XSeq$,
where every $X_i$ (with $i\in[1,\seqlength]$) is over the same integer
interval domain~$\Domain$.
This section focusses on providing the lower bound on~$\Result$.
For almost every regular expression of~\tabref{tab:patterns},
we can assign the variables of~$X$ to values in $\Domain$ in a way
that the~signature of~$X$ does not contain any occurrence of the regular
expression~$\pattern$.
The only two exceptions are the~$\SteadyPatternName = \SteadyPattern$
and the~$\SteadySequencePatternName = \SteadySequencePattern$ regular
expressions when~$\DomainMin = \DomainMax$.
The next theorem, namely~\thref{th:nb-patterns-lower-bound}, provides
a sharp lower bound on~$\Result$ assuming the property that we now
introduce holds.

\begin{property}
\label{prop:nb-patterns-lb}
A regular expression~$\pattern$ has the~\emph{\NbWidthLowerSimple}
property for an integer interval
domain~$\Domain$ if~$\pattern$ is a disjunction of
disjunction-capsuled regular expressions
and if at least one of the following conditions holds:
\begin{enumerate}[(i)]
  \item\label{cond:nb-patterns-lb1}
    Every inducing word of~$\pattern$ includes at least one letter that is different from $\reg{=}$.
  \item\label{cond:nb-patterns-lb2}
    Every inducing word of~$\pattern$ includes at least one~$\reg{=}$, and~$\DomainMax>\DomainMin$.
  \end{enumerate}
\end{property}

\begin{theorem}
  \label{th:nb-patterns-lower-bound}
  Consider a~$\Constraint{nb}\_\pattern(X,\Result)$ time-series constraint
  with $X=\XSeq$, where every $X_i$ (with $i\in[1,\seqlength]$)
  is over the same integer interval domain~$\Domain$,
  and, where~$\pattern$ is a disjunction of
  disjunction-capsuled regular expressions.
  If~$\pattern$ has the~\NbWidthLowerSimple~property for~$\Domain$,
  then a sharp lower bound on~$\Result$ is~$0$.
\end{theorem}

\begin{proof}
If~Condition~(\ref{cond:nb-patterns-lb1}) of~\propref{prop:nb-patterns-lb} is satisfied,
then by definition of an inducing word,
every word of~$\Language{\pattern}$ contains at least one letter that
is not~$\reg{=}$.
Hence, the time series $X$, where all variables are assigned to the same value,
has no occurrences of~$\pattern$ in its signature, and thus a sharp
lower bound on~$\Result$ is~$0$.

If~Condition~(\ref{cond:nb-patterns-lb2})
of~\propref{prop:nb-patterns-lb} is satisfied, then every word 
in~$\Language{\pattern}$ contains at least one~$\reg{=}$.
The ground time series of length~$\seqlength$ with alternating~$\DomainMin$ and~$\DomainMin
+ 1$ has no equalities in its signature, and thus
no occurrences of~$\pattern$.
Hence, a sharp lower bound on~$\Result$
equals~$0$. \shiftedqed

\end{proof}

Every regular expression in~\tabref{tab:patterns}
has the~\NbWidthLowerSimple~property for any integer
interval domain~$\Domain$, except~$\SteadyPatternName$
and~$\SteadySequencePatternName$ for the domain~$\Domain$ such
that~$\DomainMin = \DomainMax$.
We now consider the cases of~$\SteadyPatternName$
and~$\SteadySequencePatternName$
where neither condition of \propref{prop:nb-patterns-lb}
holds, which means that~\thref{th:nb-patterns-lower-bound} cannot be applied
for computing a sharp lower bound on~$\Result$.

\begin{proposition}
  \label{propos:lower-bound-steady}
  Consider a~$\Constraint{nb}\_\pattern(\XSeq,\Result)$ time-series
  constraint with~$\pattern$ being the $\SteadyPatternName$ regular
  expression, and with  every~$X_i$ ranging over the same integer interval
  domain~$\Domain$ such that~$\DomainMin = \DomainMax$.
  A sharp lower bound on $\Result$ equals~$\seqlength - 1$.
\end{proposition}

\begin{proof}
  Since~$\DomainMin=\DomainMax$, there exists a single ground time
  series~$\TimeSeries$ of length~$\seqlength$ over~$\Domain$.
  All the time-series variables of $\TimeSeries$ have the same value,
  namely~$\DomainMin$, and thus its signature consists of~$\seqlength
  -1$ equalities.
  The number of occurrences of~$\pattern$ in the signature
  of~$\TimeSeries$ equals~$\seqlength - 1$, which is thus a
  sharp lower bound on~$\Result$.
\shiftedqed
\end{proof}

\begin{sloppypar}
\begin{proposition}
  \label{propos:lower-bound-steady_sequence}
  Consider a~$\Constraint{nb}\_\pattern(\XSeq,\Result)$ time-series
  constraint with~$\pattern$ being the $\SteadySequencePatternName$ regular
  expression, and with every~$X_i$ ranging over the same integer interval
  domain~$\Domain$ such that~$\DomainMin = \DomainMax$.
  A sharp lower bound on $\Result$ equals~$1$.
\end{proposition}
\end{sloppypar}

\begin{proof}
 Since~$\DomainMin=\DomainMax$, there exists a single ground time
  series~$\TimeSeries$ of length~$\seqlength$ over~$\Domain$.
  All the time-series variables of $\TimeSeries$ have the same value,
  namely~$\DomainMin$, and thus its signature consists of~$\seqlength
  -1$ equalities.
  The number of occurrences of~$\pattern$ in the signature
  of~$\TimeSeries$ equals~$1$, which is thus is a
  sharp lower bound on~$\Result$.
\shiftedqed
\end{proof}

\subsection{Step~$1$: A Not Necessarily Sharp Upper Bound}
\label{sec:nb-patterns-upper-bound-not-sharp}

Consider a~$\Constraint{nb}\_\pattern(\XSeq, \Result)$ time-series
constraint with every~$X_i$ ranging over the same integer interval
domain~$\Domain$.
\lemref{lemma:nb-of-occurrences} of this section provides an upper bound, not
necessarily sharp, on~$\Result$. 
Intuitively, to get a maximal number of~$\pattern$\nobreakdash-patterns,
every~extended~$\pattern$\nobreakdash-pattern should be as short as possible
and every two consecutive~extended~$\pattern$\nobreakdash-patterns
should have a maximal number of time-series variables in common.
Although, it is not sharp in general,
it is sharp for all regular expressions
in~\tabref{tab:patterns}, except~$\DecreasingPatternName$,
 $\IncreasingPatternName$, $\DecreasingTerracePatternName$, and
 $\IncreasingTerracePatternName$.

We first define the notion of~\emph{interval without restart}, in order to
identify a subseries such that every two
consecutive extended~$\pattern$-patterns within this subseries
have~$\CharPar{\Overlap}{\pattern}{\DomainMin, \DomainMax} $ common
time-series variables.
This notion will be reused in~\secref{sec:nb-patterns-upper-bound} for
deriving a sharp upper bound on~$\Result$. 

\begin{definition}
  \label{def:interval-wo-restart}
  Consider a regular expression~$\pattern$ and a ground
  time series~$X=\XSeq$ over $\Domain$.
  An \emph{interval without restart} of $X$ is any interval~$[\alpha,\beta]$
  (with $1\leq\alpha\leq\beta\leq\seqlength$),
  such that all the following conditions hold:
  \begin{enumerate}[(1)]
  \item\label{def:interval-wo-restart1}
   Every~$X_k$ (with~$k\in[\alpha,\beta]$) belongs
    to at least one extended~$\pattern$-pattern for which all
    time-series variables have indices in~$[\alpha,\beta]$.
  \item\label{def:interval-wo-restart2}
    The width of every extended~$\pattern$-pattern whose time-series variable
    indices are in~$[\alpha,\beta]$ is equal to~$\Char{\width}{\pattern}+1$.
  \item\label{def:interval-wo-restart3}
    Every pair of consecutive extended~$\pattern$-patterns,
    whose time-series variables indices are in~$[\alpha,\beta]$,
    share~$\CharPar{\Overlap}{\pattern}{\DomainMin, \DomainMax}$ time-series variables.
  \item\label{def:interval-wo-restart4}
    When~$\CharPar{\Overlap}{\pattern}{\DomainMin, \DomainMax} > 0$ every
    extended~$\pattern$-pattern,
    whose time-series variables indices are in~$[\alpha,\beta]$,
    has no common time-series variables with any extended~$\pattern$-pattern
    that has an index outside~$[\alpha,\beta]$.
  \end{enumerate}
\end{definition}

Note that, by Condition~\emph{(\ref{def:interval-wo-restart4})}
of~\defref{def:interval-wo-restart},
the intervals without restart of a ground time series are always disjoint.
Consequently two consecutive extended~$\pattern$-patterns
belonging to distinct intervals without restart
do not share any time-series variable.

\begin{sloppypar}
  \begin{example}
  \ExStyle
  \label{ex:interval-wo-restart}
  We consider an example of intervals without restart
  for the~$\pattern = \DecreasingTerracePatternName $ regular
  expression.
  For the time series~$X=\Tuple{4,3,3,2,2,1,4,2,2,1}$,
  the intervals~$[1,6]$ and~$[7,10]$ are  intervals without
  restart corresponding to the subseries~$\TimeSeries_1 =
  \Tuple{4,3,3,2,2,1}$ and~$\TimeSeries_2 = \Tuple{4,2,2,1}$, because:
  \renewcommand{\labelitemi}{$\ast$}
  \begin{itemize}
  \item 
    Each $X_i$ (with
    $i\in[1,6]$ or $i\in[7,10]$) belongs to at least one
    extended~$\pattern$-pattern  (Condition~\emph{(\ref{def:interval-wo-restart1})}
    of~\defref{def:interval-wo-restart}).
  \item
    The subseries~$\TimeSeries_1$
    (respectively~$\TimeSeries_2$)~contains~$2$ (respectively~$1$)
    extended~$\pattern$\nobreakdash-patterns of shortest length~$4$  (Condition~\emph{(\ref{def:interval-wo-restart2})}
    of~\defref{def:interval-wo-restart}).
  \item
    The two consecutive extended~$\pattern$\nobreakdash-patterns
    of~$\TimeSeries_1$ 
    have~$\CharPar{\Overlap}{\pattern}{1,4} = 2$ time-series
    variables in common (Condition~\emph{(\ref{def:interval-wo-restart3})}
    of~\defref{def:interval-wo-restart}).
  \item
    There is no extended~$\pattern$-pattern straddling
    between~$[1,6]$ and~$[7,10]$ (Condition~\emph{(\ref{def:interval-wo-restart4})}
    of~\defref{def:interval-wo-restart}).
\qedexample
    \end{itemize}
\end{example}
\end{sloppypar}

\begin{lemma}
  \label{lemma:nb-of-occurrences}
  Consider a regular expression~$\pattern$, and a time series~$X =
  \XSeq$, with every~$X_i$ ranging 
  over the same integer interval domain~$\Domain$ such
  that~$\CharPar{\Overlap}{\pattern}{\DomainMin, \DomainMax} \leq \Char{\width}{\pattern}$.
  \begin{enumerate}[(i)]
  \item\label{lemma:nb-of-occurrences1}
  The number of~$\pattern$\nobreakdash-patterns in~$X$ is
  bounded 
  by~$\Frac{\max(0,\seqlength-\CharPar{\Overlap}{\pattern}{\DomainMin,
      \DomainMax})}{\Char{\width}{\pattern} + 1-\CharPar{\Overlap}{\pattern}{\DomainMin,
      \DomainMax}}$.
  \item\label{lemma:nb-of-occurrences2}
  In addition, if~$\seqlength \leq \Char{\width}{\pattern}$ or there
  exists at least one ground time series of length~$\seqlength$
  over~$\Domain$ that contains a single interval
  without restart longer
  than~$\seqlength-\Char{\width}{\pattern} -1 +
  \CharPar{\Overlap}{\pattern}{\DomainMin, \DomainMax}$, then
  the mentioned upper bound is sharp.
  \end{enumerate}
\end{lemma}

\begin{proof} 
 Since~$\CharPar{\Overlap}{\pattern}{\DomainMin, \DomainMax} \leq
 \Char{\width}{\pattern}$, the denominator  $\Char{\width}{\pattern} + 1-\CharPar{\Overlap}{\pattern}{\DomainMin,
      \DomainMax}$ of the considered bound, is always positive.
When $\seqlength \leq \Char{\width}{\pattern}$ the formula
  $\Frac{\max(0,\seqlength-\CharPar{\Overlap}{\pattern}{\DomainMin,
      \DomainMax})}{\Char{\width}{\pattern} + 1-\CharPar{\Overlap}{\pattern}{\DomainMin,
      \DomainMax}}$ gives~$0$ as the result,
  which is the upper bound on~$\Result$.
  Consider now the case when $\seqlength >\Char{\width}{\pattern}$.
  
 \noindent  [Proof of (\ref{lemma:nb-of-occurrences1})]
  We construct a time series~$\TimeSeries$ that we prove to have the maximum
  number of~$\pattern$\nobreakdash-patterns among all ground time series of
  length~$\seqlength$ without considering any domain restrictions.
\PlanStyle
  \begin{itemize}
  \item
  We assume that the constructed
  time series~$\TimeSeries$
  has a single interval without restart, which is longer
  than~$\seqlength-\Char{\width}{\pattern} -1 + \CharPar{\Overlap}{\pattern}{\DomainMin,
    \DomainMax}$.
  Note that such a time series may not be feasible over~$\Domain$.
  \item
  By definition of an interval without restart, every pair of consecutive
  extended~$\pattern$\nobreakdash-patterns of~$\TimeSeries$
  has~$\CharPar{\Overlap}{\pattern}{\DomainMin, \DomainMax}$ common time-series variables.
  In addition, every extended~$\pattern$\nobreakdash-pattern has exactly 
  $\Char{\width}{\pattern} + 1$ time-series variables and
  every~time-series variable whose indice is in the interval without restart belongs to at
  least one extended~$\pattern$\nobreakdash-pattern.
  \item
  We now prove that, for any ground time series,
  the number of~$\pattern$\nobreakdash-patterns cannot exceed the number
  of~$\pattern$\nobreakdash-patterns of the constructed time series~$\TimeSeries$.
  \begin{itemize}
  \item
  Assume that this is not true, then there exists a ground time series
  whose extended~$\pattern$\nobreakdash-patterns are either strictly shorter
  than~$\Char{\width}{\pattern} + 1$ or have a number common
  time-series variables
  strictly greater 
  than~$\CharPar{\Overlap}{\pattern}{\DomainMin, \DomainMax}$.
  \item
  Neither of this statements can be possible by construction
  of~$\TimeSeries$ and the definitions of~$\Char{\width}{\pattern}$
  and~$\CharPar{\Overlap}{\pattern}{\DomainMin, \DomainMax}$.
  \item
    Since the smallest length of an extended~$\pattern$-pattern
    equals~$\Char{\width}{\pattern} + 1$, and since the number of time\nobreakdash-series
    variables outside the interval without restart of~$\TimeSeries$ is strictly
    smaller than~$\Char{\width}{\pattern} + 1 -
    \CharPar{\Overlap}{\pattern}{\DomainMin, \DomainMax}$, the time
    series~$\TimeSeries$ does not have any $\pattern$\nobreakdash-pattern outside
    of its single interval without restart.
  \item
  Hence,~$\TimeSeries$ has the maximum number of~$\pattern$\nobreakdash-patterns
  compared to all ground time series of length~$\seqlength$.
  \end{itemize}
  \end{itemize}
  Let us now estimate the maximum number~$\nbocc$
  of potential~$\pattern$\nobreakdash-patterns in the time series~$\TimeSeries$.
  From the construction of $\TimeSeries$ we have

  \begin{equation}
    \label{eq:num-occurrence-1}
    \underbrace{\Char{\width}{\pattern} + 1 -  \CharPar{\Overlap}{\pattern}{\DomainMin, \DomainMax}+
      \Char{\width}{\pattern} + 1 -  \CharPar{\Overlap}{\pattern}{\DomainMin, \DomainMax} + \dots +
      \Char{\width}{\pattern} + 1 -  \CharPar{\Overlap}{\pattern}{\DomainMin, \DomainMax}}_\text{$\nbocc-1$
      times}  + \underbrace{\Char{\width}{\pattern} + 1}_\text{$1$
      time} + \seqlength_r = \seqlength,
  \end{equation}
  
  where~$\seqlength_r$ is the number of time-series variables outside the 
  interval without restart of~$\TimeSeries$.  
  From~\equalref{eq:num-occurrence-1} and from~$\seqlength_r < \Char{\width}{\pattern} + 1 
    -\CharPar{\Overlap}{\pattern}{\DomainMin, \DomainMax}$ we obtain
 
  \begin{equation}
    \label{eq:num-occurrence-2}
    \nbocc \cdot (\Char{\width}{\pattern} +1) - (\nbocc-1) \cdot
    \CharPar{\Overlap}{\pattern}{\DomainMin, \DomainMax}  +
    \seqlength_r = \seqlength ~~\Rightarrow~~
    \nbocc = \Frac{\seqlength-\CharPar{\Overlap}{\pattern}{\DomainMin,
        \DomainMax}}{\Char{\width}{\pattern}+1-\CharPar{\Overlap}{\pattern}{\DomainMin, \DomainMax}}
 \end{equation}
 
 From the right-hand side of Implication~(\ref{eq:num-occurrence-2}) we have that
 the maximum number
 of~$\pattern$-patterns among all time series of length~$\seqlength$
 over~$\Domain$ is less than or equal to~$\Frac{\seqlength 
   -\CharPar{\Overlap}{\pattern}{\DomainMin, \DomainMax}}{\Char{\width}{\pattern} + 1 
   -\CharPar{\Overlap}{\pattern}{\DomainMin, \DomainMax}}$.

\noindent  [Proof of (\ref{lemma:nb-of-occurrences2})]
If the time series~$\TimeSeries$ constructed in the first part of this
proof is feasible wrt~$\Domain$, then the obtained bound is sharp. 
 \shiftedqed
\end{proof}

\subsection{Step~$2$: Extending the Upper Bound to Get a Sharp Bound Under Some Hypothesis} 
\label{sec:nb-patterns-upper-bound}

Consider a~$\Constraint{nb}\_\pattern(X,\Result)$ time-series constraint
with $X=\XSeq$,
where every $X_i$ (with $i\in[1,\seqlength]$) is over the same integer
interval domain~$\Domain$.
This section focusses on computing a \emph{sharp} upper bound on~$\Result$ under some hypothesis on the characteristics of~$\pattern$.

\subsubsection{Required Properties of Regular Expressions}
\label{sec:nb-patterns-properties}

Building in a greedy way a time-series that maximises
the number of $\pattern$-pattern occurrences requires finding a pair
of words in~$\Language{\pattern}$
such that the superposition of these two words wrt an integer interval
domain \emph{simultaneously optimises}
several characteristics.
Depending on the value of the overlap of~$\pattern$
wrt~$\Tuple{\DomainMin, \DomainMax}$, we have
either the~\emph{\NbUpperOverlap}~property 
when~$\CharPar{\Overlap}{\pattern}{\DomainMin, \DomainMax}>0$, introduced
in~\propref{prop:nb-patterns-first}, or
the~\emph{\NbUpperNoOverlap} property 
when~$\CharPar{\Overlap}{\pattern}{\DomainMin, \DomainMax} = 0$, introduced
in~\propref{prop:nb-patterns-second}.

\ListStyle
\begin{itemize}
\item
  The~\NbUpperOverlap~property holds when there exists a pair of words
  in~$\Language{\pattern}$, whose lengths and
  heights are minimum, and both the overlap and the smallest variation of
  maxima are reached for a superposition of these words, which is not
  a factor of any word in~$\Language{\pattern}$.
\item
  The~\NbUpperNoOverlap~property holds when there exists a word
  in~$\Language{\pattern}$, whose length and height are minimum, and
  this word can be a maximal occurrence
  of~$\pattern$ in the signature of a time series over~$\Domain$.
\end{itemize}

\begin{property}
  \label{prop:nb-patterns-first}
  A regular expression~$\pattern$ has
  the~\emph{\NbUpperOverlap}~property for an integer interval
  domain~$\Domain$,
 if there exists a pair of not necessarily distinct words~$\wordv$ and~$\word$
  in~$\Language{\pattern}$,
  and there exists a superposition~$\wordz_1$ (respectively~$\wordz_2$)
  of~$\wordv$ and~$\word$ (respectively~$\word$ and~$\wordv$)
  wrt~$\Tuple{\DomainMin, \DomainMax}$,
  i.e.,~$\CharPar{\Overlap}{\pattern}{\DomainMin, \DomainMax}>0$,
  such that the following conditions are all
  satisfied:
  \begin{enumerate}[(i)]
  \item 
\label{prop:nb-overlap1}
    $\CharArg{\Height}{\pattern}{\wordv}
    =
    \CharArg{\Height}{\pattern}{\word}
    =
    \Char{\Height}{\pattern}$,
   i.e., $\wordv$ and~$\word$ have their heights being equal to the height of~$\pattern$.
  \item 
\label{prop:nb-overlap2}
    $|\wordv| = |\word| = \Char{\width}{\pattern}$, i.e.,
   $\wordv$ and $\word$ are shortest words
        in~$\Language{\pattern}$.

  \item 
\label{prop:nb-overlap3}
    $|\wordv| +  |\word| - |\wordz_1| + 1= |\word| +  |\wordv| -
    |\wordz_2| + 1= \CharPar{\Overlap}{\pattern}{\DomainMin,
      \DomainMax} \leq \Char{\width}{\pattern}$, i.e.,
   the overlap between $\wordv$ and $\word$ (respectively~$\word$ and
     $\wordv$) wrt~$\Tuple{\DomainMin,   \DomainMax}$ is maximum, and
     its value is bounded by the width of~$\pattern$.

\item
  \label{prop:nb-overlap4}
  Both superpositions~$\wordz_1$ and~$\wordz_2$ are not factors
  of any word in~$\Language{\pattern}$.

  \item
    \label{prop:nb-overlap5}
\begin{equation*}
  \CharPar{\VariationOfMax}{\pattern}{\DomainMin,
    \DomainMax} = 
  \begin{cases}
    \CharArgUp{\Shift}{\pattern}{\wordz_1, \wordv, 1} -
      \CharArgUp{\Shift}{\pattern}{\wordz_1, \word, 1} =
    \CharArgUp{\Shift}{\pattern}{\wordz_2, \word, 1} -
      \CharArgUp{\Shift}{\pattern}{\wordz_2, \wordv, 1}, &
    \textnormal{~if~} \wordv \neq \word \\
    \CharArgUp{\Shift}{\pattern}{\wordz_1, \wordv, 1} -
      \CharArgUp{\Shift}{\pattern}{\wordz_1, \word, 2}, &
    \textnormal{~if~} \wordv = \word,\\   
  \end{cases}
\end{equation*}

i.e., the smallest variation of 
  maxima of superpositions of $\wordv$ and~$\word$
  (respectively~$\word$ and~$\wordv$)
  wrt~$\Tuple{\DomainMin, \DomainMax}$ is reached for their
  superposition~$\wordz_1$ (respectively~$\wordz_2$), and is equal to
  the smallest variation of maxima of $\pattern$
  wrt~$\Tuple{\DomainMin, \DomainMax}$.

\item
\label{prop:nb-overlap5a}
 $\CharArg{\Height}{\pattern}{\wordz_1} =
 \CharArg{\Height}{\pattern}{\wordz_2} = \Char{\Height}{\pattern} +
 \abs{\CharPar{\VariationOfMax}{\pattern}{\DomainMin, \DomainMax}}$, i.e.,
 the height of each of these two superpositions~$\wordz_1$ and~$\wordz_2$ is the
 height of~$\pattern$ plus the absolute value of the smallest variation of maxima
 of~$\pattern$ wrt~$\Tuple{\DomainMin, \DomainMax}$.

\item
\label{prop:nb-overlap6}
 If~$\CharPar{\VariationOfMax}{\pattern}{\DomainMin, \DomainMax} > 0$
  (respectively~$\CharPar{\VariationOfMax}{\pattern}{\DomainMin,
    \DomainMax} < 0$), then neither~$\reg{\wordv<}$
  (respectively~$\reg{\wordv>}$) nor~$\reg{\word<}$ (respectively~$\reg{\word>}$)
  is a factor of any word in~$\Language{\pattern}$.

\end{enumerate}
\end{property}

Every regular expression~$\pattern$ in~\tabref{tab:patterns} has
the~\NbUpperOverlap~property for
any integer interval domain~$\Domain$
such that~$\CharPar{\Overlap}{\pattern}{\DomainMin,
  \DomainMax}>0$.

\begin{example}
  \label{ex:nb-patterns-property-one}

  We now illustrate the~\NbUpperOverlap~property on two
  regular expressions.

  \ExFullStyle
  \begin{itemize}
  
\item
    The~$\pattern = \DecreasingTerracePatternName$ regular expression
    has the~\NbUpperOverlap~property for the integer interval domain $\Domain$ such
    that~$\DomainMax - \DomainMin \geq
    3$, because there exists a pair of
    words~$\wordv = \word = \reg{>=>}$
    in~$\Language{\pattern}$ and their superposition ~$\wordz
    = \reg{>=>=>}$ wrt~$\Tuple{\DomainMin, \DomainMax}$, such that all
    the following conditions are satisfied:
    \begin{itemize}
    \item
      {\makebox[10cm][l]{$\CharArg{\Height}{\pattern}{\wordv} =
      \CharArg{\Height}{\pattern}{\word} = \Char{\Height}{\pattern} =
      2$.} \hfill (Cond.~(\ref{prop:nb-overlap1})~of~\Property~\ref{prop:nb-patterns-first})~}
    \item 
      {\makebox[10cm][l]{$|\wordv| = |\word| = \Char{\width}{\pattern}
          = 3$.} \hfill  (Cond.~(\ref{prop:nb-overlap2})
        of~\Property~\ref{prop:nb-patterns-first})}
    \item 
      {\makebox[10cm][l]{$|\wordv| +  |\word| - |\wordz| + 1
      =\CharPar{\Overlap}{\pattern}{\DomainMin, \DomainMax} = 2 \leq
      \Char{\width}{\pattern} = 3$.}\hfill 
    (Cond.~(\ref{prop:nb-overlap3})
    of~\Property~\ref{prop:nb-patterns-first})~}
\item 
      Since any word in~$\Language{\pattern}$ 
      contains only consecutive equalities, the word $\wordz$ is not a
      factor of any 
      {\makebox[8cm][l]{word
          in~$\Language{\pattern}$.}\hfill  (Cond.~(\ref{prop:nb-overlap4})  of~\Property~\ref{prop:nb-patterns-first})~}
    \item
      {\makebox[10cm][l]{$\CharPar{\VariationOfMax}{\pattern}{\DomainMin,
        \DomainMax} = \CharAllUp{\Shift}{\pattern}{\DomainMin,
        \DomainMax}{\wordz, \wordv, 1} - \CharAllUp{\Shift}{\pattern}{\DomainMin,
        \DomainMax}{\wordz, \word, 2} = -1$.}
    \hfill (Cond.~(\ref{prop:nb-overlap5})
    of~\Property~\ref{prop:nb-patterns-first})~}
\item
   {\makebox[10cm][l]{The height of~$\wordz$ is~$3$, which equals
       $\Char{\Height}{\pattern} + \abs{\CharPar{\VariationOfMax}{\pattern}{\DomainMin,
         \DomainMax}}$.} \hfill (Cond.~(\ref{prop:nb-overlap5a})
    of~\Property~\ref{prop:nb-patterns-first})~}
\item
   {\makebox[10cm][l]{No word in~$\Language{\pattern}$ has~$\reg{<}$, thus 
  $\reg{\wordv<}$ is not a
  factor of any 
 word in~$\Language{\pattern}$.} \hfill (Cond.~(\ref{prop:nb-overlap6})
    of~\Property~\ref{prop:nb-patterns-first})~}
    \end{itemize}
  \item
    The~$\pattern = \SteadySequencePatternName $
    regular expression does
    not have the~\NbUpperOverlap~property for any integer interval
    domain~$\Domain$, because for any pair of
    words~$\wordv$,~$\word$ 
    in~$\Language{\pattern}$, the set of 
    superpositions of~$\wordv$ and~$\word$ wrt~$\Tuple{\DomainMin,
      \DomainMax}$ is empty, and
    thus~$\CharPar{\Overlap}{\pattern}{\DomainMin, \DomainMax} = 0$.
\qedexample
  \end{itemize}
\end{example}

\begin{property}
  \label{prop:nb-patterns-second}
  A regular expression~$\pattern$ has the~\NbUpperNoOverlap~property for
  an integer interval domain $\Domain$,
 if~$\CharPar{\Overlap}{\pattern}{\DomainMin,
      \DomainMax}=0$ and if there exists a word~$\wordv$ in~$\Language{\pattern}$
    such that all the following conditions are satisfied:
 \begin{enumerate}[(i)]
 \item
   \label{prop:nb-patterns-second1}
   $|\wordv| = \Char{\width}{\pattern}$, i.e.,~$\wordv$ is a shortest
   word in~$\Language{\pattern}$.
 \item    
   \label{prop:nb-patterns-second2}
   $\CharArg{\Height}{\pattern}{\wordv} =
   \Char{\Height}{\pattern}$, i.e.,~$\wordv$ has a minimum height
   among all words in~$\Language{\pattern}$.

\item
  \label{prop:nb-patterns-second3}
  Either both words~$\reg{\wordv>}$ and
  $\reg{\wordv<}$ are not
  factors of any word in~$\Language{\pattern}$, or at least one
  of the three words~$\Curly{\reg{\wordv>\wordv}, \reg{\wordv<\wordv}, \reg{\wordv=\wordv}}$ is not a factor of any word
  in~$\Language{\pattern}$, and its height is equal to~$\Char{\Height}{\pattern}$.

\item \label{prop:nb-patterns-second4}
    For any integer~$\seqlength > \Char{\width}{\pattern}$,
    there exists at least one ground time series of
    length~$\seqlength$ over~$\Domain$, whose
    signature contains~$\wordv$
    as a maximal occurrence of~$\pattern$.
\end{enumerate}
\end{property}

Any regular expression~$\pattern$ in~\tabref{tab:patterns} has the~\NbUpperNoOverlap~property 
for any integer interval domain~$\Domain$ such
that~$\CharPar{\Overlap}{\pattern}{\DomainMin, \DomainMax} = 0$, except
the~$\SteadySequencePatternName$ regular
expression for~$\Domain$ such that~$\DomainMin = \DomainMax$.
The case of~$\SteadySequencePatternName$ when~$\DomainMin=\DomainMax$
is discussed in~\exref{ex:nb-patterns-property-two}.

\begin{example}
  \ExFullStyle
  \label{ex:nb-patterns-property-two}
  We illustrate the~\NbUpperNoOverlap~property on two regular
  expressions.
  \begin{itemize}
  \item
    The~$\pattern = \DecreasingTerracePatternName$ regular expression has the~\NbUpperNoOverlap~property
    for any integer interval domain $\Domain$ such
    that~$\DomainMax - \DomainMin = 2$ because (1)~as shown in \exref{ex:superpositions},
    for any two  words of~$\Language{\pattern}$,
    the set of their superpositions wrt~$\Tuple{\DomainMin,
      \DomainMax}$ is empty, and (2)~there exists a word~$\wordv = \reg{>=>}$
      in~$\Language{\pattern}$ that satisfies all the following conditions:
      \begin{itemize}
      \item
        {\makebox[10cm][l]{$|\wordv| = \Char{\width}{\pattern} = 3$.}
          \hfill  (Cond.~(\ref{prop:nb-patterns-second1}) of~\Property~\ref{prop:nb-patterns-second})~}
      \item
        \hspace*{2pt}{\makebox[10cm][l]{$\CharArg{\Height}{\pattern}{\wordv} =
            \Char{\Height}{\pattern} = 2$.}
         \hfill  (Cond.~(\ref{prop:nb-patterns-second2})
          of~\Property~\ref{prop:nb-patterns-second})~}
      \item
        {\makebox[10cm][l]{The word~$\reg{\wordv<\wordv}$ is not a factor of any word
        in~$\Language{\pattern}$, and its height
        is~$2$.}
          \hfill (Cond.~(\ref{prop:nb-patterns-second3}) of~\Property~\ref{prop:nb-patterns-second})~}
      \item
        For any integer~$\seqlength > \Char{\width}{\pattern}$, there
        exists a ground time series of length~$\seqlength$
        over~$\Domain$ whose signature  {\makebox[10cm][l]{contains~$\wordv$ as a maximal
        occurrence of~$\pattern$.}
         \hfill  (Cond.~(\ref{prop:nb-patterns-second4}) of~\Property~\ref{prop:nb-patterns-second})~}
    \end{itemize}
  \item
    Consider the~$\pattern = \SteadySequencePatternName$ regular expression.
    \begin{itemize}
    \item
      First,~$\pattern$ does not have the~\NbUpperNoOverlap~property
      for an integer interval domain~$\Domain$
      such that $\DomainMax-\DomainMin=0$,
      since~Condition~(\ref{prop:nb-patterns-second4}) of~\propref{prop:nb-patterns-second}
      is violated:
      the shortest word of~$\Language{\pattern}$,
      namely~$ \wordu = \reg{=}$ cannot be a maximal occurrence
      of~$\pattern$ in the signature of any ground time series longer than~$2$ over~$\Domain$;
      indeed, for any time-series length, there exists a single ground time series with all equal
      values, thus its signature contains only 
      equalities, which prevents~$\wordu$ to be a maximal occurrence of~$\pattern$. 
    \item
      Second,~$\pattern$ has the~\NbUpperNoOverlap~property for an
      integer interval domain~$\Domain$
      such that $\DomainMax-\DomainMin > 0$ 
     because there exists a word~$\wordv = \reg{=}$
     in~$\Language{\pattern}$ that satisfies all the following conditions:
     \begin{itemize}
      \item
        {\makebox[9.8cm][l]{$|\wordv| = \Char{\width}{\pattern} = 1$.}
          \hfill (Cond.~(\ref{prop:nb-patterns-second1}) of~\Property~\ref{prop:nb-patterns-second})}
      \item
        {\makebox[9.8cm][l]{$\CharArg{\Height}{\pattern}{\wordv} =
            \Char{\Height}{\pattern} = 0$.}
         \hfill  (Cond.~(\ref{prop:nb-patterns-second2})
          of~\Property~\ref{prop:nb-patterns-second})}
      \item
        No word of~$\Language{\pattern}$ contains~$\reg{>}$
        or~$\reg{<}$, hence neither~$\reg{\wordv>}$,
        nor~$\reg{\wordv<}$ 
        are factors of any word \\
        {\makebox[9.8cm][l]{ in~$\Language{\pattern}$.}
         \hfill  (Cond.~(\ref{prop:nb-patterns-second3}) of~\Property~\ref{prop:nb-patterns-second})}
      \item
        For any integer~$\seqlength > \Char{\width}{\pattern}$, there
        exists a ground time series of length~$\seqlength$
        over~$\Domain$ whose  {\makebox[9.0cm][l]{ signature contains~$\wordv$ as a maximal
        occurrence of~$\pattern$.}
         \hfill  (Cond.~(\ref{prop:nb-patterns-second4})
          of~\Property~\ref{prop:nb-patterns-second})}
\qedexample
\end{itemize}
\end{itemize}
\end{itemize}
\end{example}

\subsubsection{Structure of a \ekaterina{Maximal} Time Series} 
\label{sec:nb-patterns-upper-bound-optimal-structure}

Consider a~$\Constraint{nb}\_\pattern(\Tuple{X_1,X_2,\dots,X_\seqlength}, \Result)$
time-series constraint with every~$X_i$ having the same integer interval domain~$\Domain$.
\lemref{lemma:nb-patterns-optimal-solution-structure}
describes the structure of a \ekaterina{maximal time series for
  $\Constraint{nb}\_\pattern(\Tuple{X_1,\dots,X_\seqlength},
  \Result)$} under the hypothesis
that~$\pattern$ has
either the~\NbUpperOverlap~or the~\NbUpperNoOverlap~property for~$\Domain$.

\begin{lemma}
  \label{lemma:nb-patterns-optimal-solution-structure}
\begin{sloppypar}
  Consider a regular expression~$\pattern$ that has
  either the~\NbUpperOverlap~or the \NbUpperNoOverlap~property for an integer
  interval domain~$\Domain$.
  Then for any integer number~$\seqlength > \Char{\width}{\pattern}
  $, there exists a word~$\wordq$ such that any
  ground time series~$\TimeSeries$ of length~$\seqlength$
  over~$\Domain$ whose signature contains~$\wordq$ has the maximum
  number of $\pattern$-patterns among all ground time series of
  length~$\seqlength$ over~$\Domain$.
\end{sloppypar}
\end{lemma}

\noindent $\mathit{Proof}$~We first construct a word~$\wordq$ and we show
  that there is at least one time series of length~$\seqlength$
  over~$\Domain$ whose signature contains~$\wordq$.
  Then, we prove
  that any time series~$\TimeSeries$ of length~$\seqlength$
  over~$\Domain$ whose signature contains~$\wordq$ is \ekaterina{maximal} for
  the~$\Constraint{nb}\_\pattern(\XSeq, \Result)$ time-series
  constraint with every~$X_i$ ranging over~$\Domain$.

  \begin{wrapfigure}[7]{r}{6cm}
\scalebox{1}{
\begin{tikzpicture}
\begin{scope}[yshift=30cm]

\node at (0,0) (FirstV)  {};
\node at (1,0)  (EndFirstV) {};
 \node at (0.7,0.2)  (FirstW){};
\node at (1.7,0.2)  (EndFirstW) {};
 \node at (1.4,0)  (Shift){};
\node at (0,0.1) (Name)  {};

\draw[<->] (0,0) -- (1,0);
\node at ($.5*(FirstV) + .5*(EndFirstV) + (Name)$) {\normalsize $\wordv$};

 \draw[<->] (0.7,0.2) -- (1.7,0.2);
\node at ($.5*(FirstW) + .5*(EndFirstW) + (Name)$) {\normalsize $\word$};

\draw[<->] ($(FirstV) + (Shift)$) -- ($(EndFirstV) + (Shift)$);
\node at ($.5*(FirstV) + .5*(EndFirstV) + (Name) + (Shift)$) {\normalsize $\wordv$};

 \draw[<->] ($(FirstW) + (Shift)$) -- ($(EndFirstW) + (Shift)$); 
\node at ($.5*(FirstW) + .5*(EndFirstW) + (Name) + (Shift)$) {\normalsize $\word$};

\draw[<->] ($(FirstV) + (Shift) + (Shift)$) -- ($(EndFirstV)  + (Shift) + (Shift)$);
\node at ($.5*(FirstV) + .5*(EndFirstV) + (Name) + (Shift) + (Shift)$) {\normalsize $\wordv$};

\draw[<->] ($(FirstW) + (Shift)  + (Shift)$) -- ($(EndFirstW) + (Shift)  + (Shift)$); 
\node at ($.5*(FirstW) + .5*(EndFirstW) + (Name) + (Shift) + (Shift)$) {\normalsize $\word$};

\draw[dotted] (0,0) -- (0,-0.5);
\draw[dotted] (1.7,0.5) -- (1.7,-0.5);
\draw[<->, line width=0.8pt] (0,-0.5) -- (1.7,-0.5);
\node at (0.85,-0.35) {\normalsize  $\wordz_1$};

\draw[dotted] (0.7,0.2) -- (0.7,0.7);
\draw[dotted] (2.4,0) -- (2.4,0.7);
\draw[<->] (0.7,0.7) -- (2.4,0.7);
\node at (1.55,0.85) {\normalsize$\wordz_2$};

\draw[dotted] (1,0) -- (1,-0.8);
\draw[dotted] (1.7,-0.5) -- (1.7,-0.8);
\draw[<->] (1,-0.8) -- (1.7,-0.8);
\node at (1.35,-1.0) {\normalsize$\word_2$};

\draw[dotted] (1.7,0.2) -- (1.7,1);
\draw[dotted] (2.4,0.7) -- (2.4,1);
\draw[<->, line width=0.8pt] (1.7,1) -- (2.4,1);
\node at (2.05,1.15) {\normalsize$\word_1$};

\draw[dotted] (2.4,0) -- (2.4,-0.8);
\draw[dotted] (3.1,0.2) -- (3.1,-0.8);
\draw[<->, line width=0.8pt] (2.4,-0.8) -- (3.1,-0.8);
\node at (.5*2.4 + .5*3.1,-1.0) {\normalsize$\word_2$};

\draw[dotted] (3.1,0.2) -- (3.1,1);
\draw[dotted] (3.8,0) -- (3.8,1);
\draw[<->, line width=0.8pt] (3.1,1) -- (3.8,1);
\node at (.5*3.1 + .5*3.8,1.15) {\normalsize$\word_1$};

\draw[dotted] (3.8,0) -- (3.8,-0.8);
\draw[dotted] (4.5,0.2) -- (4.5,-0.8);
\draw[<->, line width=0.8pt] (3.8,-0.8) -- (4.5,-0.8);
\node at (.5*3.8 + .5*4.5,-1) {\normalsize$\word_2$};
\end{scope}
\end{tikzpicture}
}
\caption{\scriptsize Illustration of the
  word~$\wordz_1\word_1\word_2\word_1\word_2$ belonging to the language
of$~\reg{\wordv
       \hspace*{2pt} | \hspace*{2pt} \wordz_1(\word_1 \word_2)^*(\word_1 \hspace*{2pt}  |  \hspace*{2pt} \varepsilon)}$}
\end{wrapfigure}

 \noindent {\bf Case~(1)}:~\emph{$\pattern$ has
   the~\NbUpperOverlap~property for~$\Domain$.} \\
 \noindent Then there exist two words~$\wordv$ and~$\word$ of~$\Language{\pattern}$
  and a superposition~$\wordz_1$ (respectively~$\wordz_2$) of~$\wordv$
  and~$\word$ (respectively $\word$ and~$\wordv$) wrt~$\Tuple{\DomainMin, \DomainMax}$
  such that all the six conditions of \propref{prop:nb-patterns-first} are satisfied.
  Let~$\word_1$ and~$\word_2$ be the words such that~$\wordz_1 =
  \wordv \word_2$ and~$\wordz_2 = \word \word_1$.
  The figure on the right shows the relations between the
  words~$\wordz_1$, $\wordz_2$, $\wordv$, $\word$, $\word_1$, and~$\word_2$.
 \renewcommand{\labelitemi}{$\circ$}
 \renewcommand{\labelitemii}{$\ast$}
\begin{itemize}
\item
{\bf Step 1}: \emph{Construction of the word~$\wordq$.}

When constructing the word~$\wordq$ we
 consider two cases.

\begin{itemize}
\item  
{\bf Case~(1.1)}: \emph{The variation of maxima~$\CharPar{\VariationOfMax}{\pattern}{\DomainMin,
    \DomainMax}$ equals zero.}
  
  In this case,~$\TimeSeries$ has a single interval without restart
  that contains all~$\pattern$-patterns of~$\TimeSeries$.
  We construct the signature~$\wordq$ of this interval without
  restart by imposing the following conditions:

  \begin{enumerate}[(a)]
  \item
 \label{cond:lemma-nb-structure-1-1}
    The word~$\wordq$ is in the language of~the~$\reg{\wordv
       \hspace*{2pt} | \hspace*{2pt} \wordz_1(\word_1 \word_2)^*(\word_1 \hspace*{2pt}  |  \hspace*{2pt} \varepsilon)}$
    regular expression.
  \item
 \label{cond:lemma-nb-structure-1-2}
    The length of~$\wordq$ is less than~$\seqlength$.
   \item
 \label{cond:lemma-nb-structure-1-3}
    The length of~$\wordq$ is maximum
    among all words satisfying
    Conditions~(\ref{cond:lemma-nb-structure-1-1}),
    and~(\ref{cond:lemma-nb-structure-1-2}).
 \setcounter{ItemCounter}{\value{enumi}}
  \end{enumerate}

  By condition~(\ref{prop:nb-overlap1})
  of~\propref{prop:nb-patterns-first}, the heights of both~$\wordv$
  and~$\word$ equal~$\Char{\Height}{\pattern}$, the height of~$\pattern$.
  Since~$\CharPar{\VariationOfMax}{\pattern}{\DomainMin,
    \DomainMax} = 0$, by~Conditions~(\ref{prop:nb-overlap5})
  and~(\ref{prop:nb-overlap5a}) of~\propref{prop:nb-patterns-first}, the height of~both
  words~$\wordz_1$ and~$\wordz_2$ is~$\Char{\Height}{\pattern}$.
  Hence, the height of~$\wordq$
  is also~$\Char{\Height}{\pattern}$, thus~$\wordq$ indeed appears in
  the signature of some ground time series of length~$\seqlength$
  over~$\Domain$, and~$\TimeSeries$ is feasible.

\item  
{ \bf{Case~(1.2)}}: \emph{The variation of maxima~$\CharPar{\VariationOfMax}{\pattern}{\DomainMin,
    \DomainMax}$ does not equal zero.}
  
  For brevity, we consider only the case when~$\CharPar{\VariationOfMax}{\pattern}{\DomainMin,
    \DomainMax} > 0$, the case of a negative~$\CharPar{\VariationOfMax}{\pattern}{\DomainMin,
    \DomainMax}$ being symmetric.  
  The time series~$\TimeSeries$ may have~$p \geq 1$ intervals without
  restart, hence in order to construct~$\wordq$ we first construct
  the signature~$\wordb$ of every, except possibly the last one, interval
  without restart of~$\TimeSeries$ by imposing the following conditions:
  \begin{enumerate}[(a)]
 \setcounter{enumi}{\value{ItemCounter}}
  \item
    \label{cond:lemma-nb-structure-2-1-1}
    The word~$\wordb$ is in the language of
    the~$\reg{\wordv \hspace*{2pt} |\hspace*{2pt} \wordz_1(\word_1
      \word_2)^*(\word_1 \hspace*{2pt} | \hspace*{2pt} \varepsilon)}$
    regular expression.
  \item
    \label{cond:lemma-nb-structure-2-1-2}
    The set of supporting time series of~$\wordb$
    wrt~$\Tuple{\DomainMin, \DomainMax}$ is not empty.
  \item
     \label{cond:lemma-nb-structure-2-1-3}
     The length of~$\wordb$ is less than~$\seqlength$.
  \item
\label{cond:lemma-nb-structure-2-1-4}
    The length of~$\wordb$ is maximum
    among all words satisfying
    Conditions~(\ref{cond:lemma-nb-structure-2-1-1}),~(\ref{cond:lemma-nb-structure-2-1-2})
    and~(\ref{cond:lemma-nb-structure-2-1-3}).
 \setcounter{ItemCounter}{\value{enumi}}
  \end{enumerate}

  Note that~$\wordb$ always exists, since there is at least one word,
  namely~$\wordv$, satisfying
  Conditions~(\ref{cond:lemma-nb-structure-2-1-1}),~(\ref{cond:lemma-nb-structure-2-1-2})
    and~(\ref{cond:lemma-nb-structure-2-1-3}).
  Then, the word~$\wordq$ must satisfy the following conditions:

  \begin{enumerate}[(a)]
 \setcounter{enumi}{\value{ItemCounter}}
  \item
    \label{cond:lemma-nb-structure-2-2-1}
    The word~$\wordq$ belongs to the language of the~$\reg{(\wordb
    >)^*\tilde{q}_\mathit{rest}}$ regular
    expression, where $\tilde{q}_\mathit{rest}$ is a word in the language
    of the~$\reg{ \wordv
      \hspace*{2pt} |\hspace*{2pt} \wordz_1(\word_1
      \word_2)^*(\word_1 \hspace*{2pt} | \hspace*{2pt} \varepsilon)}$
    regular expression such that $|\tilde{q}_\mathit{rest} | \leq |\wordb|$.
  \item
     \label{cond:lemma-nb-structure-2-2-2}
    The length of~$\wordq$ is less than~$\seqlength$.
  \item
    \label{cond:lemma-nb-structure-2-2-3}
    The length of~$\wordq$ is maximum among all words satisfying
    Conditions~(\ref{cond:lemma-nb-structure-2-2-1}) and~(\ref{cond:lemma-nb-structure-2-2-2}).
\setcounter{ItemCounter}{\value{enumi}}
  \end{enumerate}
  Since~$\CharPar{\VariationOfMax}{\pattern}{\DomainMin, \DomainMax} >
  0$, by~\lemref{lemma:variation-of-maxima}, and by construction
  of~$\wordb$, the word~$\wordb$ does not contain any~$\reg{>}$.
  Then, the concatenation of~$\wordb$ and~$\reg{>}$ has the same 
  height as~$\wordb$.
  Hence, the height of~$\wordq$
  equals the height of~$\wordb$, whose set of supporting time
  series wrt~$\Tuple{\DomainMin, \DomainMax}$ is not empty,
  thus~$\wordq$ indeed appears in
  the signature of some ground time series of length~$\seqlength$
  over~$\Domain$, and~$\TimeSeries$ is feasible.

\end{itemize}

\item 
{\bf Step 2}: \emph{\ekaterina{Maximality} of any time series~$\TimeSeries$ whose signature
   contains~$\wordq$.}

 We now prove that~$\TimeSeries$ is
 a \ekaterina{maximal time series for~$\Constraint{nb}\_\pattern(\XSeq, \Result)$.}

\begin{itemize}
 \item 
   First, we show that the number~$\NbPat$ of~$\pattern$-patterns of~$\TimeSeries$
   equals the number of occurrences of the words~$\wordv$ and~$\word$ in
   its signature.
   By~Conditions~(\ref{prop:nb-overlap4}) and~(\ref{prop:nb-overlap6})
   of~\propref{prop:nb-patterns-first}, the words~$\wordv$ and~$\word$
   appearing in~$\wordq$ cannot be factors of any other
   occurrence of~$\pattern$ in~$\wordq$, hence~$\NbPat$ is not less
   than the number of occurrences of the words~$\wordv$ and~$\word$
   in~$\wordq$.
   By~Conditions~(\ref{prop:nb-overlap3})
   of~\propref{prop:nb-patterns-first}, no extended~$\pattern$-pattern
   can straddle between two other extended~$\pattern$-patterns.
   In addition, by the maximality of the length of~$\wordq$ there is no occurrence
   of~$\pattern$ in the part of the signature of~$\TimeSeries$ that is
   not~$\wordq$.
   Hence, neither is $\NbPat$ greater than the number of occurrences of the words~$\wordv$ and~$\word$
   in~$\wordq$, and thus these values are equal.
   
\item 
  Second, we prove that~$\TimeSeries$ is \ekaterina{maximal for~$\Constraint{nb}\_\pattern(\XSeq, \Result)$}.
  Suppose that~$\TimeSeries$ is not \ekaterina{maximal for~$\Constraint{nb}\_\pattern(\XSeq, \Result)$} and there exists a time
  series~$\TimeSeries'$ of length~$\seqlength$ over~$\Domain$ that has
  a number of~$\pattern$-patterns 
  strictly greater than~$\TimeSeries$.
  Then at least one of the following conditions must be satisfied:
  
  \begin{enumerate}[(a)]
    \setcounter{enumi}{\value{ItemCounter}}
     \item
        \label{cond:lemma-opimal-structure1}
        There is a smaller number of intervals without restart of the
        same total length.
    \item
      \label{cond:lemma-opimal-structure2}
      Some extended~$\pattern$-patterns of such a time series are of length shorter
      than~$\Char{\width}{\pattern} + 1$.
    \item
       \label{cond:lemma-opimal-structure3}
       Some pairs of consecutive extended~$\pattern$-patterns have
       more common time-series variables than~$\CharPar{\Overlap}{\pattern}{\DomainMin,
         \DomainMax}$.
    \item
        \label{cond:lemma-opimal-structure4}
        There is an extended~$\pattern$-pattern that straddles between
        two other extended~$\pattern$-patterns.
          \setcounter{ItemCounter}{\value{enumi}}
  \end{enumerate}
   
  Assumption~(\ref{cond:lemma-opimal-structure1})
  contradicts~Condition~(\ref{prop:nb-overlap5}) 
  of~\propref{prop:nb-patterns-first} and the construction of the
  signature of intervals without restart.
  Assumptions~(\ref{cond:lemma-opimal-structure2})
  and~(\ref{cond:lemma-opimal-structure3}) contradict 
  Conditions~(\ref{prop:nb-overlap2}) and~(\ref{prop:nb-overlap3})
  of~\propref{prop:nb-patterns-first}.
 Finally, Assumption~(\ref{cond:lemma-opimal-structure4}) is not
 possible because of the bound imposed on the value of the overlap in
 Condition~(\ref{prop:nb-overlap3}) 
  of~\propref{prop:nb-patterns-first}.
 Hence,~$\TimeSeries$ has the maximum number of~$\pattern$-patterns
 among all ground time series of the same length over~$\Domain$.

\end{itemize}

\end{itemize}

 \noindent {\bf Case~(2)}:~\emph{$\pattern$ has the~\NbUpperNoOverlap~property for~$\Domain$}.
  
  \noindent There exists a word~$\wordv$ such that all the conditions
  of~\propref{prop:nb-patterns-second} are satisfied.
  The construction of~$\wordq$ is similar to Case~(1), but the
  word~$\wordq$ will always be the signature of a single interval
  without restart. 
  The word~$\wordq$ is built using the following rules:
  \begin{enumerate}[(a)]
 \setcounter{enumi}{\value{ItemCounter}}
  \item
    \label{cond:lemma-opt-structure-case-2-1}
    If both words~$\reg{\wordv>}$ and~$\reg{\wordv<}$ are not proper factors of
    any word in~$\Language{\pattern}$, then~$\wordq$ is in
    the language of the~$\reg{(\wordv>\wordv<)^* \wordv}$ regular expression.
  \item
 \label{cond:lemma-opt-structure-case-2-2}
    If at least one word~$\word$ in~$\Curly{\reg{\wordv>},
      \reg{\wordv=}, \reg{\wordv<}}$ is not a proper factor of any word
    in~$\Language{\pattern}$, and its height
    equals~$\Char{\Height}{\pattern}$, then~$\wordq$ is in
    the language of the~$\reg{\word^* \wordv}$ regular expression.
  \item
 \label{cond:lemma-opt-structure-case-2-3}
    The length of~$\wordq$ is less than~$\seqlength$.
  \item
    The length of~$\wordq$ is maximum among all words satisfying
    Conditions~(\ref{cond:lemma-opt-structure-case-2-1}),
    (\ref{cond:lemma-opt-structure-case-2-2}),
    and~(\ref{cond:lemma-opt-structure-case-2-3}).
  \end{enumerate}
  Since all the conditions of~\propref{prop:nb-patterns-second} are
  satisfied, it can be shown that the height of~$\wordq$ is not
  greater than~$\DomainMax - \DomainMin$, and thus at
  least one time series of length~$\seqlength$ over~$\Domain$ contains~$\wordq$ in its signature.
  Then, in a similar fashion as in Case~(1), one can prove that
  any time series whose signature contains~$\wordq$ is \ekaterina{maximal for~$\Constraint{nb}\_\pattern(\XSeq, \Result)$}.
  \shiftedqed

\subsubsection{A Sharp Upper Bound on the Number of Occurrences of
  Regular Expression} 
\label{sec:nb-patterns-sharp-upper-bound}

Consider a~$\Constraint{nb}\_\pattern(\XSeq, \Result)$ time-series
constraint with every~$X_i$ ranging over the same integer interval
domain~$\Domain$.
First,~\lemref{lemma:nb-patterns-max-length} gives an upper bound on the maximum length of
an interval without restart in a time series over~$\Domain$.
Second, based on this upper bound and the structure of a
\ekaterina{maximal time series for~$\Constraint{nb}\_\pattern(\XSeq, \Result)$} 
showed in~\lemref{lemma:nb-patterns-optimal-solution-structure},
 \thref{th:nb-patterns} provides a sharp upper bound on~$\Result$ under
some hypothesis on the characteristics of the regular expression~$\pattern$.

\begin{lemma}
  \label{lemma:nb-patterns-max-length}
  Consider a regular expression~$\pattern$ and
  an integer interval domain~$\Domain$ such that one of the following
  conditions is satisfied:

\begin{enumerate}[(i)]
\item
  \label{cond:lemma-max-length-1}
  The value of~$\CharPar{\VariationOfMax}{\pattern}{\DomainMin,
    \DomainMax}$ equals zero.
\item
   \label{cond:lemma-max-length-2}
  The value of~$\CharPar{\VariationOfMax}{\pattern}{\DomainMin,
    \DomainMax}$ does not equal zero and~$\pattern$
  has the~\NbUpperOverlap~property.
\end{enumerate}
  
  Then, the \emph{maximum length of an interval without
    restart} of any ground time series over~$\Domain$ is bounded
  by

   \begin{align*}
    \label{eq:plength}     
    \CharPar{\plength}{\pattern}{\DomainMin, \DomainMax}
     &= 
       \begin{cases}
       \Frac{\DomainMax-\DomainMin -
            \Char{\Height}{\pattern} + \abs{\CharPar{\VariationOfMax}{\pattern}{\DomainMin,
          \DomainMax}}}{\abs{\CharPar{\VariationOfMax}{\pattern}{\DomainMin,
              \DomainMax}}} \cdot
          \Parentheses{\Char{\width}{\pattern}+1-\CharPar{\Overlap}{\pattern}{\DomainMin,
              \DomainMax}}+
         \CharPar{\Overlap}{\pattern}{\DomainMin,
              \DomainMax}, & \If \CharPar{\VariationOfMax}{\pattern}{\DomainMin,
              \DomainMax} \neq 0,  \\ 
            + \infty, & \Otherwise. 
        \end{cases}
    \end{align*}
\end{lemma}

\begin{proof}
  {\bf Case~(1)}: \emph{Condition~(\ref{cond:lemma-max-length-1}) is satisfied.}
  
  Since~$\CharPar{\VariationOfMax}{\pattern}{\DomainMin,
    \DomainMax} = 0$, the condition that the maximum
  length of an interval without restart is bounded
  by~$+\infty$ is trivially satisfied.
  This upper bound reflects the fact that when~$\CharPar{\VariationOfMax}{\pattern}{\DomainMin,
    \DomainMax} = 0$, the maximum length of an interval without restart does not depend on
  the domain~$\Domain$.
  
   {\bf Case~(2)}: \emph{Condition~(\ref{cond:lemma-max-length-2}) is satisfied.}
  
  Consider now the case when~$\CharPar{\VariationOfMax}{\pattern}{\DomainMin,
    \DomainMax} \neq 0$ and~$\pattern$ has the \NbUpperOverlap~property.
  Let~$\wordb$ be a word such that 
  (1)~$\wordb$ is the signature of an interval without restart of
  maximum length constructed
  in~\lemref{lemma:nb-patterns-optimal-solution-structure} for a time
  series of some length~$\seqlength$ over~$\Domain$; 
  (2)~for any time series of length~$\seqlength' > \seqlength$ over~$\Domain$,~$\wordb$
  is also the signature of an interval
  without restart of maximum length.
  Note that such~$\wordb$ necessarily exists by condition that the
  set of supporting time series of~$\wordb$ wrt~$\Tuple{\DomainMin,
    \DomainMax}$ must not be empty. 
  Then, there exists a ground time series~$\TimeSeries$  of length~$\seqlength$ over~$\Domain$
  whose signature is~$\wordb$.
  By construction of~$\wordb$,
  the maximum of every
  extended~$\pattern$-pattern of~$\TimeSeries$, except the first one,
  is~$\abs{\CharPar{\VariationOfMax}{\pattern}{\DomainMin, \DomainMax}}$ units
  smaller or greater, depending on the sign
  of~$\CharPar{\VariationOfMax}{\pattern}{\DomainMin, \DomainMax}$,
  compared to the maximum of the preceding extended~$\pattern$\nobreakdash-pattern.
  Thus, the maxima of these extended~$\pattern$-patterns form a
  monotonously decreasing (respectively increasing) sequence
  of integer numbers.
  By~Conditions~(\ref{prop:nb-overlap1}), (\ref{prop:nb-overlap3}), (\ref{prop:nb-overlap4})
  and~(\ref{prop:nb-overlap5}) of~\propref{prop:nb-patterns-first}, 
  the number of elements of such a sequence is bounded by
  $\Frac{\DomainMax - \DomainMin 
      - \Char{\Height}{\pattern} +
      \abs{\CharPar{\VariationOfMax}{\pattern}{\DomainMin, \DomainMax}}
    }{\abs{\CharPar{\VariationOfMax}{\pattern}{\DomainMin, \DomainMax}}}$.
  Since every extended~$\pattern$\nobreakdash-pattern is of
  length~$\Char{\width}{\pattern} + 1$, has a 
  height~$\Char{\Height}{\pattern}$, and the number of common
  time-series variable between two extended~$\pattern$-patterns
  equals~$\CharPar{\Overlap}{\pattern}{\DomainMin, \DomainMax}$,
  the value~$\Frac{ \DomainMax - \DomainMin -
      \Char{\Height}{\pattern} +
      \abs{\CharPar{\VariationOfMax}{\pattern}{\DomainMin,
        \DomainMax}}}{\abs{\CharPar{\VariationOfMax}{\pattern}{\DomainMin,
        \DomainMax}}}\cdot (\Char{\width}{\pattern} + 1 - \CharPar{\Overlap}{\pattern}{\DomainMin,
        \DomainMax})+ \CharPar{\Overlap}{\pattern}{\DomainMin,
        \DomainMax}$ is the maximum length of an interval
  without restart of a ground time series among all ground time
  series over~$\Domain$.
  
  \shiftedqed
\end{proof}

\begin{theorem}
  \label{th:nb-patterns}
  Consider a~$\Constraint{nb}\_\pattern(\XSeq, \Result)$  time-series constraint
  with every~$X_i$ ranging over the same integer interval
  domain~$\Domain$.
  If~$\pattern$ has either the~\NbUpperOverlap~or
  the \NbUpperNoOverlap~properties for~$\Domain$,
  then a sharp upper bound on~$\Result$ is

{\small
  \begin{equation}
    \label{eq:nb-patterns-number-of-occurrences}
    \underbrace{\Frac{\max(0,m -
          \CharPar{\Overlap}{\pattern}{\DomainMin,
        \DomainMax})} { \Char{\width}{\pattern} +1 -
          \CharPar{\Overlap}{\pattern}{\DomainMin,
        \DomainMax}}}_\text{$A$}
    \cdot \underbrace{\vphantom{\Frac{\max(0,m -
          \CharPar{\Overlap}{\pattern}{\DomainMin,
        \DomainMax})} { \Char{\width}{\pattern} +1 -
          \CharPar{\Overlap}{\pattern}{\DomainMin,
        \DomainMax}}} 
        \Frac{\seqlength}{m}}_\text{$B$} 
    + \underbrace{ \Frac{\max(0, (\seqlength
            \bmod m) -  \CharPar{\Overlap}{\pattern}{\DomainMin,
        \DomainMax})}
            { \Char{\width}{\pattern} + 1 - \CharPar{\Overlap}{\pattern}{\DomainMin,
        \DomainMax}
          }}_\text{$C$}
  \end{equation}
}
  where:
  \vspace*{-6pt}
  \ListStyle
  \begin{itemize}
  \item
  $m = \min(\seqlength, \max(1, \CharPar{\plength}{\pattern}{\DomainMin,
    \DomainMax}))$, where~$\CharPar{\plength}{\pattern}{\DomainMin,
        \DomainMax}$ is the upper bound on the maximum length of an
      interval without restart in a time series over~$\Domain$,
      introduced by \lemref{lemma:nb-patterns-max-length}.
\item
  $A$ is the maximum number of~$\pattern$-patterns in an interval without restart of maximum length.
  \item
  $B$ is the number of intervals without restart of maximum length in
  a \ekaterina{maximal time series for the
    $\Constraint{nb}\_\pattern(\XSeq, \Result)$ time-series constraint}.
  \item
  $C$ is the maximum number of~$\pattern$-patterns in an 
  interval without restart of non-maximum length in a \ekaterina{maximal time series for $\Constraint{nb}\_\pattern(\XSeq, \Result)$}.
  
  \end{itemize}
\end{theorem}

\begin{proof}
  \lemref{lemma:nb-patterns-optimal-solution-structure} showed the
  existence of a word~$\wordq$ such that any time series~$\TimeSeries$ of
  length~$\seqlength$ over~$\Domain$ whose signature contains~$\wordq$
  is \ekaterina{maximal for $\Constraint{nb}\_\pattern(\XSeq, \Result)$}.
  Hence, a sharp upper bound on~$\Result$ can be obtained by counting
  the number of occurrences of~$\pattern$ in~$\wordq$.

 Case~(a): $\CharPar{\plength}{\pattern}{\DomainMin,
   \DomainMax} \geq \seqlength - \Char{\width}{\pattern} + \CharPar{\Overlap}{\pattern}{\DomainMin,
   \DomainMax}$.
 Then,~$\TimeSeries$ contains a
 single interval without restart longer than~$\seqlength -
 \Char{\width}{\pattern} + \CharPar{\Overlap}{\pattern}{\DomainMin,
   \DomainMax}$.
 Further, the value of~$\min(\seqlength, \max(1,\CharPar{\plength}{\pattern}{\DomainMin,
   \DomainMax}))$ equals~$\seqlength$, and
 the components~$B$ and~$C$ become respectively equal to~$1$ and~$0$,
 thus Formula~(\ref{eq:nb-patterns-number-of-occurrences}) simplifies to~$A$.
 By~\lemref{lemma:nb-of-occurrences},
 the obtained value is a sharp upper bound on~$\Result$.

Case~(b):~$\CharPar{\plength}{\pattern}{\DomainMin,
   \DomainMax} < \seqlength - \Char{\width}{\pattern} + \CharPar{\Overlap}{\pattern}{\DomainMin,
   \DomainMax} $.
 Then~$\TimeSeries$ may contain multiple intervals
 without restart.
 Furthermore, the length of all intervals without restart
 of~$\TimeSeries$, except maybe the last one,
 equals~$\plength_\pattern^{\Domain}$.
 By~\lemref{lemma:nb-of-occurrences}, the maximum number of~$\pattern$-patterns within every interval
 without restart of maximum length is~$\Frac{
   \max(0,m- \CharPar{\Overlap}{\pattern}{\DomainMin,
   \DomainMax})}{
     \Char{\width}{\pattern} + 1 - \CharPar{\Overlap}{\pattern}{\DomainMin,
   \DomainMax}}$, i.e., the term~$A$.
 The number of intervals without restart of maximum length is
 $\Frac{ \seqlength}{
     m} $, i.e., the term~$B$.
 The last interval without restart of~$\TimeSeries$ may be shorter
 than~$\CharPar{\plength}{\pattern}{\DomainMin,
   \DomainMax}$, then its length is computed as~$\seqlength \bmod
 m$, and the number of~$\pattern$-patterns
 in the last interval without restart is computed as~$
   \Frac{ \max(0,(\seqlength \bmod m) - \CharPar{\Overlap}{\pattern}{\DomainMin,
   \DomainMax})}{\Char{\width}{\pattern}  + 1- \CharPar{\Overlap}{\pattern}{\DomainMin,
   \DomainMax}}$, which is~$C$.
\shiftedqed
\end{proof}

\begin{example}
\ExStyle

 Consider
  a~$\Constraint{nb}\_\pattern(\XSeq, \Result)$ time-series
  constraint with every~$X_i$ ranging over the same integer interval
  domain~$\Domain$.
  Let~$\pattern$  be the~$\DecreasingTerracePatternName$ regular expression.
  \begin{itemize}
  \item
  First, assume that~$\DomainMax - \DomainMin = 2$, and recall
    some of the computed characteristics,
    namely~$\CharPar{\Overlap}{\pattern}{\DomainMin,
      \DomainMax} = 0$,
    $\Char{\width}{\pattern}=3$
    and~$\CharPar{\VariationOfMax}{\pattern}{\DomainMin,
   \DomainMax}=0$.
 It was shown in~\exref{ex:nb-patterns-property-two} that~$\pattern$
 has the~\NbUpperNoOverlap~property for~$\Domain$,
 thus~\thref{th:nb-patterns} can be applied for computing a sharp
 upper bound on~$\Result$.
  By~\lemref{lemma:nb-patterns-max-length}, we have that~$\CharPar{\plength}{\pattern}{\DomainMin,
   \DomainMax} = + \infty$, and thus a sharp
  upper bound on~$\Result$
  is~$\Frac{\max(0, \min(\seqlength, \max(1,\CharPar{\plength}{\pattern}{\DomainMin,
   \DomainMax})) - \CharPar{\Overlap}{\pattern}{\DomainMin,
   \DomainMax})}{ \Char{\width}{\pattern} +1 -
      \CharPar{\Overlap}{\pattern}{\DomainMin,
   \DomainMax}}= \Frac{\max(0,\min(\seqlength, \max(1,+\infty)) -0)}{ 3 +1 -0} = 
  \Frac{\seqlength}{ 4}$.

  \item
  Second, assume~$\DomainMax - \DomainMin \geq 3$,
  then~$\CharPar{\Overlap}{\pattern}{\DomainMin,
   \DomainMax}$ is now equal to~$2$, and~$\CharPar{\VariationOfMax}{\pattern}{\DomainMin,
   \DomainMax}$ is equal to~$-1$.
 It was shown in~\exref{ex:nb-patterns-property-two} that~$\pattern$
 has the~\NbUpperOverlap~property for~$\Domain$,
 thus~\thref{th:nb-patterns} can be applied for computing a sharp
 upper bound on~$\Result$, and a sharp upper bound on~$\Result$
  is equal
  to~$\Frac{\max(0,m - 2)}{2} 
  \cdot \Frac{\seqlength}{m} + \Frac{\max\Parentheses{0, (\seqlength
        \bmod m) -  2}}{2}$ where~$m = \min( \seqlength, \max(1,\CharPar{\plength}{\pattern}{\DomainMin,
   \DomainMax})) = \min(\seqlength, \max(1,(\DomainMax-\DomainMin-1))\cdot
   2+2)$, computed by using 
 \lemref{lemma:nb-patterns-max-length}. \qedexample
  \end{itemize}

\end{example}

\begin{sloppypar}
All the~$22$ regular expression in~\tabref{tab:patterns} have
either the~\NbUpperOverlap~or the~\NbUpperNoOverlap~property for any integer interval
domain~$\Domain$, except the~$\SteadySequencePatternName$ regular
expression when~$\DomainMin = \DomainMax$.
A sharp upper bound on the result variable of a time-series constraint
in this case is given in~\proposref{propos:nb-ub-steady-sequence}.
\end{sloppypar}

\subsection{A Sharp Upper Bound: Special Case}
\label{sec:nb-patterns-exceptions}

\proposref{propos:nb-ub-steady-sequence} provides a sharp upper
bound on the number of occurrences of the~$\SteadySequencePatternName$
regular expression in the signature of a time series over an integer
interval domain~$[\DomainMin, \DomainMax]$ such that~$\DomainMin = \DomainMax$.

\begin{sloppypar}
\begin{proposition}
\label{propos:nb-ub-steady-sequence}
Consider a~$\Constraint{nb}\_\pattern(\XSeq, \Result)$
time-series constraint with~$\pattern$ being
the $\SteadySequencePatternName$ regular expression and with
every~$X_i$ ranging over the  same integer
interval domain~$[\DomainMin, \DomainMax]$ such that~$\DomainMin = \DomainMax$.
A sharp upper bound on~$\Result$ equals~$1$.
\end{proposition}
\end{sloppypar}

\begin{proof}
Since~$\DomainMin = \DomainMax$, there exists a single time series of
length~$\seqlength$ over~$\Domain$, and all its time-series variables
have the same value, namely~$\DomainMin$.
The entire signature of this time series is the word
in~$\Language{\pattern}$, thus a sharp upper bound on~$\Result$
equals~$1$.
\shiftedqed
\end{proof}

\section{Time-Series Constraints with Feature \textsc{width}}
\label{sec:width-constraints}

We now consider 
the~$\Constraint{g\_width}\_\pattern(\XSeq,
\Result)$ family of time-series constraints with every~$X_i$ ranging
over the same integer interval domain~$\Domain$,
i.e., the case when the feature is~$\Width$, $\Constraint{g}$ is
in the set~$\Curly{\MaxAggr, \MinAggr, \SumAggr}$
 and~$\pattern$ is a non-fixed length regular
expression.
\secref{sec:width-properties} defines Properties~\ref{prop:max-width}
and~\ref{prop:sum-width} of regular expressions that we use to
obtain sharp upper bounds on~$\Result$.
All the regular expressions in~\tabref{tab:patterns} have both
Properties~\ref{prop:max-width} and~\ref{prop:sum-width}. 
Based on these properties,~\secref{sec:max-width} (respectively~\secref{sec:sum-width})
provides a sharp upper bound on~$\Result$ when~$\Constraint{g}$
is~$\MaxAggr$ (respectively~$\SumAggr$). 
Finally,~\secref{sec:min-width} gives a sharp lower bound on~$\Result$
when~$\Constraint{g}$ is~$\SumAggr$. 
Note that we do not consider a lower (respectively upper) bound for
the case when the aggregator is~$\MaxAggr$ (respectively~$\MinAggr$),
since when~$\pattern$ has the~\NbWidthLowerSimple~property
(see~\propref{prop:nb-patterns-lb}) for~$\Domain$,
there exists a time series of length~$\seqlength$
over~$\Domain$ that
has no~$\pattern$-patterns, and thus yields the default value
of~$\Result$, namely~$-\infty$ (respectively~$+ \infty$).
Among the~$22$ regular expressions in~\tabref{tab:patterns} only
the~$\SteadyPatternName$ and the~$\SteadySequencePatternName$ regular
expressions do not have the~\NbWidthLowerSimple~property for a domain with a
single element, i.e.,~$\DomainMin = \DomainMax$.

\subsection{Properties of Regular Expressions}
\label{sec:width-properties}

\propref{prop:max-width} is used for deriving a sharp
upper bound on~$\Result$ for
a~$\Constraint{max\_width}\_\pattern(\XSeq, \Result)$ time-series
constraint.
\propref{prop:max-width} requires the range of a
regular expression be a monotonically increasing linear function of~$\seqlength$.

\begin{property}
  \label{prop:max-width}
  A regular expression~$\pattern$ has the~\emph{\WidthUpper}~property if the
  following conditions are all satisfied:
  \begin{enumerate}[(i)]
  \item
    \label{cond:max-width-1}
    There exists a shortest word in~$\Language{\pattern}$
    whose height equals~$\Char{\Height}{\pattern}$, the height of~$\pattern$.
  \item
    \label{cond:max-width-2}
    For every time-series length~$\seqlength > \Char{\width}{\pattern}
    + 1
    $, the range of~$\pattern$
    wrt~$\Tuple{\seqlength}$,~$\CharPar{\Range}{\pattern}{\seqlength}$, is defined and
    equals $\Char{\FullExtLinCoeff}{\pattern}
    \cdot ( \seqlength - 1 - \Char{\Height}{\pattern}) +
  \Char{\FullExtFreeCoeff}{\pattern} + \Char{\Height}{\pattern} $ with
   $ \Tuple{\Char{\FullExtLinCoeff}{\pattern},
     \Char{\FullExtFreeCoeff}{\pattern} } \in \{\Tuple{0,0},
   \Tuple{0,1}, \Tuple{1,0}\}$.
  \end{enumerate}
\end{property}

\propref{prop:sum-width} is used for deriving a sharp
upper bound on~$\Result$ for
a~$\Constraint{sum\_width}\_\pattern(\XSeq,$ $\Result)$ time-series
constraint.

\begin{property}
  \label{prop:sum-width}
  A regular expression~$\pattern$ has
  the~\emph{\WidthUpperSum}~property for an integer interval domain~$\Domain$ if the
  following conditions are all satisfied:
  \begin{enumerate}[(i)]
  \item
    \label{cond:sum-width-1}
    $\CharPar{\Overlap}{\pattern}{\DomainMin, \DomainMax} \leq
    \Char{\After}{\pattern} + \Char{\Before}{\pattern}$.
  \item
    \label{cond:sum-width-2}
    If for every time-series length~$\seqlength >
    \Char{\width}{\pattern} + 1$, the range of~$\pattern$
    wrt~$\Tuple{\seqlength}$,~$\CharPar{\Range}{\pattern}{\seqlength}$,
    equals~$ \seqlength - 1 $, then~$a_\pattern$,~$b_\pattern$
    and~$\CharPar{\Overlap}{\pattern}{\DomainMin, \DomainMax} $ are
    all equal to~$0$, and~$\Char{\width}{\pattern}$, the width
    of~$\pattern$, is equal to~$1$.
  \end{enumerate}
\end{property}

Condition~(\ref{cond:sum-width-1}) of~\propref{prop:sum-width} withdraws from
consideration a regular expression~$\pattern$
whose~$\pattern$\nobreakdash-patterns overlap, i.e., some time-series variables
belong simultaneously to two~$\pattern$\nobreakdash-patterns, which
will be formalised in~\lemref{lem:sum-width}. 
Condition~(\ref{cond:sum-width-2}) of~\propref{prop:sum-width}
restricts further a class
of regular expressions whose range depends linearly on~$\seqlength$.

\subsection{Upper Bound for $\Constraint{max\_width}\_\pattern$\label{sec:max-width}}

We first consider the case when the aggregator is~$\MaxAggr$, i.e., the~$\Constraint{max\_width}\_\pattern(\XSeq,\Result)$ family
of time-series constraints
with~$\pattern$ being a non-fixed length regular expression and
every~$X_i$ ranging over the same integer interval
domain~$\Domain$.
To compute a sharp upper bound on~$\Result$,
we maximise the width of a~$\pattern$-pattern in~$X = \XSeq$.
We do so by detecting a longest word in~$\Language{\pattern}$ 
that may appear in the signature of~$X$.
The transition from the length of a~$\pattern$-pattern to the length
of the corresponding word in~$\Language{\pattern}$ is sound because
the width of the~$\pattern$-pattern is the width of
the corresponding word plus $1$ and minus the sum
of~$\Char{\After}{\pattern}$ and~$\Char{\Before}{\pattern}$, which are
constant parameters of~$\pattern$, introduced in~\tabref{tab:patterns}.

A trivial but, possibly not sharp upper bound on~$\Result$
is~$\seqlength - \Char{\After}{\pattern} - \Char{\Before}{\pattern}$.
Further, for regular expressions that have the~\WidthUpper~property,
we show that the sharpness of the mentioned upper bound depends only on the
difference between~$\DomainMax$ and~$\DomainMin$.

The idea for computing a sharp upper bound on~$\Result$
when~$\pattern$ has the \WidthUpper~property is to
identify the minimum value~$\MinDiff$ of~$\DomainMax - \DomainMin$ such that the
bound~$\seqlength - \Char{\After}{\pattern}
- \Char{\Before}{\pattern}$ is still sharp.
When~$\DomainMax - \DomainMin$ is
smaller than~$\MinDiff$ we need to find the maximum value of~$\WordLength <
\seqlength$, such that~$\WordLength - \Char{\After}{\pattern} -
\Char{\Before}{\pattern}$ is a sharp upper bound on~$\Result$ for
a~$\Constraint{max\_width}\_\pattern(\Tuple{X_1, X_2,\dots,X_\WordLength},\Result)$
time-series constraint
with every~$X_i$ ranging over~$\Domain$.

The next theorem provides a sharp
upper bound on~$\Result$ when the regular expression~$\pattern$
has the \WidthUpper \\ property.

\begin{theorem}
  \label{th:max-width}

  Consider a~$\Constraint{max\_width}\_\pattern ( \XSeq,
  \Result)$ time-series constraint with~$\pattern$ being a non-fixed
  length regular expression, and all~$X_i$ ranging over the same integer interval domain
  $\Domain$.
  If~$\pattern$ has the~\WidthUpper~property, then a sharp upper bound
  on~$\Result$ is
\setcounter{equation}{0} 
\begin{numcases}{}
  \seqlength - \Char{\After}{\pattern}  -
   \Char{\Before}{\pattern} & \textnormal{if} 
  $\DomainMax - \DomainMin \geq
  \CharPar{\Range}{\pattern}{\seqlength}$  \label{formula:max-width-1} \\
 \Char{\FullExtLinCoeff}{\pattern} \cdot (\DomainMax - \DomainMin +1 - \Char{\After}{\pattern}  -\Char{\Before}{\pattern}  ) +
  \Char{\FullExtFreeCoeff}{\pattern}  \cdot  (\Char{\width}{\pattern}
  + 1 - \Char{\After}{\pattern}  - \Char{\Before}{\pattern} ) &
  \textnormal{if} $\DomainMax - \DomainMin <
   \CharPar{\Range}{\pattern}{\seqlength}$  \label{formula:max-width-2}
\end{numcases}

where $\Char{\FullExtLinCoeff}{\pattern}$
    and
  $\Char{\FullExtFreeCoeff}{\pattern} $ are
  parameters of the regular expression~$\pattern$, introduced
  in~\propref{prop:max-width}.

\end{theorem}

\begin{proof}
  When the regular expression~$\pattern$ has the~\WidthUpper~property,
  the range~$\CharPar{\Range}{\pattern}{\seqlength}$ of $\pattern$
  wrt~$\Tuple{\seqlength}$ is a monotonically increasing function
  of~$\seqlength$.
  It implies that, if the bound $\seqlength - \Char{\After}{\pattern} -
  \Char{\Before}{\pattern}$ is sharp for some interval integer
  domain $[\DomainMin_1, \DomainMax_1]$, then it is also sharp for any
  interval integer domain $[\DomainMin_2, \DomainMax_2]$ such that
  $\DomainMax_2 - \DomainMin_2  >\DomainMax_1 - \DomainMin_1$.
  Hence, the sharpness of the upper  bound
  $\seqlength - \Char{\After}{\pattern} - \Char{\Before}{\pattern}$ depends
  only on~$\DomainMax - \DomainMin$.
  
  \noindent[Case~(\ref{formula:max-width-1}): $\DomainMax -
  \DomainMin \geq \CharPar{\Range}{\pattern}{\seqlength }$].
  By definition of~$\CharPar{\Range}{\pattern}{\seqlength }$, we have
  that if~$\DomainMax -
  \DomainMin \geq \CharPar{\Range}{\pattern}{\seqlength }$, then there
  exists a word in~$\Language{\pattern}$
  of length~$\seqlength - 1$ whose height is not greater than~$\DomainMax - \DomainMin$.
  Hence,~$\seqlength -\Char{\After}{\pattern} - \Char{\Before}{\pattern}$ is a sharp
  upper bound on~$\Result$.
  
  \noindent[Case~(\ref{formula:max-width-2}): $\DomainMax -
  \DomainMin < \CharPar{\Range}{\pattern}{\seqlength }$].
  This case requires a more detailed analysis than~Case~(\ref{formula:max-width-1}).
  Let us consider the three distinct pairs of
  $\Tuple{\Char{\FullExtLinCoeff}{\pattern},
    \Char{\FullExtFreeCoeff}{\pattern}}$ from
  Condition~(\ref{cond:max-width-2}) of~\propref{prop:max-width}:

  \begin{enumerate}[(a)]
  \item
    The case of $\Tuple{\Char{\FullExtLinCoeff}{\pattern},
    \Char{\FullExtFreeCoeff}{\pattern}}$ being
    $\Tuple{0,0}$.
    Since~$\DomainMax - \DomainMin <  0
    \cdot ( \seqlength - 1 - \Char{\Height}{\pattern}) +  0 + \Char{\Height}{\pattern} =
    \Char{\Height}{\pattern}$, the~\NecessaryCondition,
    i.e.,~\propref{prop:necessary-condition}, is
    not satisfied, thus~$\Result$ is equal to its default
    value, namely~$0$.
  \item
    The case of $\Tuple{\Char{\FullExtLinCoeff}{\pattern},
    \Char{\FullExtFreeCoeff}{\pattern}}$ being
    $\Tuple{0,1}$.
    Since~$\DomainMax - \DomainMin <  0
    \cdot ( \seqlength - 1 - \Char{\Height}{\pattern}) +  1 + \Char{\Height}{\pattern}
    = \Char{\Height}{\pattern} + 1$, the only words
    in~$\Language{\pattern}$ that can appear in the signature of a
    ground time series over~$\Domain$ are the
    ones with the minimum height, namely~$\Char{\Height}{\pattern}$.
    For every time-series length~$\seqlength > \Char{\width}{\pattern} +
    1$, we have that~$\CharPar{\Range}{\pattern}{\seqlength} =
    \Char{\Height}{\pattern} + 1$, which implies that for every word
    in~$\Language{\pattern}$ of length strictly greater
    than~$\Char{\width}{\pattern}$, the height is at
    least~$\Char{\Height}{\pattern} + 1$.  
    Hence, only a word of length~$\Char{\width}{\pattern}$ and of
    height~$\Char{\Height}{\pattern}$ can be an occurrence
    of~$\pattern$ in the signature of a ground time series over~$\Domain$.
    By Condition~(\ref{cond:max-width-1}) of~\propref{prop:max-width},
    such a word exists in~$\Language{\pattern}$
    and thus, a sharp upper
    bound on~$\Result$ is $\Char{\width}{\pattern}  +1
    - \Char{\After}{\pattern}  - \Char{\Before}{\pattern} $.
  \item
    The case of $\Tuple{\Char{\FullExtLinCoeff}{\pattern},
    \Char{\FullExtFreeCoeff}{\pattern}}$ being
    $\Tuple{1,0}$.
    Since~$\DomainMax - \DomainMin <  1
    \cdot ( \seqlength - 1 - \Char{\Height}{\pattern} )+  0 + \Char{\Height}{\pattern} =
    \seqlength - 1$, we have that~$\DomainMax - \DomainMin < \seqlength - 1$.
    Hence, we aim at finding the longest
    time-series length~$\WordLength < \seqlength$ such
    that~$\DomainMax - \DomainMin = \WordLength - 1$, and a sharp
    upper bound on~$\Result$ will be~$\WordLength -
    \Char{\After}{\pattern} - \Char{\Before}{\pattern}$.
    The largest value of such~$\WordLength$ equals~$\DomainMax -
    \DomainMin + 1$, thus a sharp upper
    bound on~$\Result$ is $\DomainMax - \DomainMin +1
    - \Char{\After}{\pattern} - \Char{\Before}{\pattern}$.
    \shiftedqed
  \end{enumerate}

\end{proof}

\begin{example}
  \label{ex:max-width}
  Consider a~$\Constraint{max\_width}\_\pattern(\XSeq, \Result)$ time-series
  constraint with every $X_i$ having the same integer interval
  domain~$\Domain$.
  The three items of this example cover each value of~$\Tuple{\Char{\FullExtLinCoeff}{\pattern},
    \Char{\FullExtFreeCoeff}{\pattern}}$ in the set~$\Curly{\Tuple{0,0},
    \Tuple{0,1}, \Tuple{1,0}}$.

 \ExFullStyle

  \begin{itemize}

  \item
    Consider the~$\pattern = \InflexionPatternName$ regular expression.
    Recall that both~$\Char{\After}{\pattern}$ and~$\Char{\Before}{\pattern}$
    are equal to~$1$, the width $\Char{\width}{\pattern}$
    of~$\pattern$ is equal to~$2$, the height $\Char{\Height}{\pattern}$
    of~$\pattern$ is equal to~$1$, and for any time-series
    length~$\seqlength > \Char{\width}{\pattern} + 1$, the range
    $\CharPar{\Range}{\pattern}{\seqlength}$ of~$\pattern$
    wrt~$\Tuple{\seqlength}$ is equal to~$
     \Char{\FullExtLinCoeff}{\pattern} \cdot (\seqlength - 1 -
    \Char{\Height}{\pattern}) + \Char{\FullExtFreeCoeff}{\pattern} +
    \Char{\Height}{\pattern} = \Char{\Height}{\pattern} = 1$.
    Since there exists a word, namely~$\wordv = \reg{<>}$,
    in~$\Language{\pattern}$ whose length equals~$2$ and whose height
    is equal to~$1$, and~$\Tuple{\Char{\FullExtLinCoeff}{\pattern},
      \Char{\FullExtFreeCoeff}{\pattern}}$
    is~$\Tuple{0,0}$,~$\pattern$ has the~\WidthUpper~property. 
    Hence, we apply~\thref{th:max-width} for computing a sharp
    upper bound on~$\Result$.

    \begin{itemize}
    \item
      If~$\DomainMax - \DomainMin \geq \CharPar{\Range}{\pattern}{\seqlength} =
      1$,
      then a sharp upper bound on~$\Result$ is equal to
      $\seqlength - \Char{\After}{\pattern} -
      \Char{\Before}{\pattern} = \seqlength - 2$.
    \item
      If~$\DomainMax - \DomainMin < \CharPar{\Range}{\pattern}{\seqlength} = 1$,
      then a sharp upper bound on~$\Result$ is equal to
      $0$.
    \end{itemize} 

  \item
    Consider the~$\pattern = \GorgePatternName$ regular expression.
    Recall that both~$\Char{\After}{\pattern}$ and~$\Char{\Before}{\pattern}$
    are equal to~$1$, the width $\Char{\width}{\pattern}$
    of~$\pattern$ is equal to~$2$, the height $\Char{\Height}{\pattern}$
    of~$\pattern$ is equal to~$1$, and for any time-series
    length~$\seqlength > \Char{\width}{\pattern} + 1$, the range
    $\CharPar{\Range}{\pattern}{\seqlength}$ of~$\pattern$
    wrt~$\Tuple{\seqlength}$ is equal to~$
     \Char{\FullExtLinCoeff}{\pattern} \cdot (\seqlength - 1 -
    \Char{\Height}{\pattern}) + \Char{\FullExtFreeCoeff}{\pattern} +
    \Char{\Height}{\pattern} = \Char{\Height}{\pattern} + 1= 2$.
    Since there exists a word, namely~$\wordv = \reg{><}$,
    in~$\Language{\pattern}$ whose length equals~$2$ and whose height
    is equal to~$1$, and~$\Tuple{\Char{\FullExtLinCoeff}{\pattern},
      \Char{\FullExtFreeCoeff}{\pattern}}$
    is~$\Tuple{0,1}$,~$\pattern$ has the~\WidthUpper~property. 
    Hence, we apply~\thref{th:max-width} for computing a sharp
    upper bound on~$\Result$.

    \begin{itemize}
    \item
      If~$\DomainMax - \DomainMin \geq 2$,
      then a sharp upper bound on~$\Result$ is equal to
      $\seqlength - \Char{\After}{\pattern} -
      \Char{\Before}{\pattern} = \seqlength - 2$.
    \item
      If~$\DomainMax - \DomainMin <  2$,
      then a sharp upper bound on~$\Result$ is equal to
      $\Char{\width}{\pattern} + 1 -
      \Char{\After}{\pattern} -
      \Char{\Before}{\pattern} = 1$.
    \end{itemize} 

  \item
    Consider the~$\pattern = \StrictlyDecreasingSequencePatternName$ regular expression.
    Recall that both~$\Char{\After}{\pattern}$ and~$\Char{\Before}{\pattern}$
    are equal to~$0$, the width $\Char{\width}{\pattern}$
    of~$\pattern$ is equal to~$1$, the height $\Char{\Height}{\pattern}$
    of~$\pattern$ is equal to~$1$, and for any time-series
    length~$\seqlength > \Char{\width}{\pattern} + 1$, the range
    $\CharPar{\Range}{\pattern}{\seqlength}$ of~$\pattern$
    wrt~$\Tuple{\seqlength}$ is equal to~$
     \Char{\FullExtLinCoeff}{\pattern} \cdot (\seqlength - 1 -
    \Char{\Height}{\pattern}) + \Char{\FullExtFreeCoeff}{\pattern} +
    \Char{\Height}{\pattern} = \seqlength - 1 -
    \Char{\Height}{\pattern} + \Char{\Height}{\pattern} = \seqlength - 1 $.
    Since there exists a word, namely~$\wordv = \reg{>}$,
    in~$\Language{\pattern}$ whose length is equal to~$1$ and whose height
    is equal to~$1$, and~$\Tuple{\Char{\FullExtLinCoeff}{\pattern},
      \Char{\FullExtFreeCoeff}{\pattern}}$
    is~$\Tuple{1,0}$,~$\pattern$ has the~\WidthUpper~property. 
    Hence, we apply~\thref{th:max-width} for computing a sharp
    upper bound on~$\Result$.
    
    \begin{itemize}
    \item
      If~$\DomainMax - \DomainMin \geq
      \CharPar{\Range}{\pattern}{\seqlength} = \seqlength - 1$,
      then a sharp upper bound on~$\Result$ is equal to
      $\seqlength - \Char{\After}{\pattern} -
      \Char{\Before}{\pattern} = \seqlength$.
    \item
      If~$\DomainMax - \DomainMin <
      \CharPar{\Range}{\pattern}{\seqlength} = \seqlength - 1$,
      then a sharp upper bound on~$\Result$ 
      is equal to~$\DomainMax - \DomainMin + 1$.
\qedexample
    \end{itemize} 

    \end{itemize} 
  \end{example}
  
\subsection{Upper Bound for~$\Constraint{sum\_width}\_\pattern$ \label{sec:sum-width}}

We now consider the~$\Constraint{sum\_width}\_\pattern(\XSeq,\Result)$
family of time-series constraints
with~$\pattern$ being a non-fixed length regular expression and with
every~$X_i$ ranging over the same integer interval domain~$\Domain$.
Under some hypothesis on the overlap of~$\pattern$
wrt~$\Tuple{\DomainMin, \DomainMax}$,
\lemref{lem:sum-width} provides an upper bound on~$\Result$ and a
condition when this bound is sharp.
Then, \thref{th:sum-width} extends the bound of~\lemref{lem:sum-width}
and gives a more general condition under which the extended bound
on~$\Result$ is sharp.

\begin{lemma}
  \label{lem:sum-width}
  Consider a
  $\Constraint{sum\_width}\_\pattern(\XSeq, \Result)$ time-series
  constraint with every~$X_i$ ranging over the same integer interval
  domain~$\Domain$, and  with
  $\pattern$ being a non\nobreakdash-fixed length regular
  expression.
  
  \begin{enumerate}[(i)]
  \item
    \label{cond:lemma-sum-width-1}
    If~$\CharPar{\Overlap}{\pattern}{\DomainMin, \DomainMax} \leq
    \Char{\After}{\pattern} + \Char{\Before}{\pattern}$ then~$\seqlength
    - \Char{\After}{\pattern} - \Char{\Before}{\pattern}$ is an upper
    bound on~$\Result$.
  \item
    \label{cond:lemma-sum-width-2}
    If, in addition, 
$\DomainMax -
    \DomainMin \geq \CharPar{\Range}{\pattern}{\seqlength}$, then
    this bound is sharp.
  \end{enumerate}
\end{lemma}

\begin{proof}
\setcounter{equation}{0} 
  
[Proof of~(\ref{cond:lemma-sum-width-1})]
  Let us consider a time series~$\TimeSeries$ of length~$\seqlength$
  over~$\Domain$ that has~$\NbPat>1$~$\pattern$-patterns.
  Let~$\Char{\width}{i}$ be the length
  of the~$\pattern$-pattern~$i$ (with~$i$ in~$[1,\NbPat]$);
  let~$\SeqRest$ be the number of time-series
  variables that are not in any extended $\pattern$-pattern of~$\TimeSeries$; 
  and let~$\Char{\Overlap}{i}$ be the number of common time-series variables 
  of the extended~$\pattern$-patterns~$i$ and~$i+1$.
  Then, the following equality holds
 
  \begin{equation}
    \label{eq:sum-width-1}
    \seqlength = 
    \Char{\width}{1} + \Char{\After}{\pattern} +
    \Char{\Before}{\pattern} + 
    \sum \limits_{i=1}^{\NbPat-1s} (\Char{\width}{i+1} +
    \Char{\After}{\pattern} + \Char{\Before}{\pattern} - 
    \Char{\Overlap}{i}) + \SeqRest 
  \end{equation}

  The time series~$\TimeSeries$ yields~$\sum \limits_{i=1}^{\NbPat} \Char{\width}{i} $ as
  the value of~$\Result$, thus we express this quantity
  from~\equalref{eq:sum-width-1} and obtain

  \begin{equation}
    \label{eq:sum-width-2}
    \Result = \seqlength -
    \SeqRest  -
    \NbPat \cdot (\Char{\After}{\pattern} +
    \Char{\Before}{\pattern}) +  \sum \limits_{i=1}^{\NbPat-1}\Char{\Overlap}{i}
  \end{equation}

  In order to prove that $\seqlength - a_\pattern- b_\pattern$ is a
  valid upper bound on~$\Result$, we show that the
  difference between~$\seqlength -
  a_\pattern - b_\pattern$ and the right-hand side
  of~\equalref{eq:sum-width-2}  is always non-negative
  if~$\CharPar{\Overlap}{\pattern}{\DomainMin, \DomainMax} \leq
  \Char{\After}{\pattern} + \Char{\Before}{\pattern}$.
  
  \begin{equation}
    \label{eq:sum-width-3}
    \seqlength - (\Char{\After}{\pattern} + \Char{\Before}{\pattern}) - \seqlength +
    \SeqRest  +
    \NbPat \cdot (\Char{\After}{\pattern} +
    \Char{\Before}{\pattern}) - \sum
    \limits_{i=1}^{\NbPat-1}\Char{\Overlap}{i}
    = \SeqRest  + (\NbPat-1) \cdot
    (\Char{\After}{\pattern}+\Char{\Before}{\pattern}) - \sum
    \limits_{i=1}^{\NbPat-1}\Char{\Overlap}{i}
  \end{equation}
 
  The value of~$\SeqRest$ is non-negative, and by the definition
    of~$\CharPar{\Overlap}{\pattern}{\DomainMin, \DomainMax}$, every
    $\Char{\Overlap}{i}$ is not
  greater than~$\CharPar{\Overlap}{\pattern}{\DomainMin, \DomainMax}$.
  In addition, we have the following inequality
  $\CharPar{\Overlap}{\pattern}{\DomainMin, \DomainMax} \leq
  \Char{\After}{\pattern} + \Char{\Before}{\pattern}$.
  Hence, a lower estimate of the
  right-hand side of~\equalref{eq:sum-width-3} is given by the
  following inequality

 \begin{equation}
    \label{eq:sum-width-4}
    \SeqRest  + (\NbPat-1) \cdot
    (\Char{\After}{\pattern}+\Char{\Before}{\pattern}) - \sum
    \limits_{i=1}^{\NbPat-1}\Char{\Overlap}{i} \geq 0 +  (\NbPat-1)
    \cdot (\Char{\After}{\pattern}+\Char{\Before}{\pattern}) -
    (\NbPat-1) \cdot  (\Char{\After}{\pattern}+\Char{\Before}{\pattern}) = 0
  \end{equation}

  By~\ineqref{eq:sum-width-4} we obtain that,
  when~$\CharPar{\Overlap}{\pattern}{\DomainMin, \DomainMax}
  \leq \Char{\After}{\pattern} + \Char{\Before}{\pattern}$, the
  difference between~$\seqlength - \Char{\After}{\pattern} -
  \Char{\Before}{\pattern}$ and the value of~$\Result$ is always
  non-negative.
  Hence,~$\seqlength
  - \Char{\After}{\pattern} - \Char{\Before}{\pattern}$ is an upper
  bound on~$\Result$.

 [Proof of~(\ref{cond:lemma-sum-width-2})]
  We now show that~$\seqlength - \Char{\After}{\pattern} -
  \Char{\Before}{\pattern}$ is a sharp upper bound on~$\Result$, when~$\DomainMax -
  \DomainMin \geq \CharPar{\Range}{\pattern}{\seqlength}$.
  By definition of~$\CharPar{\Range}{\pattern}{\seqlength}$, the range
  of~$\pattern$ wrt~$\Tuple{\seqlength}$, there exists
  a word~$\wordv$ of length~$\seqlength-1$ in~$\Language{\pattern}$ whose
  height is at most~$\DomainMax - \DomainMin$.
  Hence, there exists at least one ground time series of length~$\seqlength$
  over~$\Domain$ whose signature is~$\wordv$, all its time-series variable belong to a
  single extended~$\pattern$-pattern.
  For such a time series, the value of~$\NbPat$ equals~$1$, and~$\SeqRest$ equals~$0$.
  By the right-hand side of~\equalref{eq:sum-width-2}, we have
  that~$\Result$ equals~$\seqlength - \Char{\After}{\pattern}  - \Char{\Before}{\pattern}  - 0 - (1-1)( \Char{\After}{\pattern}  +
  \Char{\Before}{\pattern} ) = \seqlength - \Char{\After}{\pattern}  -
  \Char{\Before}{\pattern} $, which was proved to be an upper bound. 
  Hence, in this case $\seqlength - \Char{\After}{\pattern}  - \Char{\Before}{\pattern} $ is a
  sharp upper bound on~$\Result$.
\shiftedqed

\end{proof}

\setcounter{equation}{0} 
\begin{theorem}
  \label{th:sum-width}
  Consider a~$\Constraint{sum\_width}\_\pattern (\XSeq,
  \Result)$ time-series constraint with~$\pattern$ being a\
  non-fixed-length regular expression and every~$X_i$ ranging over the same
  integer interval domain~$\Domain$.
  If~$\pattern$ has both the~\WidthUpper~property
  and the~\WidthUpperSum~property for~$\Domain$,
   then a sharp upper bound on~$\Result$ is
\begin{numcases}{}
   \seqlength - \Char{\After}{\pattern} -
    \Char{\Before}{\pattern} & \textnormal{if}
   $\DomainMax - \DomainMin \geq
    \CharPar{\Range}{\pattern}{\seqlength}$   \label{formula:th-sum-width-1}\\
    \Char{\FullExtLinCoeff}{\pattern} \cdot (\seqlength -
    \CharPar{\Rest}{\pattern}{\DomainMin, \DomainMax, \seqlength} ) +
    \Char{\FullExtFreeCoeff}{\pattern} \cdot (\Char{\width}{\pattern}
    + 1 - \Char{\After}{\pattern } - \Char{\Before}{\pattern }) \cdot \CharPar{\NbOcc}{\pattern}{\DomainMin, \DomainMax,
      \seqlength} & \textnormal{if}~$\DomainMax - \DomainMin <
    \CharPar{\Range}{\pattern}{\seqlength}$  \label{formula:th-sum-width-2}
\end{numcases}
  where:

\ListStyle
\begin{itemize}
\item
   $\Char{\FullExtLinCoeff}{\pattern}$
    and
  $\Char{\FullExtFreeCoeff}{\pattern} $ are
  parameters of the regular expression~$\pattern$, introduced
  in~\propref{prop:max-width}.
\item 
$\CharPar{\Rest}{\pattern}{\DomainMin, \DomainMax,
    \seqlength}$ equals~$\min(1, \max(0, \Char{\Height}{\pattern} + 1
  -(\DomainMax - \DomainMin))) \cdot (\seqlength \mod 2)$.
\item
  $\CharPar{\NbOcc}{\pattern}{\DomainMin, \DomainMax, \seqlength}$
  is the maximum number of~$\pattern$-patterns of shortest length in a 
  time series among all ground time series of length~$\seqlength$
  over~$\Domain$.
\end{itemize}

\end{theorem}

\begin{proof}
  When a regular expression~$\pattern$ has the~\WidthUpperSum~property for~$\Domain$,
  Condition~(\ref{cond:lemma-sum-width-1}) of~\lemref{lem:sum-width} is satisfied and
  thus,~$\seqlength - \Char{\After}{\pattern} -
  \Char{\Before}{\pattern}$ is an upper bound on~$\Result$.
  
  \noindent [Case~(\ref{formula:th-sum-width-1}):~$\DomainMax -\DomainMin \geq
  \CharPar{\Range}{\pattern}{\seqlength}$].
  Since~Condition~(\ref{cond:lemma-sum-width-2})
  of~\lemref{lem:sum-width} is also satisfied, by~\lemref{lem:sum-width},
  $\DomainMax -\DomainMin \geq \CharPar{\Range}{\pattern}{\seqlength}$
  is a sharp upper bound on~$\Result$.
  
  \noindent [Case~(\ref{formula:th-sum-width-2}):~$\DomainMax -\DomainMin <
  \CharPar{\Range}{\pattern}{\seqlength}$].
  Let us consider the three potential values of
  $\Tuple{\Char{\FullExtLinCoeff}{\pattern},
    \Char{\FullExtFreeCoeff}{\pattern}}$ from
  Condition~(\ref{cond:max-width-2}) of~\propref{prop:max-width}:

  \begin{enumerate}[(a)]
    \item
     The case of~$\Tuple{\Char{\FullExtLinCoeff}{\pattern},  \Char{\FullExtFreeCoeff}{\pattern} }$
     being~$\Tuple{0,0}$.
     Since~$\DomainMax - \DomainMin < \Char{\Height}{\pattern}$, the
     necessary-sufficient condition,
     i.e.,~\propref{prop:necessary-condition}, is not satisfied, and
     thus no word
     of~$\Language{\pattern}$ can occur
     in the signature of~$\XSeq$.
     Hence,~$\Result$ is equal to
     its default value, namely~$0$.
   \item
     The case of~$\Tuple{\Char{\FullExtLinCoeff}{\pattern},  \Char{\FullExtFreeCoeff}{\pattern} }$
     being~$\Tuple{0,1}$.
     Since~$\DomainMax - \DomainMin \leq \Char{\Height}{\pattern}$, only a
     shortest word with a height being~$\Char{\Height}{\pattern}$ may occur in
     a signature of~$\XSeq$, as it was shown in the proof of~\thref{th:max-width}.
     By Condition~(\ref{cond:max-width-1}) of~\propref{prop:max-width}, such a word exists,
     and thus a sharp upper
     bound on~$\Result$ is equal to~$\Char{\width}{\pattern}  +1
     - \Char{\After}{\pattern}  - \Char{\Before}{\pattern} $.
     Hence, any~$\pattern$-pattern of any ground time series of
     length~$\seqlength$ over~$\Domain$ is of
     length~$\Char{\width}{\pattern} + 1 -
     \Char{\After}{\pattern}  - \Char{\Before}{\pattern} $.
     Since it is not possible to increase the length of
     any~$\pattern$-patterns, in order to maximise~$\Result$, it is necessary to maximise the number
     of~$\pattern$-patterns of shortest length in a time series of length~$\seqlength$ over~$\Domain$.
     Since~$\CharPar{\NbOcc}{\pattern}{\DomainMin, \DomainMax, \seqlength}$
     is the maximum number of~$\pattern$-patterns of
     minimum length, a sharp upper bound on~$\Result$
     equals~$(\Char{\width}{\pattern} + 1 - \Char{\After}{\pattern}  -
     \Char{\Before}{\pattern}) \cdot
     \CharPar{\NbOcc}{\pattern}{\DomainMin, \DomainMax,
       \seqlength} $.
   \item
     The case of~$\Tuple{\Char{\FullExtLinCoeff}{\pattern},
       \Char{\FullExtFreeCoeff}{\pattern} }$
     being~$\Tuple{1,0}$.
     When~$\pattern$ has the~\WidthUpperSum-property
      for~$\Domain$, it belongs to the following class of
      regular expressions: $\Char{\After}{\pattern}$, 
      $\Char{\Before}{\pattern}$,
      $\CharPar{\Overlap}{\pattern}{\DomainMin, \DomainMax}$ are all
      equal to~$0$, and~$\Char{\width}{\pattern}$ is equal to~$1$.
      Consider a time series~$\TimeSeries$ of length~$\seqlength$
      over~$\Domain$ with~$\NbPat \geq 1$~$\pattern$-patterns, where
      $\Char{\width}{i}$ is the length of
      the~$\pattern$-pattern~$i$,~$\Char{\Overlap}{i}$ is the overlap
      of the extended~$\pattern$-patterns~$i$ and~$i+1$,
      and~$\CharPar{\Rest}{\pattern}{\DomainMin, \DomainMax, \seqlength}$ is the number of time-series
      variables of~$\TimeSeries$ that do not belong to any
      extended~$\pattern$\nobreakdash-pattern of~$\TimeSeries$. 
      Then, the following equality holds
      \begin{equation*}
        \label{eq:sum-width-th}
        \Result = \seqlength - \CharPar{\Rest}{\pattern}{\DomainMin, \DomainMax, \seqlength}
         -  \NbPat \cdot (\Char{\After}{\pattern} +
        \Char{\Before}{\pattern}) +  \sum \limits_{i=1}^{\NbPat-1}\Char{\Overlap}{i}
      \end{equation*}
      
      In this equality we replace~$\Char{\After}{\pattern}$, and~$\Char{\Before}{\pattern}$
      with
      their actual values, namely~$0$, which gives a simplified equality
      $\Result = \seqlength - \CharPar{\Rest}{\pattern}{\DomainMin, \DomainMax, \seqlength}$.
      Since the smaller~$\CharPar{\Rest}{\pattern}{\DomainMin,
        \DomainMax, \seqlength}$, the larger is~$\Result$, the aim is
      to find a time series for
      which~$\CharPar{\Rest}{\pattern}{\DomainMin, \DomainMax,
        \seqlength}$ is minimum.
      Assume that in such a time series~$\NbPat$ equals the maximum number of~$\pattern$-patterns in a
      time series among all ground time series of length~$\seqlength$ over~$\Domain$.
      Then,~$\CharPar{\Rest}{\pattern}{\DomainMin, \DomainMax,
        \seqlength}$ is strictly less than~$\Char{\width}{\pattern}
      + 1 = 2$, otherwise there would be a contradiction with the maximality of~$\NbPat$.
      Hence,~$\TimeSeries$
      has at most one time-series variable that is outside of any
      extended~$\pattern$\nobreakdash-pattern of~$\TimeSeries$.
      By definition of~$\CharPar{\Range}{\pattern}{\seqlength}$, the number of
      time-series variables in any extended~$\pattern$-pattern is at most~$\DomainMax -
      \DomainMin + 1$, thus if~$\TimeSeries$ contains at
        least one~$\pattern$\nobreakdash-pattern shorter than~$\DomainMax -
      \DomainMin + 1$ the value
      of~$\CharPar{\Rest}{\pattern}{\DomainMin, \DomainMax,
        \seqlength}$ can be decreased by extending this
      $\pattern$-pattern with one time-series variable.
      Furthermore, if~$\DomainMax - \DomainMin \geq \Char{\Height}{\pattern} +1$,
      then~$\CharPar{\Rest}{\pattern}{\DomainMin, \DomainMax,
        \seqlength} = 0$, otherwise~$\CharPar{\Rest}{\pattern}{\DomainMin, \DomainMax,
        \seqlength}  =
      \seqlength      \mod 2$.
      Hence, the minimum value
      of~$\CharPar{\Rest}{\pattern}{\DomainMin, \DomainMax,
        \seqlength}$ equals
      $\min(1, \max(0, \Char{\Height}{\pattern} + 1
  -(\DomainMax - \DomainMin))) \cdot (\seqlength \mod 2)$. \shiftedqed
    \end{enumerate} 
\end{proof}

Note that for the~$22$ regular expressions in~\tabref{tab:patterns}, the maximum number
of~$\pattern$-patterns of shortest length in a time series coincides
with the maximum number of~$\pattern$-patterns in
the same time series. 
Although, in the general case it may not be true.

\begin{example}
  Consider
  a~$\Constraint{sum\_width}\_\pattern(\XSeq, \Result)$ time-series
  constraint with every~$X_i$ ranging over the same integer interval
  domain~$\Domain$, and each value of~$\Tuple{\Char{\FullExtLinCoeff}{\pattern},
    \Char{\FullExtFreeCoeff}{\pattern}}$ in~$\Curly{\Tuple{0,0},
    \Tuple{0,1}, \Tuple{1,0}}$.
  
 \ExFullStyle
  \begin{itemize}

  \item
    Consider the~$\pattern = \InflexionPatternName$ regular expression.
    In~\exref{ex:max-width}, we showed that the regular
    expression~$\pattern$ has the \WidthUpper~property.
    Recall that~$\CharPar{\Overlap}{\pattern}{\DomainMin, \DomainMax}$ is
    equal to~$2$ and both~$\Char{\After}{\pattern}$
    and~$\Char{\Before}{\pattern}$ are equal to~$1$.
    Hence, Condition~(\ref{cond:sum-width-1})
    of~\propref{prop:sum-width} is also satisfied.
    Since for any time-series length greater
    than~$\Char{\width}{\pattern} + 1$, the value
    of~$\CharPar{\Range}{\pattern}{\seqlength}$
    equals~$\Char{\Height}{\pattern}$, Condition~(\ref{cond:sum-width-2})
    of~\propref{prop:sum-width} is trivially satisfied.
    Hence,~$\pattern$ has also the~\WidthUpperSum~property,
    and~\thref{th:sum-width} can be used for computing a sharp upper bound
    on~$\Result$:

    \begin{itemize}
    \item
      If~$\DomainMax - \DomainMin \geq \Char{\Height}{\pattern} =1$,
      then a sharp upper bound on~$\Result$ is equal to
      $\seqlength - \Char{\After}{\pattern} - \Char{\Before}{\pattern} = \seqlength - 2$.
    \item
      If~$\DomainMax - \DomainMin < \Char{\Height}{\pattern} = 1$,
      then a sharp upper bound on~$\Result$ is equal to
      $0$.
    \end{itemize} 

  \item
\begin{sloppypar}
     Consider the~$\pattern = \GorgePatternName$ regular expression.
    In~\exref{ex:max-width} , we showed that the regular expression~$\pattern$
    has the \WidthUpperSum~property.
    Recall that~$\CharPar{\Overlap}{\pattern}{\DomainMin, \DomainMax}$ is
    equal to~$1$ and both~$\Char{\After}{\pattern}$
    and~$\Char{\Before}{\pattern}$ are equal to~$1$.
    Hence, Condition~(\ref{cond:sum-width-1})
    of~\propref{prop:sum-width} is also satisfied.
    Since for any time-series length greater
    than~$\Char{\width}{\pattern} + 1$, the value
    of~$\CharPar{\Range}{\pattern}{\seqlength}$
    equals~$\Char{\Height}{\pattern} + 1$, Condition~(\ref{cond:sum-width-2})
    of~\propref{prop:sum-width} is trivially satisfied.
    Hence,~$\pattern$ has also the \WidthUpperSum~property,
    and~\thref{th:sum-width} can be used for computing a sharp upper bound
    on~$\Result$:
\end{sloppypar}

    \begin{itemize}
    \item
      If~$\DomainMax - \DomainMin \geq 2$,
      then a sharp upper bound on~$\Result$ equals
      $\seqlength - \Char{\After}{\pattern} - \Char{\Before}{\pattern} = \seqlength - 2$.
    \item
      If~$\DomainMax - \DomainMin < 2$,
      then a sharp upper bound on~$\Result$ is equal to
      $\CharPar{\NbOcc}{\pattern}{\DomainMin, \DomainMax, \seqlength}
      \cdot (\Char{\width}{\pattern} + 1 -
      \Char{\After}{\pattern} - \Char{\Before}{\pattern})
      = \CharPar{\NbOcc}{\pattern}{\DomainMin, \DomainMax, \seqlength} \cdot  (2 + 1 - 1 - 1) =
      \CharPar{\NbOcc}{\pattern}{\DomainMin, \DomainMax, \seqlength} $.   
    \end{itemize} 

      For this particular regular expression, 
      $\CharPar{\NbOcc}{\pattern}{\DomainMin, \DomainMax,
        \seqlength} $ equals the maximum number
      of~$\pattern$-patterns in a time series among all ground time
      series of length~$\seqlength$ over~$\Domain$,
      namely~$\Frac{\seqlength -1}{2}$, which is the upper
      bound obtained in~\secref{sec:nb-patterns}.

  \item
    Consider the~$\pattern= \StrictlyDecreasingSequencePatternName$ regular expression.
    It was shown in~\exref{ex:max-width} that~$\pattern$ has the~\WidthUpper-property.
    Recall that~$\CharPar{\Overlap}{\pattern}{\DomainMin, \DomainMax}$ is
    equal to~$0$, and both~$\Char{\After}{\pattern}$
    and~$\Char{\Before}{\pattern}$ are equal to~$0$, thus Condition~(\ref{cond:sum-width-1})
    of~\propref{prop:sum-width} is also satisfied.
    Since~$\CharPar{\Overlap}{\pattern}{\DomainMin,
      \DomainMax}$,~$\Char{\After}{\pattern}$
    and~$\Char{\Before}{\pattern}$ are all equal to~$0$,
    and~$\Char{\width}{\pattern}$ is equal to~$1$,
    Condition~(\ref{cond:sum-width-2})
    of~\propref{th:sum-width} is also satisfied.
    Hence,~$\pattern$ has the~\WidthUpperSum~property,
    and~\thref{th:sum-width} can be used for computing a sharp upper bound
    on~$\Result$:

   \begin{itemize}
    \item
      If~$\DomainMax - \DomainMin \geq \seqlength - 1$,
      then a sharp upper bound on~$\Result$ is equal to
      $\seqlength$.
     \item
       If~$\DomainMax - \DomainMin < \seqlength - 1$, then
       a sharp upper bound on~$\Result$ is equal to
      $\seqlength - \CharPar{\Rest}{\pattern}{\DomainMin, \DomainMax, \seqlength}
      =
      \seqlength - \min(1, \max(0,  (2 -(\DomainMax - \DomainMin)) \cdot (\seqlength \mod 2)))$.
\qedexample
  \end{itemize} 
 \end{itemize}
\end{example}

\subsection{Lower Bound for
  $\Constraint{min\_width}\_\pattern$ \label{sec:min-width}}

Finally, consider
the~$\Constraint{min\_width}\_\pattern(\XSeq,
\Result)$ family of time-series constraints with~$\pattern$ being a
non-fixed-length regular expression and with every~$X_i$ ranging over the same integer interval
domain~$\Domain$.
The next theorem, \thref{th:min-width}, provides a sharp lower bound on~$\Result$
assuming the property that we now introduce holds.

\begin{property}
  \label{prop:min-width}
  A non-fixed-length regular expression~$\pattern$ has the~\MinWidth~property for an
  integer interval domain~$\Domain$, if there exist a shortest word~$\wordv$
  in~$\Language{\pattern}$, i.e.,~$|\wordv| = \Char{\width}{\pattern}$, and a word~$\word$
  in~$\Curly{\wordv<,\wordv=,\wordv> }$ such that
  the following conditions are all satisfied:
 \begin{enumerate}[(i)]
  \item
    \label{cond:prop-min-width-1}
    The height of~$\wordv$ equals~$\Char{\Height}{\pattern}$, the
    height of~$\pattern$.
  \item
    \label{cond:prop-min-width-2}
    The height of~$\word$ is  less than or equal to~$\DomainMax -
    \DomainMin$.
    \item
    \label{cond:prop-min-width-3}
    The word~$\word$ is not a factor of any word in~$\Language{\pattern}$.
  \end{enumerate}
\end{property}

\begin{theorem}
  \label{th:min-width}
  Consider a~$\Constraint{min\_width}\_\pattern(\XSeq,
  \Result)$ time-series constraint with~$\pattern$ being a non-fixed-length regular
  expression, and with every~$X_i$ having the same integer interval
  domain~$\Domain$.
  If~$\pattern$ has the~\MinWidth~property for~$\Domain$,
  then a sharp lower bound on~$\Result$
  equals~$\Char{\width}{\pattern} + 1 - \Char{\After}{\pattern} - \Char{\Before}{\pattern}$.
\end{theorem}

\begin{proof}
  Since~$\Char{\width}{\pattern}$ is the length of a shortest word in~$\Language{\pattern}$,
  the length of any~$\pattern$-pattern is at least~$\Char{\width}{\pattern} + 1 -
  \Char{\After}{\pattern} - \Char{\Before}{\pattern}$, and thus it is a lower bound on~$\Result$.
 When~$\pattern$ has the~\MinWidth~property, there exists a shortest
  word~$\wordv$ in~$\Language{\pattern}$ and a word~$\word$
  in~$\Curly{\wordv<,\wordv=,\wordv> }$ such that the three conditions
  of~\propref{prop:min-width} are all satisfied.
  We now show that in this case, the bound is sharp.
  
  Case~(a):~$\seqlength = \Char{\width}{\pattern} + 1$.
  When~Condition~(\ref{cond:prop-min-width-1}) of~\propref{prop:min-width} is satisfied, there
  exists a ground time series of length~$\seqlength =
  \Char{\width}{\pattern} + 1$ over~$\Domain$ whose signature
  is~$\wordv$. Hence,~$\Char{\width}{\pattern} + 1 -
  \Char{\After}{\pattern} - \Char{\Before}{\pattern}$ is a sharp lower
  bound on~$\Result$.

  Case~(b):~$\seqlength > \Char{\width}{\pattern} + 1$.
  When~Condition~(\ref{cond:prop-min-width-2}) of~\propref{prop:min-width} is satisfied, there
  exists a ground time series~$\TimeSeries$ of length~$\seqlength$ over~$\Domain$
  whose signature is a word in the language of the~$\reg{\word=^*}$ regular expression.
  If~Condition~(\ref{cond:prop-min-width-3})
  of~\propref{prop:min-width} is also satisfied, then the~$\wordv$
  in the signature of~$\TimeSeries$ is a maximal occurrence
  of~$\pattern$, because~$\word$ is not a factor of any word
  in~$\Language{\pattern}$.
  The length of the corresponding~$\pattern$-pattern
  is~$\Char{\width}{\pattern} + 1 - \Char{\After}{\pattern} -
  \Char{\Before}{\pattern}$, thus this value is a sharp lower bound on~$\Result$.
  \shiftedqed
\end{proof}

\begin{example}
  \label{ex:min-width}
  Consider a~$\Constraint{min\_width}\_\pattern(\XSeq,
  \Result)$ time-series constraints with $\pattern$ being
  the $\InflexionPatternName$ regular
  expression and with every~$X_i$ ranging over the same
  integer interval domain~$\Domain$ such that~$\DomainMax - \DomainMin
  \geq \Char{\Height}{\pattern} = 1$.
  It was shown in~\exref{ex:non-fixed-length} that~$\pattern$ is a
  non-fixed-length regular expression.
  Furthermore, there exists a word~$\wordv = \reg{<>}$ and a
  word~$\word = \reg{<>=}$
  in~$\Curly{\reg{\wordv<},\reg{\wordv=},\reg{\wordv>} }$ such that
  the following conditions are all satisfied:
  \ExStyle
  \begin{itemize}
  \item
    {\makebox[10.9cm][l]{The height of~$\wordv$
        equals~$\Char{\Height}{\pattern} = 1$.} 
     \hfill  (Cond.~(\ref{cond:prop-min-width-1})~~of~\Property~\ref{prop:min-width})~}
  \item
    {\makebox[10.9cm][l]{The height of~$\word$ equals~$1$, and thus
    is less than or equal to~$\DomainMax - \DomainMin$. } 
     \hfill  (Cond.~(\ref{cond:prop-min-width-2}) of~\Property~\ref{prop:min-width})~}
  \item
    {\makebox[9cm][l]{The word~$\word$ is not a factor of any word in~$\Language{\pattern}$. } 
    \hfill   (Cond.~(\ref{cond:prop-min-width-3}) of~\Property~\ref{prop:min-width})~}
  \end{itemize}

  Hence,~$\pattern$ has the~\MinWidth~property for~$\Domain$, and
  by~\thref{th:min-width}, a sharp lower bound on~$\Result$
  equals~$\Char{\width}{\pattern} + 1 - \Char{\After}{\pattern} -
   \Char{\Before}{\pattern} = 2 + 1  - 1 - 1 = 1$.
\qedexample
\end{example}

All the~$22$ regular expressions in~\tabref{tab:patterns} have
the~\MinWidth-property for any integer interval domain~$\Domain$, except
the~$\SteadySequencePatternName$ regular
expression when~$\DomainMin = \DomainMax$.
This special case is considered in~\proposref{propos:min-width-steady-sequence}.

\begin{proposition}
\label{propos:min-width-steady-sequence}
Consider a~$\Constraint{min\_width}\_\pattern(\XSeq,
  \Result)$ time\nobreakdash-series constraints with $\pattern$ being
  the $\SteadySequencePatternName$ regular
  expression and with every~$X_i$ being over an
  integer interval domain~$\Domain$ such that~$\DomainMin =
  \DomainMax$.
A sharp lower bound on~$\Result$ equals~$\seqlength$.
\end{proposition}

\begin{proof}
When~$\DomainMin$ equals~$\DomainMax$, there exists a single ground
time series~$\TimeSeries$ of length~$\seqlength$ over~$\Domain$ with all time\nobreakdash-series
variables having the same value, namely~$\DomainMin$.
The signature of~$\TimeSeries$ is a sequence of~$\seqlength-1$
equalities, which is a word in~$\Language{\pattern}$.
Hence, every time-series variable of~$\TimeSeries$ belongs to a single
extended~$\pattern$-pattern of~$\TimeSeries$, and thus
a sharp lower bound on~$\Result$ equals~$\seqlength -
\Char{\After}{\pattern} - \Char{\Before}{\pattern} = \seqlength$.
\shiftedqed
\end{proof}

 \section{Synthesis}
\label{sec:synthesis}

Consider a~$\Constraint{g\_f}\_\pattern(
 \XSeq, \Result)$ time-series constraints
   with every~$X_i$ being over the same integer interval domain~$\Domain$.
\tabref{tab:synthesis} provides a synthesis of the bounds on~$\Result$
obtained in~Sections~\ref{sec:nb-patterns},~\ref{sec:width-constraints} and~\cite{CP16},
when~$\Tuple{g, f}$ is in~$\Curly{
  \Tuple{\MaxAggr, \MinFeature}, \Tuple{\MaxAggr, \Width},
  \Tuple{\MinAggr, \Width}, \Tuple{\SumAggr, \One}, \Tuple{\SumAggr, \Width}}$.
The theorems and the propositions mentioned in~\tabref{tab:synthesis} were applied for
computing sharp bounds on~$\Result$ for~$93$ time-series constraints
of Volume~II of the global constraint
catalogue~\cite{ArafailovaBeldiceanuDouenceCarlssonFlenerRodriguezPearsonSimonis16}.
An entry of~\tabref{tab:synthesis} corresponds to an upper (respectively lower)
  bound on~$\Result$ for a~$\Constraint{g\_f}\_\pattern(\XSeq, \Result
  )$ time-series constraint with every~$X_i$ ranging over the same integer interval
  domain~$\Domain$, if the
  corresponding ``Type'' column
  contains~$\overline{\Result}$ (respectively~$\underline{\Result}$).
  The ``Theorem'' column contains the theorem or the proposition providing
  the corresponding sharp bound under the hypothesis that~$\pattern$
  has the properties mentioned in
  the corresponding ``Properties'' column.
 The ``Theorem'' (respectively ``Property'') column recalls also the set of
 characteristics used in the bound of the corresponding theorem or proposition
 (respectively property).

Note that when the aggregator is~$\MaxAggr$ (respectively~$\MinAggr$)
we do not consider a lower (respectively upper) bound on~$\Result$.
When~$\pattern$ has the~\NbWidthLowerSimple~property for~$\Domain$, there
exists a time series of length~$\seqlength$ over~$\Domain$ whose
signature contains no~$\pattern$-patterns, and thus such a time series
yields the default value of~$\MaxAggr$ (respectively~$\MinAggr$),
which is~$-\infty$ (respectively~$+ \infty)$. \\

\begin{table}[!h]
\begin{tabular}{lclll}
\toprule
$\scriptsize \Tuple{\Constraint{g,f}}$ &
                                         \scriptsize Type& 
                                                           \scriptsize Theorem & 
                                                                                 \scriptsize Properties &
  \\ \toprule
\scriptsize$\Tuple{\SumAggr, \One}$ & 
\scriptsize$ \underline{\Result}$ & 
\scriptsize Theorem~\ref{th:nb-patterns-lower-bound} 
&
\scriptsize \hyperref[prop:nb-patterns-lb]{\NbWidthLowerSimple}
($\Char{\IndWordSet}{\pattern}$)
\\

& 
\scriptsize$ \underline{\Result}$ & 
\scriptsize Proposition~\ref{propos:lower-bound-steady} 
&
\scriptsize $\pattern =\SteadyPatternName$, $\DomainMax = \DomainMin$
\\

& 
\scriptsize$ \underline{\Result}$ & 
\scriptsize Proposition~\ref{propos:lower-bound-steady_sequence} 
&
\scriptsize $\pattern =\SteadySequencePatternName$, $\DomainMax = \DomainMin$
\\

&
\scriptsize$ \overline{\Result}$ & 
\scriptsize Theorem~\ref{th:nb-patterns} 
($\Char{\width}{\pattern}, \Char{\Height}{\pattern}, \CharPar{\Overlap}{\pattern}{\DomainMin, \DomainMax}, 
\CharPar{\VariationOfMax}{\pattern}{\DomainMin, \DomainMax}$)&
\scriptsize  \hyperref[prop:nb-patterns-first]{\NbUpperOverlap} or \hyperref[prop:nb-patterns-second]{\NbUpperNoOverlap}
(\scriptsize $\Char{\width}{\pattern}, \Char{\Height}{\pattern}, \CharPar{\Overlap}{\pattern}{\DomainMin, \DomainMax}, 
\CharPar{\VariationOfMax}{\pattern}{\DomainMin, \DomainMax}$) \\

& 
\scriptsize$ \overline{\Result}$ & 
\scriptsize Proposition~\ref{propos:nb-ub-steady-sequence} 
&
\scriptsize $\pattern =\SteadySequencePatternName$, $\DomainMax = \DomainMin$
 \\ \midrule

\scriptsize$ \Tuple{\MaxAggr, \Width}$ & 
\scriptsize$ \overline{\Result}$ & 
\scriptsize Theorem~\ref{th:max-width} 
($\Char{\width}{\pattern}, \Char{\Height}{\pattern},
                                   \CharPar{\Range}{\pattern}{\DomainMin,
                                   \DomainMax}$)&
 \scriptsize \hyperref[prop:max-width]{\WidthUpper}
(\scriptsize $\Char{\width}{\pattern}, \Char{\Height}{\pattern},
                                   \CharPar{\Range}{\pattern}{\DomainMin,
                                   \DomainMax}$)
 \\ \midrule

\scriptsize $\Tuple{\SumAggr, \Width}$ & 
\scriptsize $\overline{\Result}$ &  
\scriptsize Theorem~\ref{th:sum-width}
($\Char{\width}{\pattern}, \Char{\Height}{\pattern},
                                   \CharPar{\Range}{\pattern}{\DomainMin,
                                   \DomainMax}$)&
\scriptsize  \hyperref[prop:max-width]{\WidthUpper} and \hyperref[prop:sum-width]{\WidthUpperSum}
(\scriptsize $\Char{\width}{\pattern}, \Char{\Height}{\pattern},
                                   \CharPar{\Range}{\pattern}{\DomainMin,
                                   \DomainMax},
 \CharPar{\Overlap}{\pattern}{\DomainMin, \DomainMax}$)
 \\ \midrule

\scriptsize $ \Tuple{\MinAggr, \Width}$ & 
\scriptsize$ \underline{\Result}$ &  
\scriptsize Theorem~\ref{th:min-width}
($ \Char{\width}{\pattern}$)&
\scriptsize \hyperref[prop:min-width]{\MinWidth}
(\scriptsize $ \Char{\width}{\pattern},
\Char{\Height}{\pattern}$)
 \\ 

& 
\scriptsize$ \underline{\Result}$ & 
\scriptsize Proposition~\ref{propos:min-width-steady-sequence} 
&
\scriptsize $\pattern =\SteadySequencePatternName$, $\DomainMax = \DomainMin$
\\\midrule

\scriptsize $ \Tuple{\MaxAggr, \MinFeature}$ & 
\scriptsize $ \overline{\Result}$ &  
\scriptsize Theorem~1 in~\cite{CP16}&
\scriptsize The~Condition of~Theorem~1 in~\cite{CP16}
 \\ \bottomrule

\end{tabular}
\caption{A synthesis of bounds presented
  in~Sections~\ref{sec:nb-patterns},~\ref{sec:width-constraints}, and
  in~\cite{CP16}.
\label{tab:synthesis}} 
\end{table}

\begin{table*}
\begin{center}
\rotatebox{90}{
\begin{tabular}{@{}llcccccc@{}} \toprule
  name~$\pattern$                 		
  & $\Char{\width}{\pattern}$
  & $\Char{\Height}{\pattern}$
  & $\Tuple{e_\sigma,c_\sigma}$
  & $\CharPar{\CompleteExtension}{\pattern}{\seqlength}$
  & $\Char{\IndWordSet}{\pattern}$
  & $\CharPar{\Overlap}{\pattern}{\DomainMin,\DomainMax}$
  & $\CharPar{\VariationOfMax}{\pattern}{\DomainMin,\DomainMax}$ \\\midrule
  {\scriptsize$\mathtt{Bump}$}   & $5$ & $2$ & {\scriptsize\Undefined}    & $\begin{cases}  2 &\textnormal{~if~} \seqlength = 6 \\
                                                                                                    \text{\scriptsize\Undefined}  &\text{\scriptsize~otherwise~} \end{cases}$
                                                        & {\scriptsize$\Set{\BumpOnDecreasingSequenceInduced}$}
                                                        & $3$
                                                        & $0$ \\
  {\scriptsize$\mathtt{Dec}$}                 & $1$ & $1$ & {\scriptsize\Undefined}    & $\begin{cases}  1 &\textnormal{~if~} \seqlength = 2 \\
                                                                                                    \text{\scriptsize\Undefined}  &\text{\scriptsize~otherwise~} \end{cases}$
                                           & {\scriptsize$\Set{\DecreasingInduced}$}
                                           & $\begin{cases}
                                              0 & \text{if } \DomainMax - \DomainMin \leq 1 \\
                                              1 & \text{\scriptsize otherwise}
                                              \end{cases}$
                                           & $\begin{cases}
                                              0 & \text{if } \DomainMax - \DomainMin \leq 1 \\
                                              -1 & \text{\scriptsize otherwise} 
                                              \end{cases}$ \\
  {\scriptsize$\mathtt{DecSeq}$}         & $1$ & $1$ & $\Tuple{0,1}$ & $\begin{cases}  1 &\text{~if~} \seqlength = 2 \\
                                                                                       2 &\text{~if~} \seqlength > 2\end{cases}$
                                           & {\scriptsize$\Set{\DecreasingSequenceInduced}$}
                                           & $0$
                                           & $0$ \\
  {\scriptsize$\mathtt{DecTer}$}           & $3$ & $2$ & $\Tuple{0,0}$ & $2$
                                           & {\scriptsize$\Set{\DecreasingTerraceInduced}$}
                                           & $\begin{cases}
                                              0 & \text{if } \DomainMax - \DomainMin \leq 2 \\
                                              2 & \text{\scriptsize otherwise} 
                                              \end{cases}$
                                           & $\begin{cases}
                                              0 & \text{if } \DomainMax - \DomainMin \leq 2 \\
                                              -1 & \text{\scriptsize otherwise} 
                                              \end{cases}$ \\
  {\scriptsize$\mathtt{Dip}$}    & $5$ & $2$ & {\scriptsize\Undefined}    & $\begin{cases}  2 &\text{~if~} \seqlength = 6 \\
                                                                                               \text{\scriptsize\Undefined} &\text{\scriptsize~otherwise~} \end{cases}$
                                           & {\scriptsize$\Set{\DipOnIncreasingSequenceInduced}$}
                                           & $3$
                                           & $0$ \\
  {\scriptsize$\GorgePatternName$}         & $2$ & $1$ & $\Tuple{0,1}$ & $\begin{cases}  1 &\text{~if~} \seqlength = 3 \\
                                                                          2 &\text{~if~} \seqlength > 3\end{cases}$
                                           & {\scriptsize$\Set{\GorgeInduced}$}
                                           & $1$
                                           & $0$ \\
  {\scriptsize$\mathtt{Inc}$}    & $1$ & $1$ & {\scriptsize\Undefined}    & $\begin{cases}  1 &\textnormal{~if~} \seqlength = 2 \\
                                                                                       \text{\scriptsize\Undefined} &\text{\scriptsize~otherwise~} \end{cases}$
                                           & {\scriptsize$\Set{\IncreasingInduced}$}
                                           & $\begin{cases}
                                              0 & \text{if } \DomainMax - \DomainMin \leq 1 \\
                                              1 & \text{\scriptsize otherwise}
                                              \end{cases}$
                                           & $\begin{cases}
                                              0 & \text{if } \DomainMax - \DomainMin \leq 1 \\
                                              1 & \text{\scriptsize otherwise} 
                                              \end{cases}$ \\
  {\scriptsize$\mathtt{IncSeq}$}         & $1$ & $1$ & $\Tuple{0,1}$ & $\begin{cases}  1 &\text{~if~} \seqlength = 2 \\
                                                                          2 &\text{~if~} \seqlength > 2\end{cases}$
                                           & {\scriptsize$\Set{\IncreasingSequenceInduced}$}
                                           & $0$
                                           & $0$ \\
  {\scriptsize$\mathtt{IncTer}$}          & $3$ & $2$ & $\Tuple{0,0}$ & $2$
                                           & {\scriptsize$\Set{\IncreasingTerraceInduced}$}
                                           & $\begin{cases}
                                              0 & \text{if } \DomainMax - \DomainMin \leq 2 \\
                                              2 & \text{\scriptsize otherwise} 
                                              \end{cases}$
                                           & $\begin{cases}
                                              0 & \text{if } \DomainMax - \DomainMin \leq 2 \\
                                              1 & \text{\scriptsize otherwise} 
                                              \end{cases}$ \\
  {\scriptsize$\InflexionPatternName$}                  & $2$ & $1$ & $\Tuple{0,0}$ & $1$
                                           & {\scriptsize$\Set{\InflexionInduced}$}
                                           & $2$
                                           & $0$ \\
  {\scriptsize$\PeakPatternName$}                       & $2$ & $1$ & $\Tuple{0,0}$ & $1$
                                           & {\scriptsize$\Set{\PeakInduced}$}
                                           & $1$
                                           & $0$ \\
  {\scriptsize$\PlainPatternName$}                      & $2$ & $1$ & $\Tuple{0,0}$ & $1$
                                           & {\scriptsize$\Set{\PlainInduced}$}
                                           & $1$
                                           & $0$ \\
  {\scriptsize$\PlateauPatternName$}                    & $2$ & $1$ & $\Tuple{0,0}$ & $1$
                                           & {\scriptsize$\Set{\PlateauInduced}$}
                                           & $1$
                                           & $0$ \\
  {\scriptsize$\mathtt{PropPlain}$}                & $3$ & $1$ & $\Tuple{0,0}$ & $1$
                                           & {\scriptsize$\Set{\ProperPlainInduced}$}
                                           & $1$
                                           & $0$ \\
  {\scriptsize$\mathtt{PropPlateau}$}              & $3$ & $1$ & $\Tuple{0,0}$ & $1$
                                           & {\scriptsize$\Set{\ProperPlateauInduced}$}
                                           & $1$
                                           & $0$ \\
  {\scriptsize$\SteadyPatternName$}                     & $1$ & $0$ & {\scriptsize\Undefined}    & $\begin{cases}  0 &\textnormal{~if~} \seqlength = 2 \\
                                                                                                    \text{\scriptsize\Undefined}  &\text{\scriptsize~otherwise~} \end{cases}$
                                           & {\scriptsize$\Set{\SteadyInduced}$}
                                           & $1$
                                           & $0$ \\  
  {\scriptsize$\mathtt{SteadySeq}$}             & $1$ & $0$ & $\Tuple{0,0}$ & $0$
                                           & {\scriptsize$\Set{\SteadySequenceInduced}$}
                                           & $0$
                                           & $0$ \\
  {\scriptsize$\mathtt{SDecSeq}$} & $1$ & $1$ & $\Tuple{1,0}$ & $\seqlength-1$
                                           & {\scriptsize$\Set{\StrictlyDecreasingSequenceInduced}$}
                                           & $0$
                                           & $0$ \\
  {\scriptsize$\mathtt{SIncSeq}$} & $1$ & $1$ & $\Tuple{1,0}$ & $\seqlength-1$
                                           & {\scriptsize$\Set{\StrictlyIncreasingSequenceInduced}$}
                                           & $0$
                                           & $0$ \\
  {\scriptsize$\SummitPatternName$}                     & $2$ & $1$ & $\Tuple{0,1}$ & $\begin{cases}  1 &\textnormal{~if~} \seqlength = 3 \\
                                                                          2  &\textnormal{~if~} \seqlength > 3\end{cases}$
                                           & {\scriptsize$\Set{\SummitInduced}$}
                                           & $1$
                                           & $0$ \\
  {\scriptsize$\ValleyPatternName$}                     & $2$ & $1$ & $\Tuple{0,0}$ & $1$
                                           & {\scriptsize$\Set{\ValleyInduced}$}
                                           & $1$
                                           & $0$ \\
  {\scriptsize$\ZigzagPatternName$}                     & $3$ & $1$ & $\Tuple{0,0}$ & $1$
                                           & {\scriptsize$\Set{\ZigzagInduced}$}
                                           & $\begin{cases}
                                              0 & \text{if } \DomainMax - \DomainMin \leq 1 \\
                                              1 & \text{\scriptsize otherwise} 
                                              \end{cases}$
                                           & $0$ \\
\bottomrule
\end{tabular}
}
\caption{\label{tab:characteristics} Regular expression names $\pattern$ and corresponding
\emph{width} $\Char{\width}{\pattern}$,
\emph{height} $\Char{\Height}{\pattern}$,
\emph{range} $\CharPar{\CompleteExtension}{\pattern}{\seqlength}$ (for a non-fixed-length regular expression~$\pattern$ and for any~$\seqlength  > \Char{\width}{\pattern} + 1$, $\CharPar{\CompleteExtension}{\pattern}{\seqlength} = \Char{\FullExtLinCoeff}{\pattern} \cdot ( \seqlength - 1 - \Char{\Height}{\pattern}) +
  \Char{\FullExtFreeCoeff}{\pattern} + \Char{\Height}{\pattern}$),
\emph{inducing words} $\Char{\IndWordSet}{\pattern}$,
\emph{overlap} $\CharPar{\Overlap}{\pattern}{\DomainMin,\DomainMax}$, and
\emph{smallest variation of maxima} $\CharPar{\VariationOfMax}{\pattern}{\DomainMin,\DomainMax}$, where
$\mathtt{Bump}$,
$\mathtt{Dec}$,
$\mathtt{DecSeq}$,
$\mathtt{DecTer}$,
$\mathtt{Dip}$,
$\mathtt{Inc}$,
$\mathtt{IncSeq}$,
$\mathtt{IncTer}$,
$\mathtt{PropPlain}$,
$\mathtt{PropPlateau}$,
$\mathtt{SteadySeq}$,
$\mathtt{SDecSeq}$,
$\mathtt{SIncSeq}$
are respectively shortcuts for
$\BumpOnDecreasingSequencePatternName$,
$\DecreasingPatternName$,
$\DecreasingSequencePatternName$,
$\DecreasingTerracePatternName$,
$\DipOnIncreasingSequencePatternName$,
$\IncreasingPatternName$,
$\IncreasingSequencePatternName$,
$\IncreasingTerracePatternName$,
$\ProperPlainPatternName$,
$\ProperPlateauPatternName$,
$\SteadySequencePatternName$,
$\StrictlyDecreasingSequencePatternName$,
$\StrictlyIncreasingSequencePatternName$.
}
\end{center}
\end{table*}

\begin{sloppypar}
\tabref{tab:characteristics} provides for each of the regular expressions in~\tabref{tab:patterns}
the corresponding value of each regular expression characteristics.
The~$22$ regular expressions in~\tabref{tab:patterns} have
the~\NbWidthLowerSimple~property for any domain, except
the~$\SteadyPatternName$ and
the~$\SteadySequencePatternName$ regular expressions when~$\DomainMin
= \DomainMax$.
\tabref{tab:properties} classifies the~$22$ regular expressions according
to the set of properties they share.
There are three main groups, and two special ones, namely for
the~$\SteadyPatternName$ and for
the~$\SteadySequencePatternName$ regular expressions.
The partitioning into the three main groups is related to the fact that 
the entry of~\tabref{tab:synthesis}
with~\thref{th:nb-patterns}, contains a disjunction between
the~\NbUpperOverlap~and the~\NbUpperNoOverlap~properties.
Furthermore, a regular expression~$\pattern$ cannot have both
properties for the same integer interval domain~$\Domain$.
This allows to partition the~$22$ regular expressions into three
classes, namely:
\begin{enumerate}
\item 
  The regular expressions that
  have the \NbUpperOverlap~property for any~$\Domain$, i.e., the first
  group in~\tabref{tab:properties}.

\item 
  The regular expressions that
  have the \NbUpperNoOverlap~property for any~$\Domain$, i.e., the second
  group in \tabref{tab:properties}.

\item
  The regular expressions that
  have the \NbUpperNoOverlap~property for any~$\Domain$ such
  that~$\DomainMax - \DomainMin = \Char{\Height}{\pattern}$, and have
  the~\NbUpperNoOverlap~property for any other~$\Domain$, i.e., the third
  group in~\tabref{tab:properties}.
\end{enumerate}

The~$\SteadySequencePatternName$ represents a special case, because
when~$\DomainMax - \DomainMin = \Char{\Height}{\pattern}$,~$\pattern$
has neither property for~$\Domain$, and when~$\DomainMax - \DomainMin
> \Char{\Height}{\pattern}$,~$\pattern$
has the~\NbUpperNoOverlap~property for~$\Domain$.
\end{sloppypar}

\begin{table*}[!h]\centering
\setlength{\tabcolsep}{6pt}
\begin{tabular}{@{}llc@{}} \toprule
~~~~Regular Expressions &  Set of Properties &  \\\midrule
 
$
\text{\rotatebox[origin=c]{90}{\scriptsize\begin{tabular}{c}
                                            Overlapping
                                            \\Class \end{tabular}}} 
\begin{dcases}
\BumpOnDecreasingSequencePatternName \\
\DipOnIncreasingSequencePatternName \\
\GorgePatternName \\
\InflexionPatternName \\
\PeakPatternName \\
\PlainPatternName \\
\PlateauPatternName \\
\ProperPlainPatternName \\
\ProperPlateauPatternName \\
\SummitPatternName \\
\ValleyPatternName
\end{dcases}$ &

$\begin{array}{l}
\textnormal{\NbWidthLowerSimple} \\
\textnormal{\NbUpperOverlap} \\
\textnormal{\WidthUpper}  \\
\textnormal{\WidthUpperSum} \\
\textnormal{\MinWidth} \\
\textnormal{\MaxMin}  \\ 
\\ \\ \\ \\ \\ \\ \\ \\
\end{array}$
  
 \\ \midrule

$
\text{\rotatebox[origin=c]{90}{\scriptsize\begin{tabular}{c}
                                            Non-Overlapping
                                            \\Class \end{tabular}}} 
\begin{dcases} 
\DecreasingSequencePatternName \\
\IncreasingSequencePatternName \\
\StrictlyDecreasingSequencePatternName \\
\StrictlyIncreasingSequencePatternName  \\ \\ \\ 
\end{dcases}$ &

$\begin{array}{l}
\textnormal{\NbWidthLowerSimple} \\
\textnormal{\NbUpperNoOverlap} \\
\textnormal{\WidthUpper}  \\
\textnormal{\WidthUpperSum} \\
\textnormal{\MinWidth} \\
\textnormal{\MaxMin}    \\ \\ 
\end{array}
$
 
  \\ \midrule 

$
\text{\rotatebox[origin=c]{90}{\scriptsize\begin{tabular}{c}
                                            Overlapping
                                            \\Non-Overlapping Class \end{tabular}}} 
\begin{dcases} 
\DecreasingPatternName \\
\IncreasingPatternName \\
\DecreasingTerracePatternName \\
\IncreasingTerracePatternName \\ 
\ZigzagPatternName \\ \\
\end{dcases}$ &

$\begin{array}{l}
\textnormal{\NbWidthLowerSimple} \\
\textnormal{\NbUpperNoOverlap~when  $\DomainMax - \DomainMin =
    \Char{\Height}{\pattern}$} \\
\textnormal{\NbUpperOverlap~when  $\DomainMax - \DomainMin \geq
    \Char{\Height}{\pattern} + 1$} \\
\textnormal{\WidthUpper}  \\
\textnormal{\WidthUpperSum} \\
\textnormal{\MinWidth} \\
\textnormal{\MaxMin}   \\ \\ \\ 
\end{array}$
 \\ \midrule

$
\text{\rotatebox[origin=c]{90}{\scriptsize\begin{tabular}{c}
                                            Special
                                            \\Case \end{tabular}}} 
\begin{dcases} 
\SteadyPatternName \\ \\ \\ \\ \\ \\ 
\SteadySequencePatternName \\ \\ \\  \\ \\ \\
\end{dcases}$ &

$\begin{array}{l}
\textnormal{\NbWidthLowerSimple~when $\DomainMax -
                                               \DomainMin >
  \Char{\Height}{\pattern}$} \\
\textnormal{\NbUpperNoOverlap~when  $\DomainMax - \DomainMin =
    \Char{\Height}{\pattern}$} \\
\textnormal{\NbUpperOverlap} \\
\textnormal{\WidthUpper}  \\
\textnormal{\WidthUpperSum} \\
\textnormal{\MinWidth} \\
\textnormal{\MaxMin}   \\ \\ 
\textnormal{\NbWidthLowerSimple~when $\DomainMax -
                                               \DomainMin >
  \Char{\Height}{\pattern}$} \\
\textnormal{\NbUpperNoOverlap~when  $\DomainMax - \DomainMin =
    \Char{\Height}{\pattern}$} \\
\textnormal{\NbUpperOverlap} \\
\textnormal{\WidthUpper}  \\
\textnormal{\WidthUpperSum} \\
\textnormal{\MinWidth~when  $\DomainMax - \DomainMin > \Char{\Height}{\pattern}$} \\
\textnormal{\MaxMin}   
\end{array}$ \\

\bottomrule
\end{tabular}
\caption{\label{tab:properties}Classification of regular expressions: regular expression names $\pattern$,
  their properties and conditions on domain when they hold.}
\end{table*}

\clearpage

\section{Evaluation}
\label{sec:evaluation}

\begin{sloppypar}
We evaluate the impact of the methods introduced in the previous
sections on both execution time and the number of backtracks
(failures) for all the~$200$ time-series constraints for which the
glue constraint exists.
Given the time-series constraints
$\gamma(\Tuple{X_1,X_2,\dots,X_\seqlength},\Result)$,
$\gamma(\Tuple{X_1,X_2,\dots,X_i},\Result_p)$ and
$\gamma(\Tuple{X_\seqlength,X_{\seqlength-1},\dots,X_i},\Result_s)$ with $i\in[1,n]$,
the \emph{glue constraint}
links the overall result $\Result$ with the two results $\Result_p$ and $\Result_s$~\cite{ASTRA:CP14}.
\end{sloppypar}

In our first experiment, we consider a single
$\Constraint{g\_f}\_\pattern(\XSeq,\Result)$ constraint
for which we first enumerate~$\Result$ and then either find solutions
by assigning the~$X_i$
or prove infeasibility of the chosen~$\Result$.
For each constraint, we compare four variants of \emph{Automaton},
which just states the constraint, using the automaton
of~\cite{ASTRA:CPAIOR16}:
\emph{Glue} adds to~\emph{Automaton} the glue constraints~\cite{CP16},
\cite{ASTRA:CP14} for all prefixes and corresponding reversed
suffixes;
\emph{Bounds} adds to~\emph{Automaton} the bound restrictions;
\emph{Bounds}+\emph{Glue} uses both the glue constraints and the
bounds; and
\emph{Combined} adds to \emph{Bounds}+\emph{Glue} the bounds for each
prefix and corresponding reversed suffix.

\begin{figure}[t]
  \begin{subfigure}[b]{0.5\textwidth}
    \label{fig:variants-a}%
    \includegraphics[width=\textwidth]{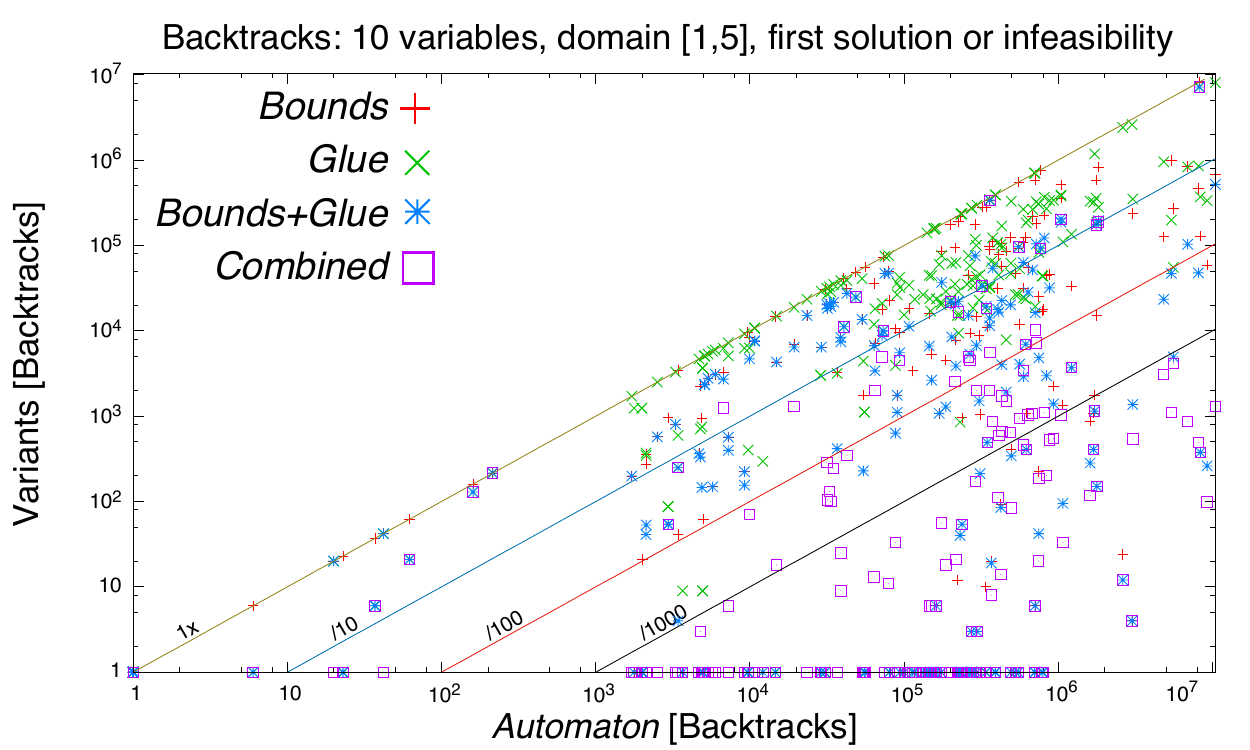}
  \end{subfigure}%
  \begin{subfigure}[b]{0.5\textwidth}
    \label{fig:variants-b}%
    \includegraphics[width=\textwidth]{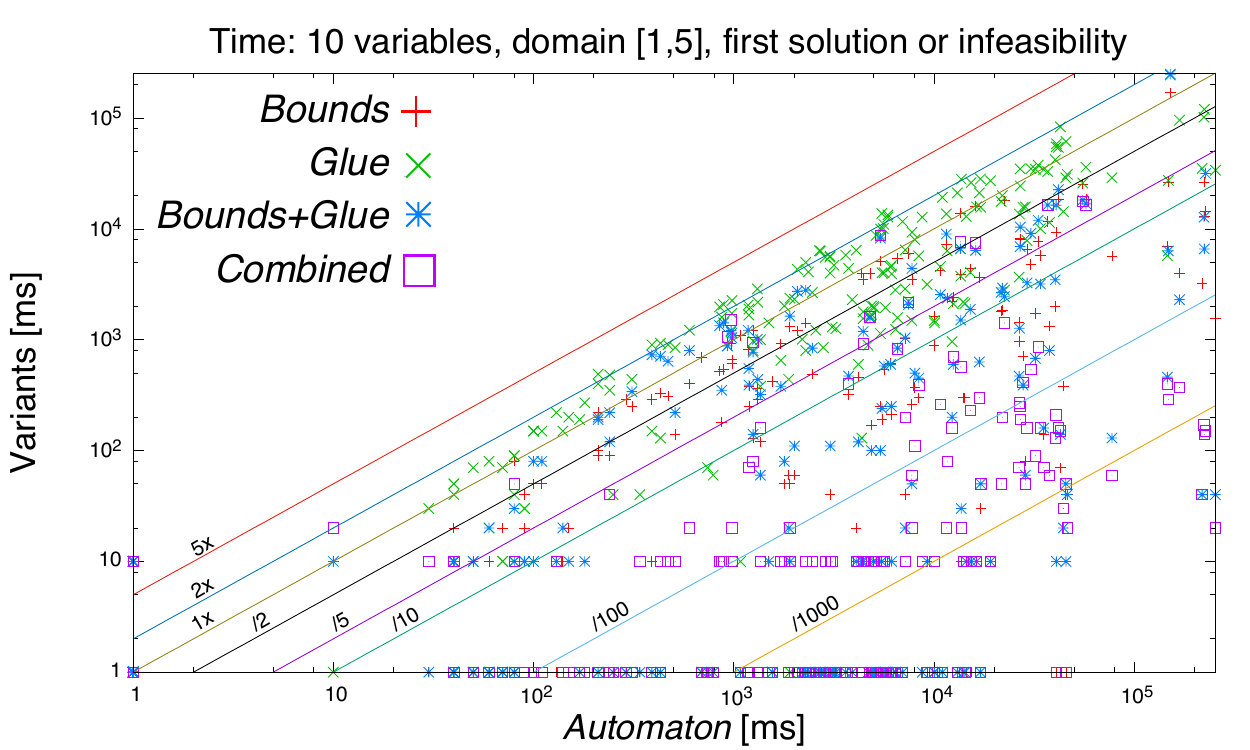}
  \end{subfigure}\\
  \begin{subfigure}[b]{0.5\textwidth}
    \label{fig:variants-c}%
    \includegraphics[width=\textwidth]{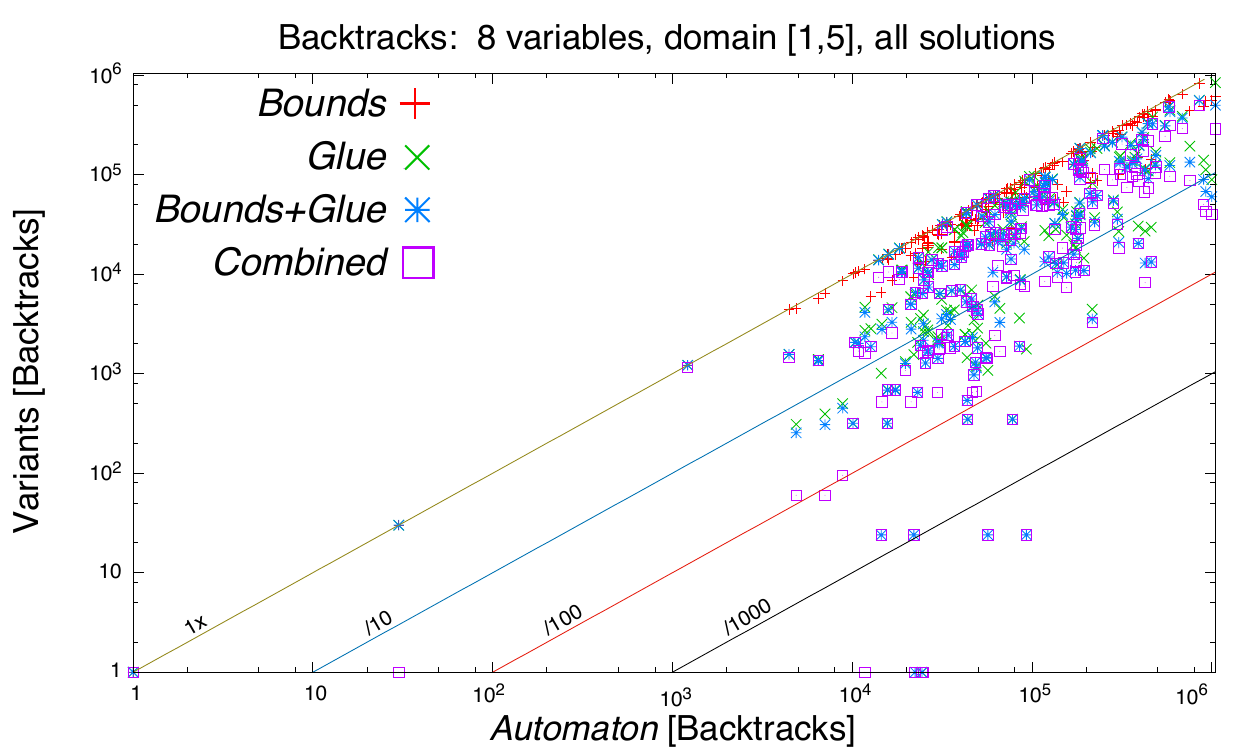}
  \end{subfigure}%
  \begin{subfigure}[b]{0.5\textwidth}
    \label{fig:variants-d}%
    \includegraphics[width=\textwidth]{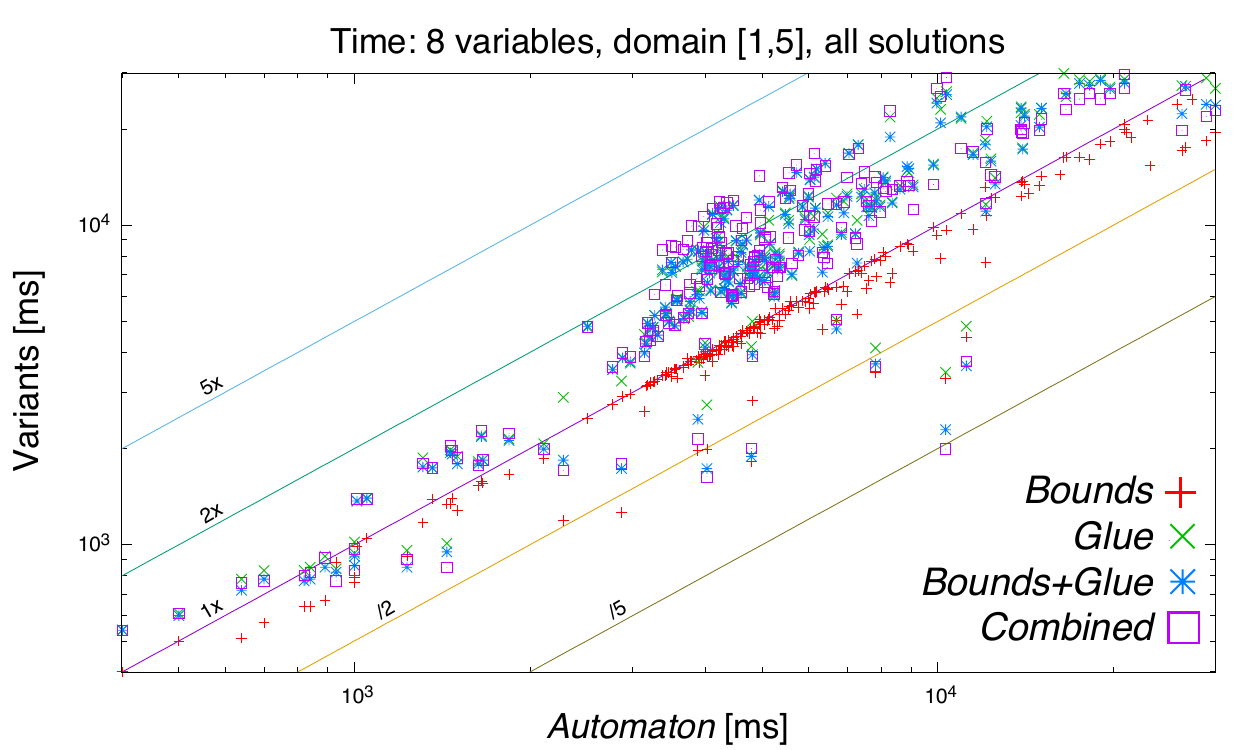}
  \end{subfigure}
  \caption{\label{fig:Variants}Comparing backtrack count and runtime
    for \emph{Automaton} and its variants for the first solution
    (length $10$) and all solutions (length $8$).}
\end{figure}

In \figref{fig:Variants}, we show results for two problems that are
small enough to perform all computations for \emph{Automaton} and all
variants within a reasonable time.
In the first problem (first row of plots), we use time series of
length~$10$ over the domain~$[1,5]$, and find, for each value
of~$\Result$, the first solution or prove infeasibility.
This would be typical for satisfaction or optimisation problems, where
one has to detect infeasibility quickly.
Our static search routine enumerates the time-series variables~$X_i$
from left to right, starting with the smallest value in the domain.
In the case of the initial domains being of the same size, this
  heuristic typically works best.
 In the second problem (second row
of plots), we consider time series of length~$8$ over the
domain~$[1,5]$, and find all solutions for each value of~$\Result$.
This allows us to verify that no solutions are incorrectly eliminated
by any of the variants, and provides a worst-case scenario exploring
the complete search tree.
Results for the backtrack count are on the left, results for the
execution time on the right.
We use log scales on both axes, replacing a zero value by one in order
to allow plotting.
All experiments were run with SICStus Prolog~$4.2.3$ on a $2011$
MacBook Pro $2.2$~GHz quadcore Intel Core i7-$950$ machine with $6$~MB
cache and $16$~GB memory using a single core.

We see that \emph{Bounds} and \emph{Glue} on their own bring good
reductions of the search space, but their combinations
\emph{Bounds}+\emph{Glue} and \emph{Combined} in many cases reduce the
number of backtracks by more than three orders of magnitude.
Indeed, for many constraints, finding the first solution requires no
backtracks.
On the other hand, there are a few constraints for which the number of
backtracks is not reduced significantly.
These are constraints
for which values of $\Result$ in the middle of the domain are
infeasible, but this is not detected by any of our variants.

The time for finding the first solution or proving infeasibility is
also significantly reduced by the combinations
\emph{Bounds}+\emph{Glue} and \emph{Combined}, even though the glue
constraints require two time-series constraints.
When finding all solutions, this overhead shows in the total time
taken for the three variants using the glue constraints.
The bounds on their own reduce the time for many constraints, but
rarely by more than a factor of ten.

\begin{figure}[t]
  \includegraphics[width=\textwidth]{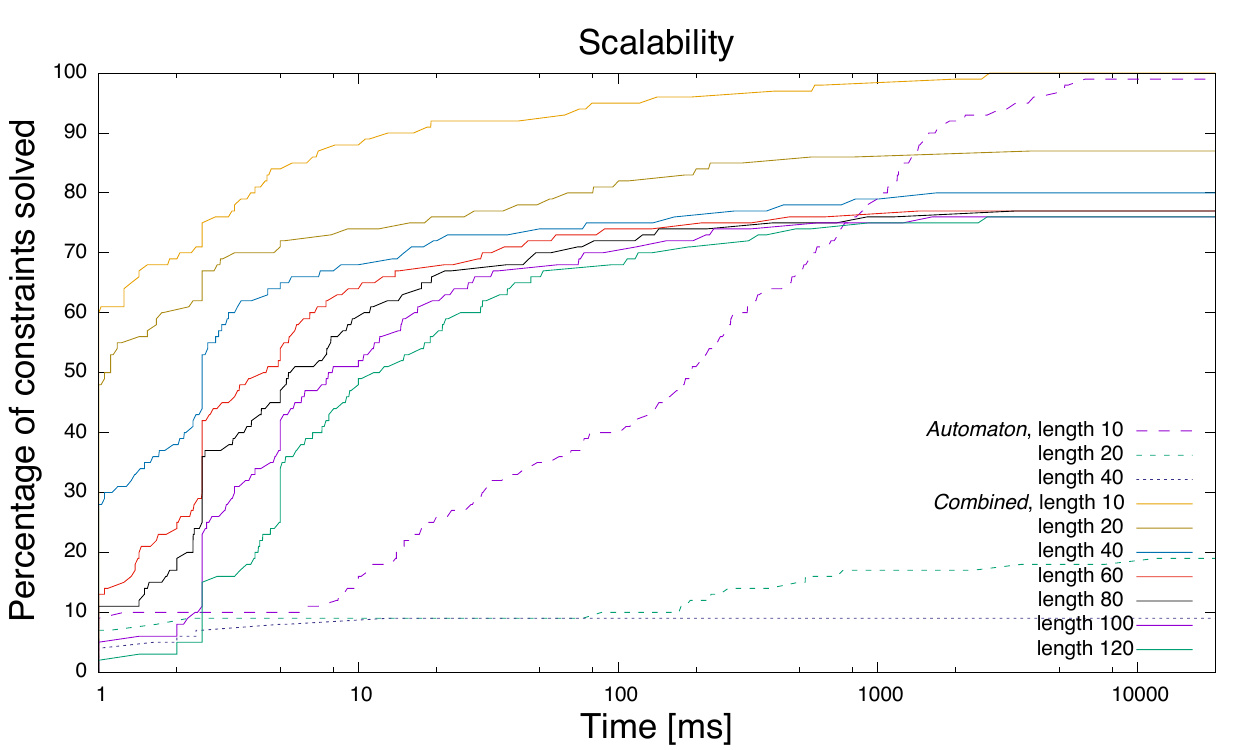}
  \caption{\label{fig:scale}Scalability results comparing time for
    \emph{Automaton} and \emph{Combined} on problems of increasing
    length.}
\end{figure}

In our second experiment, shown in \figref{fig:scale}, we want to see
whether the \emph{Combined} variant is scalable.
For this, we increase the length of the time series from~$10$ to~$120$
over the domain~$[1,5]$.
We enumerate all possible values of~$\Result$ and find a first
solution or prove infeasibility.
For each time-series constraint and value of~$\Result$, we impose a
timeout of~$20$ seconds, and we do not consider the constraint if
there is a timeout on some value of~$\Result$.
We plot the percentage of all constraints for which the average
runtime is less than or equal to the value on the horizontal axis.
For small time values, there are some quantisation effects due to the
SICStus time resolution of $10$ milliseconds.

For length~$10$, we find solutions for all values of~$\Result$ within
the timeout, and our plots for \emph{Automaton} (dashed) and
\emph{Combined} (solid) reach~$100\%$, but the average time of
\emph{Combined} is much smaller.
For \emph{Automaton}, the percentage of constraints that are solved
within the timeout drops to less than~$20\%$ for length~$20$, and less
than~$10\%$ for length~$40$.
For \emph{Combined}, we solve over~$75\%$ of all constraints within
the time limit, even for lengths~$100$ and~$120$.

The constraints that are not solved by \emph{Combined} use the
feature~$\Surf$ or the aggregator~$\SumAggr$.
The worst performance is observed for constraints combining
both~$\Surf$ and~$\SumAggr$.
This is not surprising, as we know that achieving domain consistency
for many of those constraints is NP-hard (encoding of
\emph{subset-sum}).

\begin{figure}[t]
  \begin{subfigure}[b]{0.5\textwidth}
    \label{fig:trees-a}%
    \includegraphics[width=\textwidth]{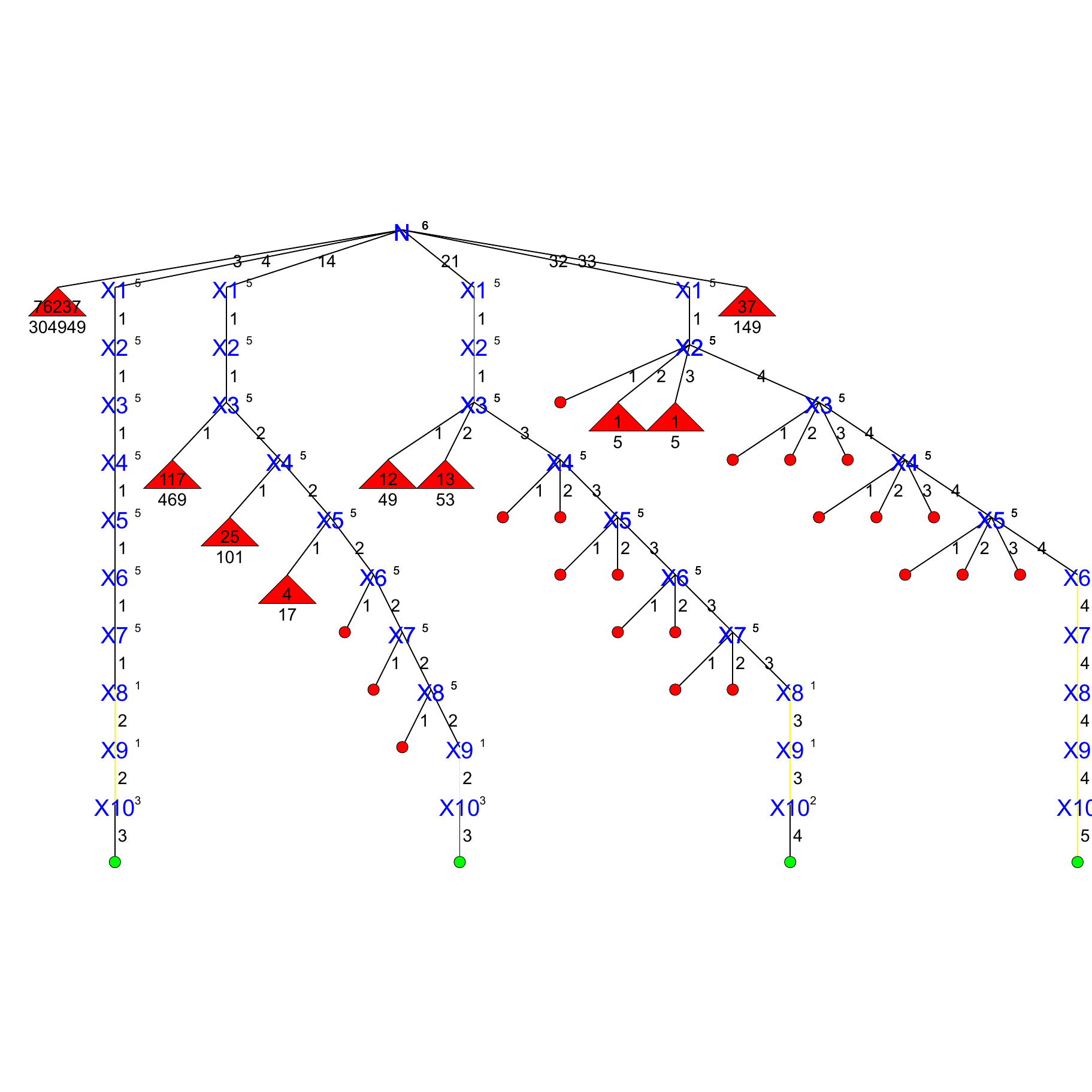}
    \caption{\emph{Automaton}}
  \end{subfigure}%
  \begin{subfigure}[b]{0.5\textwidth}
    \label{fig:treess-b}%
    \includegraphics[width=\textwidth]{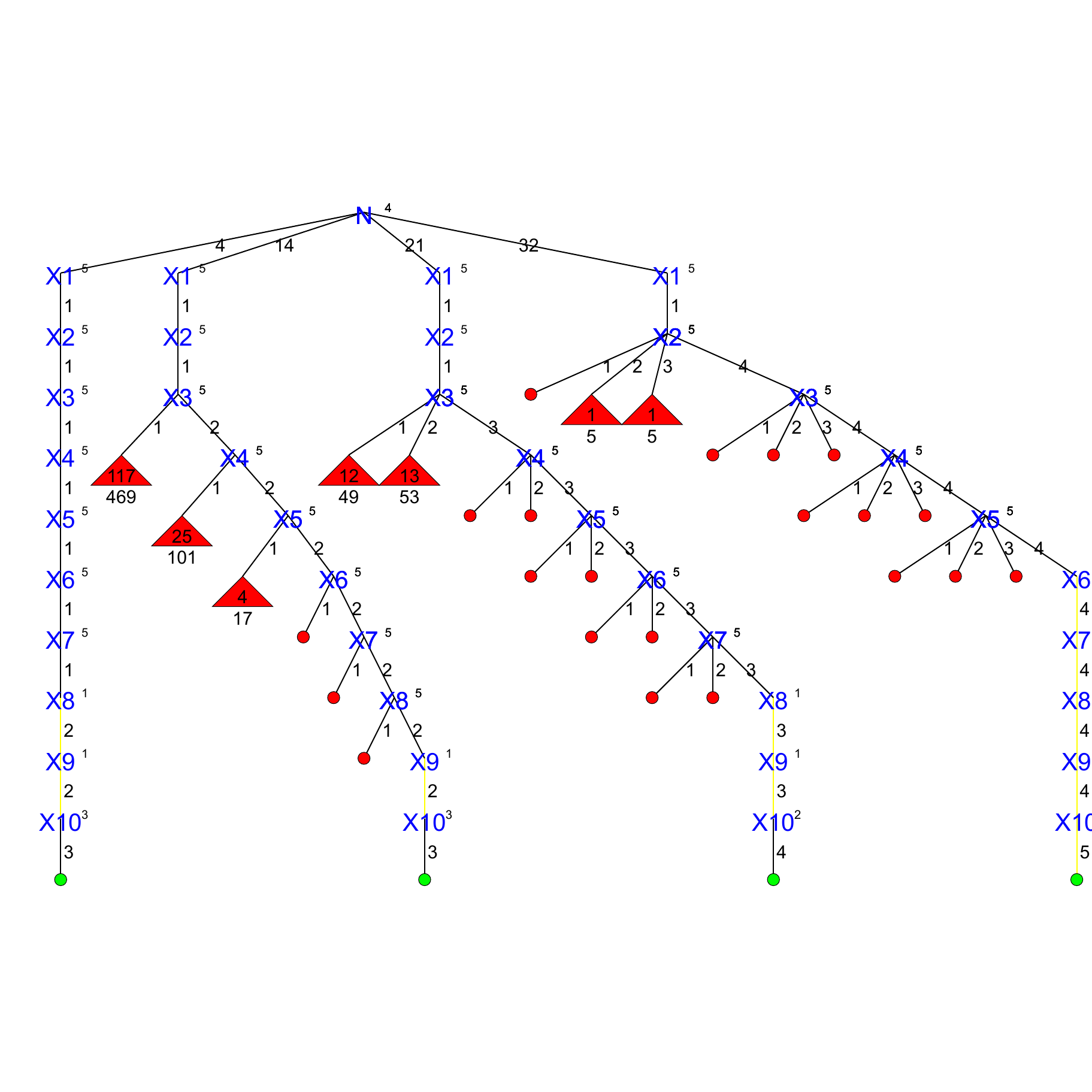}
    \caption{\emph{Bounds}}
  \end{subfigure}\\
  \begin{subfigure}[b]{0.5\textwidth}
    \label{fig:trees-c}%
    \includegraphics[width=\textwidth]{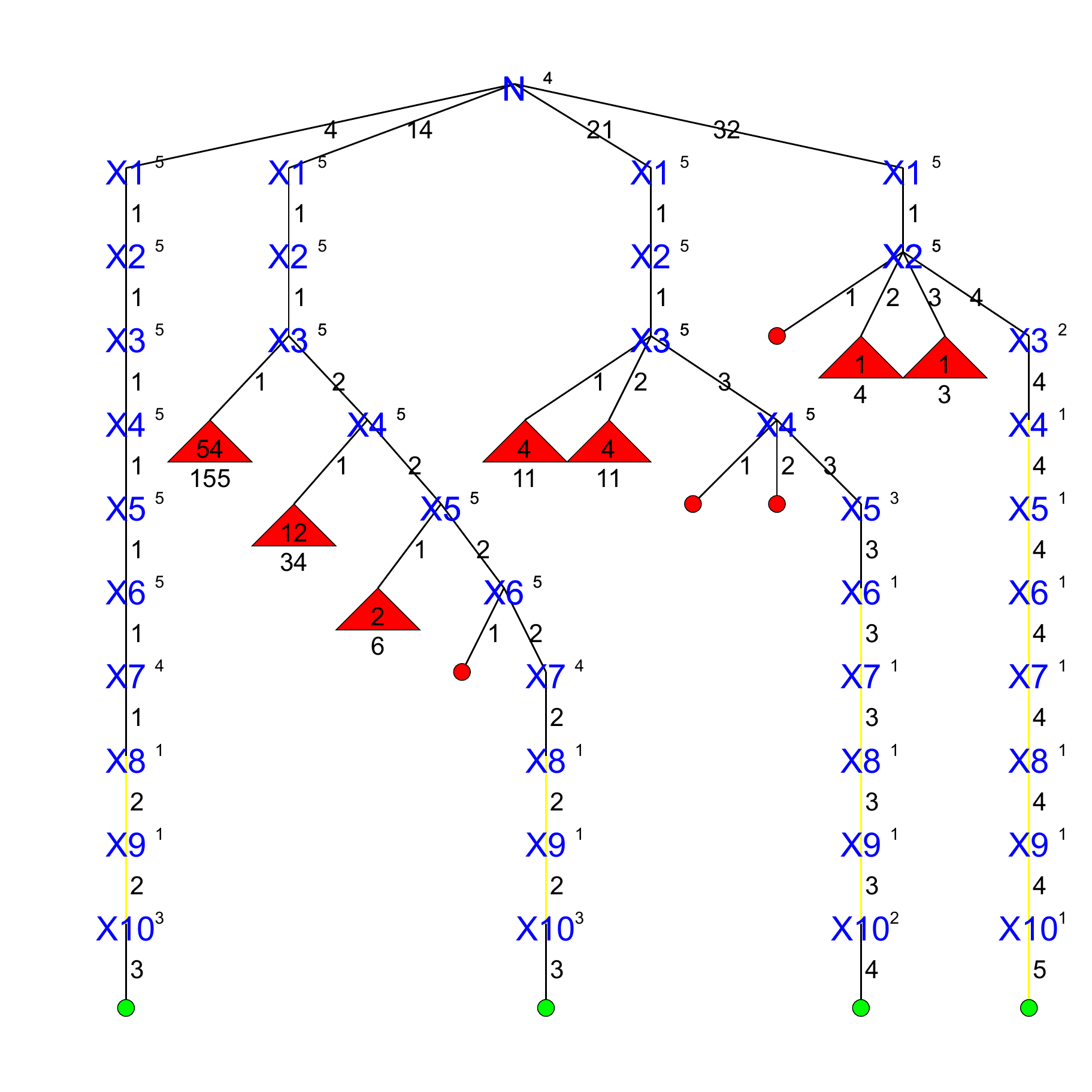}
    \caption{\emph{Bounds}+\emph{Glue}}
  \end{subfigure}%
  \begin{subfigure}[b]{0.5\textwidth}
    \label{fig:trees-d}%
    \includegraphics[width=\textwidth]{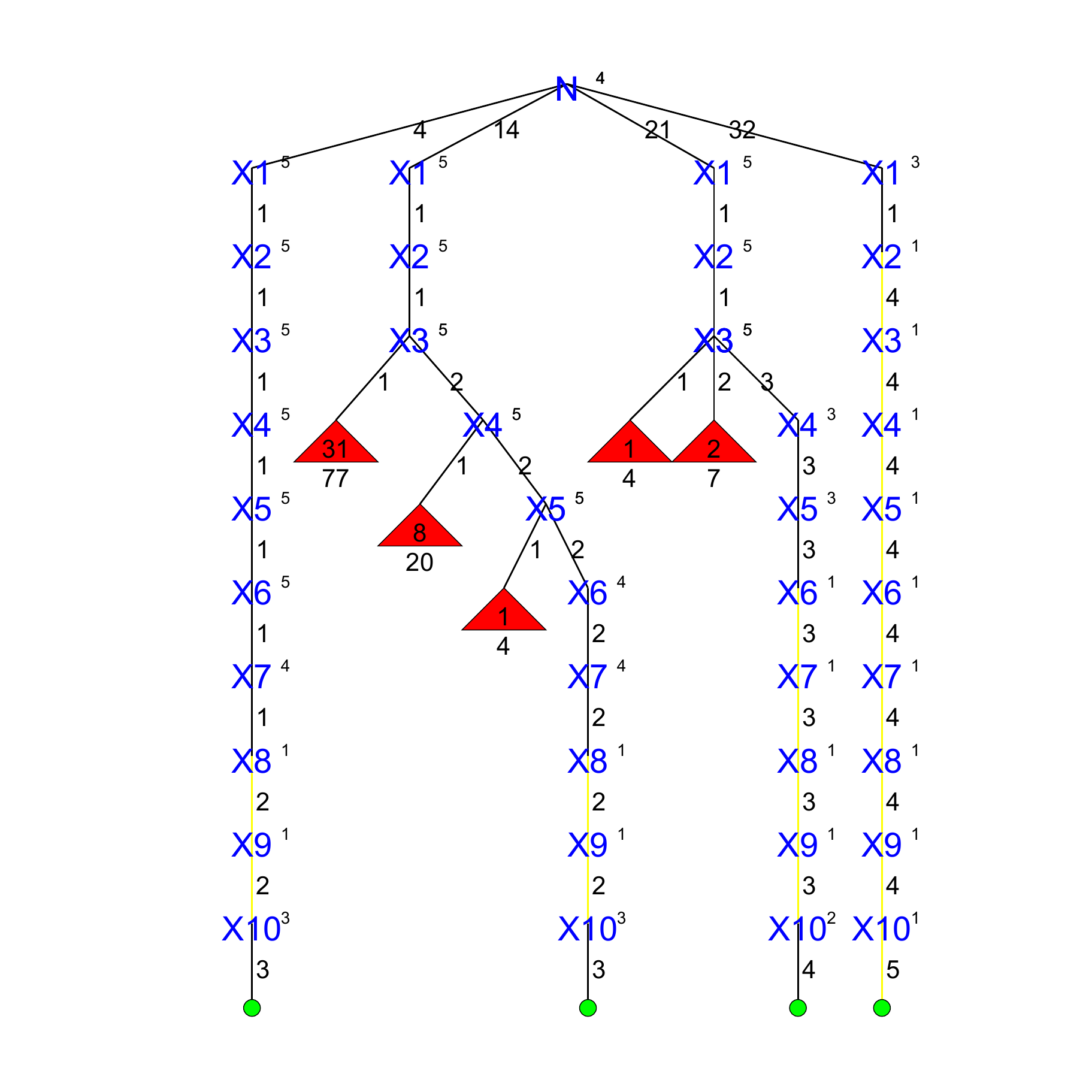}
    \caption{\emph{Combined}}
  \end{subfigure}
  \caption{\label{fig:Trees}Comparing parts of the search tree for
    max\_surf\_increasing\_terrace, finding the first solution or
    proving infeasibility for the manually selected values 3, 4, 14,
    21, 32, and 33 of variable $N$ and 10 variables $X_1, X_2, \dots,
    X_{10}$, each with domain $[1,5]$.}
\end{figure}

As a final experiment, we look at the search trees generated by four solution variants for a single constraint \Constraint{max\_surf\_increasing\_terrace}. We only display some of the values for the parameter $N$, to make the trees more legible. Figure~\ref{fig:Trees} shows the search tree produced with the help of CP-Viz~\cite{DBLP:conf/cp/SimonisDFMQC10}. Each tree shows the branches explored to find a first solution or proving infeasibility for each parameter value, with the initial choice of the value $N$ at the top, and then the assignment of ten variables with a standard left-to-right labeling. Failed subtrees are abstracted as red triangles containing two numbers, the one above is the number of internal nodes in the tree, the one below the number of failed leaf nodes. Success nodes are colored in green, while failure nodes are colored red. Internal nodes are labeled by the variable name currently being assigned, and a superscript indicating the number of values in the domain of that variable. Edges indicate choices that are explored, the number indicates the value that is assigned to the selected variable, while a yellow edge color indicates that the value had been fixed by propagation.

In all trees, a first solution for parameter value 4, the smallest feasible value, is found without backtracking. The solution chooses value 1 for $X_1$ to $X_7$, then value 2 for $X_8$ and $X_9$, and finally value 3 for variable $X_{10}$. On the other hand, in the initial automaton, a very large failed subtree is shown for the left-most parameter value 3, and a much smaller failed tree for the right-most value 33. Both of those values are infeasible, and are removed by the bounds for this constraint. The \emph{Bounds} version therefore avoids these failed sub-trees, but there are no changes for the other, feasible values. When we consider the \emph{Bounds}+\emph{Glue} version, the search for feasible solutions is reduced, with a further reduction for the \emph{Combined} variant. But we still need search to find the initial solution for some of the parameter values. This occurs since the bounds and the glue matrix reasoning only consider lower and upper bounds, and we don't detect holes in the domain of variable $N$. To get the best use of the generated bounds, we have to use the incremental combination of \emph{Bounds} with the \emph{Glue} constraint, as the bounds are then applied for each suffix of unassigned variables to maximise the information extracted. 

\section{Conclusion}
\label{sec:conclusion}

Within the context of quantitative extensions of regular languages~(QRE)
we introduce the concept of regular expression characteristics
as a way to unify combinatorial aspects of quantitative extensions of regular languages.
We illustrate that approach for time-series constraints where,
introducing six regular expression characteristics,
allows coming up with generic bounds for families of time\nobreakdash-series constraints.
We believe the introduction of regular expression characteristics is
important for the area of~QRE.

\subsubsection*{Acknowledgements.}
We thank Pierre Flener for his feedback on the notation system for
regular expression characteristics in Section~\ref{sec:notation}.

\clearpage

\bibliographystyle{splncs03}
\bibliography{bounds} 

\newpage
\appendix
\section{Appendix: Tables of Regular Expressions Characteristics}
\label{app:tables}
\normalsize

\tabref{tab:width} gives the \emph{width} characteristics for each regular expression in~\tabref{tab:patterns}.
Within \tabref{tab:width}, smallest words are obtained by (1)~first discarding from a regular expression all sub\nobreakdash-expressions containing the empty word,
and then by (2)~keeping within each disjunction the smallest length words.
For instance,
within the $\ZigzagPatternName$ regular expression $\textnormal{`}(<>)^*<><(>|\varepsilon)~|~(><)^*><>(<|\varepsilon)\textnormal{'}$,
we remove the sub\nobreakdash-expressions
$\textnormal{`}(<>)^*\textnormal{'}$,
~$\textnormal{`}(>|\varepsilon)\textnormal{'}$,
~$\textnormal{`}(><)^*\textnormal{'}$,
~$\textnormal{`}(<|\varepsilon)\textnormal{'}$
and obtain the disjunction $\textnormal{`}<><~|~><>\textnormal{'}$ containing two words of length $3$.
\\

\begin{table*}[!h]\centering
\setlength{\tabcolsep}{6pt}
\begin{tabular}{@{}lllc@{}} \toprule
  name~$\pattern$                 		
  & regular expression
  & $\Char{\width}{\pattern}$ \\\midrule
  $\BumpOnDecreasingSequencePatternName$   & {\scriptsize$\BumpOnDecreasingSequencePatternWidth$}   & $5$ \\
  $\DecreasingPatternName$                 & {\scriptsize$\DecreasingPatternWidth$}                 & $1$ \\
  $\DecreasingSequencePatternName$         & {\scriptsize$\DecreasingSequencePatternWidth$}         & $1$ \\
  $\DecreasingTerracePatternName$          & {\scriptsize$\DecreasingTerracePatternWidth$}          & $3$ \\
  $\DipOnIncreasingSequencePatternName$    & {\scriptsize$\DipOnIncreasingSequencePatternWidth$}    & $5$ \\
  $\GorgePatternName$                      & {\scriptsize$\GorgePatternWidth$}                      & $2$ \\
  $\IncreasingPatternName$                 & {\scriptsize$\IncreasingPatternWidth$}                 & $1$ \\
  $\IncreasingSequencePatternName$         & {\scriptsize$\IncreasingSequencePatternWidth$}         & $1$ \\
  $\IncreasingTerracePatternName$          & {\scriptsize$\IncreasingTerracePatternWidth$}          & $3$ \\
  $\InflexionPatternName$                  & {\scriptsize$\InflexionPatternWidth$}                  & $2$ \\
  $\PeakPatternName$                       & {\scriptsize$\PeakPatternWidth$}                       & $2$ \\
  $\PlainPatternName$                      & {\scriptsize$\PlainPatternWidth$}                      & $2$ \\
  $\PlateauPatternName$                    & {\scriptsize$\PlateauPatternWidth$}                    & $2$ \\
  $\ProperPlainPatternName$                & {\scriptsize$\ProperPlainPatternWidth$}                & $3$ \\
  $\ProperPlateauPatternName$              & {\scriptsize$\ProperPlateauPatternWidth$}              & $3$ \\
  $\SteadyPatternName$                     & {\scriptsize$\SteadyPatternWidth$}                     & $1$ \\  
  $\SteadySequencePatternName$             & {\scriptsize$\SteadySequencePatternWidth$}             & $1$ \\
  $\StrictlyDecreasingSequencePatternName$ & {\scriptsize$\StrictlyDecreasingSequencePatternWidth$} & $1$ \\
  $\StrictlyIncreasingSequencePatternName$ & {\scriptsize$\StrictlyIncreasingSequencePatternWidth$} & $1$ \\
  $\SummitPatternName$                     & {\scriptsize$\SummitPatternWidth$}                     & $2$ \\
  $\ValleyPatternName$                     & {\scriptsize$\ValleyPatternWidth$}                     & $2$ \\
  $\ZigzagPatternName$                     & {\scriptsize$\ZigzagPatternWidth$}                     & $3$ \\
\bottomrule
\end{tabular}
\caption{\label{tab:width} Regular expression names $\pattern$ and corresponding \emph{width} (see~\defref{def:width});
within each regular expression subparts corresponding to a smallest length word are highlighted in yellow.}
\end{table*}

\clearpage
\noindent
\tabref{tab:height} gives the \emph{height} characteristics for each regular expression in~\tabref{tab:patterns}.
Within \tabref{tab:height}, the `illustration' column provides for each regular expression $\pattern$ a word achieving the smallest height
among all words of $\Language{\pattern}$.
For a regular expression $\pattern$, a word $\word$ achieving the smallest height is a word of $\Language{\pattern}$ that minimises
the number of occurrences of $\textnormal{`}>\textnormal{'}$ (respectively $\textnormal{`}<\textnormal{'}$)
over all maximal occurrences of $\textnormal{`}>(=|>)^*\textnormal{'}$ (respectively $\textnormal{`}<(=|<)^*\textnormal{'}$) in $\word$.
We illustrate this for two regular expressions.
\renewcommand{\labelitemi}{$\bullet$}
\begin{itemize}
\item
For the fixed length regular expression $\pattern =
\BumpOnDecreasingSequencePatternName$, $\Language{\pattern}$ contains a single word $\word=\BumpOnDecreasingSequencePattern$.
Since $\word$ is the concatenation of three proper factors
$\textnormal{`}>>\textnormal{'}$, $\textnormal{`}<\textnormal{'}$ and $\textnormal{`}>>\textnormal{'}$
of respective length $2$, $1$ and $2$
we obtain a height of $2$.
\item
For the non\nobreakdash-fixed length regular expression $\pattern =
\DecreasingTerracePatternName$, the word
$\textnormal{`}>=>\textnormal{'} \in \Language{\pattern}$
has a height of $2$.
No word in $\Language{\pattern}$ can have a smaller height,
since any word~$\word$ in  the language of $\DecreasingTerracePattern$ contains two
occurrences of $\textnormal{`}>\textnormal{'}$, one at both extremities of $\word$,
separated by a single stretch of $\textnormal{`}=\textnormal{'}$.\\
\end{itemize}

\input{table_characteristic_height.tex}

\clearpage
\noindent
\tabref{tab:range-regular-expression} gives the \emph{range} characteristics for each regular expression in~\tabref{tab:patterns}.
Within \tabref{tab:range-regular-expression}, the `illustration' column provides for each regular expression $\pattern$
a time series whose signature is a word of the smallest height
among all words of the same length $\seqlength-1$ in $\Language{\pattern}$, i.e.~the
range of $\pattern$ wrt $\Tuple{\seqlength}$. 
We distinguish three cases:
\begin{itemize}
\item
For a fixed length regular expression $\pattern$ (e.g.~$\BumpOnDecreasingSequencePatternName$),
the range $\CharPar{\CompleteExtension}{\pattern}{\seqlength}$ is only
defined for one plus the length of the single word in~$\Language{\pattern}$,
and is equal to the height $\Char{\Height}{\pattern}$ of~$\pattern$.
\item
For a non\nobreakdash-fixed length regular expression $\pattern$, if
we can find a word of length $\seqlength-1$ in~$\Language{\pattern}$
which has the same height as the height $\Char{\Height}{\pattern}$,
we cannot have a smaller height by definition.
This is the case for many our non\nobreakdash-fixed length regular
expressions, for example~$\PeakPatternName$, $\InflexionPatternName$ or $\ZigzagPatternName$.
\item
For some non\nobreakdash-fixed length regular expressions $\pattern$ like
$\DecreasingSequencePatternName$, $\IncreasingSequencePatternName$, $\GorgePatternName$ or $\SummitPatternName$,
only the corresponding shortest word has a height of~$\Char{\Height}{\pattern}$.
Then, any longer word in~$\Language{\pattern}$, has a height of at least~$\Char{\Height}{\pattern}+1$.
\item
For $\pattern=\StrictlyDecreasingSequencePatternName$,
$\Language{\pattern}$ contains
a single word of length $\seqlength-1$,
namely a stretch of $\seqlength-1$ consecutive
$\textnormal{`}>\textnormal{'}$.
Hence, the range of~$\pattern$ wrt $\Tuple{\seqlength}$ is reached for this word and equals~$\seqlength-1$.

The same reasoning applies for $\StrictlyIncreasingSequencePatternName$.
\end{itemize}

\input{table_characteristic_range.tex}

\clearpage
\noindent
\tabref{tab:set-of-inducing-words} gives the \emph{inducing words} characteristics for each regular expression in~\tabref{tab:patterns}.
Within \tabref{tab:set-of-inducing-words}, the inducing words characteristics
is derived from the corresponding regular expression by removing all sub\nobreakdash-expressions
containing the empty word and by keeping the rest, i.e.~the part highlighted in yellow.\\

\begin{table*}[!h]\centering
\setlength{\tabcolsep}{6pt}
\begin{tabular}{@{}llc@{}} \toprule
  name~$\pattern$                 		
  & regular expression
  & $\Char{\IndWordSet}{\pattern}$ \\\midrule
  $\BumpOnDecreasingSequencePatternName$   & $\BumpOnDecreasingSequencePatternInduced$   & $\Set{\BumpOnDecreasingSequenceInduced}$   \\
  $\DecreasingPatternName$                 & $\DecreasingPatternInduced$                 & $\Set{\DecreasingInduced}$                 \\
  $\DecreasingSequencePatternName$         & $\DecreasingSequencePatternInduced$         & $\Set{\DecreasingSequenceInduced}$         \\
  $\DecreasingTerracePatternName$          & $\DecreasingTerracePatternInduced$          & $\Set{\DecreasingTerraceInduced}$          \\
  $\DipOnIncreasingSequencePatternName$    & $\DipOnIncreasingSequencePatternInduced$    & $\Set{\DipOnIncreasingSequenceInduced}$    \\
  $\GorgePatternName$                      & $\GorgePatternInduced$                      & $\Set{\GorgeInduced}$                      \\
  $\IncreasingPatternName$                 & $\IncreasingPatternInduced$                 & $\Set{\IncreasingInduced}$                 \\
  $\IncreasingSequencePatternName$         & $\IncreasingSequencePatternInduced$         & $\Set{\IncreasingSequenceInduced}$         \\
  $\IncreasingTerracePatternName$          & $\IncreasingTerracePatternInduced$          & $\Set{\IncreasingTerraceInduced}$          \\
  $\InflexionPatternName$                  & $\InflexionPatternInduced$                  & $\Set{\InflexionInduced}$                  \\
  $\PeakPatternName$                       & $\PeakPatternInduced$                       & $\Set{\PeakInduced}$                       \\
  $\PlainPatternName$                      & $\PlainPatternInduced$                      & $\Set{\PlainInduced}$                      \\
  $\PlateauPatternName$                    & $\PlateauPatternInduced$                    & $\Set{\PlateauInduced}$                    \\
  $\ProperPlainPatternName$                & $\ProperPlainPatternInduced$                & $\Set{\ProperPlainInduced}$                \\
  $\ProperPlateauPatternName$              & $\ProperPlateauPatternInduced$              & $\Set{\ProperPlateauInduced}$              \\
  $\SteadyPatternName$                     & $\SteadyPatternInduced$                     & $\Set{\SteadyInduced}$                     \\  
  $\SteadySequencePatternName$             & $\SteadySequencePatternInduced$             & $\Set{\SteadySequenceInduced}$             \\
  $\StrictlyDecreasingSequencePatternName$ & $\StrictlyDecreasingSequencePatternInduced$ & $\Set{\StrictlyDecreasingSequenceInduced}$ \\
  $\StrictlyIncreasingSequencePatternName$ & $\StrictlyIncreasingSequencePatternInduced$ & $\Set{\StrictlyIncreasingSequenceInduced}$ \\
  $\SummitPatternName$                     & $\SummitPatternInduced$                     & $\Set{\SummitInduced}$                     \\
  $\ValleyPatternName$                     & $\ValleyPatternInduced$                     & $\Set{\ValleyInduced}$                     \\
  $\ZigzagPatternName$                     & $\ZigzagPatternInduced$                     & $\Set{\ZigzagInduced}$                     \\
\bottomrule
\end{tabular}
\caption{\label{tab:set-of-inducing-words} Regular expression names $\pattern$ and corresponding \emph{inducing words} (see~\defref{def:set-of-inducing-words}).}
\end{table*}

\clearpage
\noindent
\begin{sloppypar}
\tabref{tab:overlap} gives the \emph{overlap} characteristics for each regular expression in~\tabref{tab:patterns}.
Within \tabref{tab:overlap} we distinguish the following cases for computing the overlap characteristics:
\begin{itemize}
\item
Consider a fixed length regular expression $\pattern$ whose regular language contains a single word $\word$.
Then, we compute the length of the maximum overlap $o$ between $\word$
and itself for which $o<|\word|$.
	\begin{itemize}
	\item
	If such overlap exists the corresponding overlap characteristics
	$\CharPar{\Overlap}{\pattern}{\DomainMin,\DomainMax}$ is equal to $o+1$,
	e.g.~for $\pattern=\BumpOnDecreasingSequencePatternName$
	the maximum overlap of $\BumpOnDecreasingSequencePattern$ with itself is $2$
	assuming the two words do not completely overlap,
	leading to $\CharPar{\Overlap}{\pattern}{\DomainMin,\DomainMax}=3$.
	\item
	If such overlap does not exist, depending whether the difference $\DomainMax-\DomainMin$
	is big enough or not, we can concatenate $\word$ with itself or not,
	leading to an overlap of $1$ (one time-series variable is shared) or to an overlap of $0$.
	This is the case for $\DecreasingPatternName$ and $\IncreasingPatternName$ where,
	depending whether the difference $\DomainMax-\DomainMin$ is strictly greater than $1$ or not,
	we get an overlap $\CharPar{\Overlap}{\pattern}{\DomainMin,\DomainMax}$ of $1$ or $0$.
	\end{itemize}
\item
Consider a regular expression $\pattern$ for which the set of superpositions of any pair of words of $\Language{\pattern}$ is empty;
then the corresponding overlap $\CharPar{\Overlap}{\pattern}{\DomainMin,\DomainMax}$ is equal to $0$.
This is the case for $\DecreasingSequencePatternName$, $\IncreasingSequencePatternName$,
$\SteadySequencePatternName$, $\StrictlyDecreasingSequencePatternName$,
and $\StrictlyIncreasingSequencePatternName$, because Condition~\ref{Cond:i}  of
\defref{def:superposition} is always violated.
\item
Given a regular expression $\pattern$ for which
(1)~the set of superpositions of any pair of words of $\Language{\pattern}$ is limited to
the concatenation of the pair of corresponding words, and
(2)~any pair of word of $\Language{\pattern}$ starts with a $\textnormal{`<'}$ and ends up with a $\textnormal{`>'}$
(or conversely starts with a $\textnormal{`>'}$ and ends up with a $\textnormal{`<'}$),
then we can concatenate them so that they share one time-series variable regardless the value of $\DomainMax-\DomainMin$;
we get an overlap $\CharPar{\Overlap}{\pattern}{\DomainMin,\DomainMax}$ of $1$.
This is the case for
$\GorgePatternName$,
$\PeakPatternName$,
$\PlainPatternName$,
$\PlateauPatternName$,
$\ProperPlainPatternName$,
$\ProperPlateauPatternName$,
$\SummitPatternName$,
and $\ValleyPatternName$.
\item
  Consider the $\pattern = \DecreasingTerracePatternName$ regular expression.
  For any two words~$\wordv = \reg{>=^{(k)}>}$ and~$\word= \reg{>=^{(l)}>}$
  in~$\Language{\pattern}$ with~$k, l$ being positive integers, the set of their superpositions
  wrt~$\Tuple{\DomainMin, \DomainMax}$ contains at most two words,
  namely~$\wordz_1 = \reg{>=^{(k)}>>=^{(l)}>}$ and~$\wordz_2 =
  \reg{>=^{(k)}>=^{(l)}>}$.
  The value of overlap achieved for~$\wordz_1$ and for~$\wordz_2$
  is~$2$ and~$1$, respectively.
  \begin{itemize}
  \item
    When~$\DomainMax - \DomainMin = \Char{\Height}{\pattern} = 2$, neither
    $\wordz_1$ nor $\wordz_2$ can appear in the signature of a ground
    time series over~$\Domain$, thus the set of superpositions of~$\pattern$
    wrt~$\Tuple{\DomainMin, \DomainMax}$ is empty,
    and~$\CharPar{\Overlap}{\pattern}{\DomainMin,\DomainMax} = 0$.
  \item
    When~$\DomainMax - \DomainMin > \Char{\Height}{\pattern} = 2$, the word
    $\wordz_1$ is always in the set of superpositions of~$\pattern$
    wrt~$\Tuple{\DomainMin, \DomainMax}$, and 
    thus~$\CharPar{\Overlap}{\pattern}{\DomainMin,\DomainMax} = 2$.
  \end{itemize}
The same reasoning applies for the~$\IncreasingTerracePatternName$
regular expression.
\item
  Consider the $\pattern = \InflexionPatternName$ regular expression.
  Any word in~$\Language{\pattern}$ belongs to the language of either
  $\pattern_1 = \reg{< (< | =)^*>}$ or  $\pattern_2 = \reg{> (> |
    =)^*<}$.
\begin{itemize}
  \item
    For any two words~$\wordv, \word \in \Language{\pattern_1}$ 
    (respectively~$\wordv, \word \in
    \Language{\pattern_2}$), their overlap wrt~$\DomTuple$ is at most
    $1$, since their only possible superposition is~$\wordv\word$.
  \item
    For any two words~$\wordv \in \Language{\pattern_1}$ and~$\word \in
    \Language{\pattern_2}$, their overlap wrt~$\DomTuple$ is at most
    $2$, since the maximum length of a suffix of~$\wordv$ that is also
    a prefix of~$\word$ is~$1$.
    Hence,~$\CharPar{\Overlap}{\pattern}{\DomainMin,\DomainMax} \leq 2$.
\end{itemize}
The overlap of the words~$\reg{><}$ and $\reg{<>}$ wrt~$\DomTuple$ such
that~$\DomainMax - \DomainMin \geq \Char{\Height}{\pattern}$ is~$2$,
which is maximum.
Hence, $\CharPar{\Overlap}{\pattern}{\DomainMin,\DomainMax} = 2$.
\item
  Consider the $\pattern = \ZigzagPatternName$ regular expression.
  Any word in~$\Language{\pattern}$ belongs to the regular language either of
  $\pattern_1 = \reg{(<>)^+<(>|\varepsilon)}$ or  of $\pattern_2 = \reg{(><)^+>(<|\varepsilon)}$.
\begin{itemize}
  \item
    For any two words~$\wordv \in \Language{\pattern_1}$ and~$\word \in
    \Language{\pattern_2}$, their overlap wrt~$\DomTuple$ is always
    $0$, since their set of superpositions wrt~$\DomTuple$ is empty,
    because~\condref{Cond:i} of~\defref{def:superposition} is
    violated. 
  \item
    For any two words~$\wordv, \word \in \Language{\pattern_1}$ 
    (respectively~$\wordv, \word \in
    \Language{\pattern_2}$), their overlap wrt~$\DomTuple$ is at most
    $1$, since their only possible superposition is~$\wordv\word$.
    Note that the height of every word in~$\Language{\pattern}$ is
    $\Char{\Height}{\pattern} = 1$, then the height of~$\wordv\word$ is~$2$, since $\wordv$ last
    symbol coincides with the $\word$ first symbol, and it is not~$\reg{=}$.
\end{itemize}
Hence, when~$\DomainMax-\DomainMin=\Char{\Height}{\pattern}=1$,
the overlap of~$\pattern$ wrt~$\DomTuple$ is~$0$, and
when $\DomainMax-\DomainMin\geq\Char{\Height}{\pattern}+1$,
the overlap of~$\pattern$ wrt~$\DomTuple$ is~$1$.
\end{itemize}
\end{sloppypar}

\input{table_characteristic_overlap.tex}

\clearpage

\noindent
\tabref{tab:variation-of-maxima} gives the \emph{smallest variation of maxima} characteristics
for each regular expression in~\tabref{tab:patterns}.
Within \tabref{tab:variation-of-maxima} we distinguish the following
cases for computing the smallest variation of maxima characteristics:
\begin{sloppypar}
\begin{itemize}
\item
 When $\CharPar{\Overlap}{\pattern}{\DomainMin,\DomainMax}$ is~$0$, the
 quantity $\CharPar{\VariationOfMax}{\pattern}{\DomainMin,\DomainMax}$ is also~$0$
 by definition.
 This is for example the case of $\DecreasingSequencePatternName$ and
 $\ZigzagPatternName$ when~$\DomainMax-\DomainMin=1$.
\item
  When $\CharPar{\Overlap}{\pattern}{\DomainMin,\DomainMax}$ is not~$0$, 
  and we can give a pair of words~$\wordv$, $\word$ in
  $\Language{\pattern}$ such that their set of superpositions
  wrt~$\Tuple{\DomainMin, \DomainMax}$ is not empty and
  $\CharAll{\VariationOfMax}{\pattern}{\DomainMin,\DomainMax}{\wordv,
    \word}$ is~$0$, the value of $\CharPar{\VariationOfMax}{\pattern}{\DomainMin,\DomainMax}$ is
  also $0$.
  Note that by definition
  $\CharPar{\VariationOfMax}{\pattern}{\DomainMin,\DomainMax}$ has the
  minimum absolute value.
  Hence, if the value of zero is reached for at least one
  pair of words, then
  $\CharPar{\VariationOfMax}{\pattern}{\DomainMin,\DomainMax}$ is
  zero.
\item
  When $\CharPar{\Overlap}{\pattern}{\DomainMin,\DomainMax}$ is not~$0$, and
  when the regular language of~$\pattern$ contains a single word,
  $\CharPar{\VariationOfMax}{\pattern}{\DomainMin,\DomainMax}$ is
  reached for a superposition of this word with itself.
  See, for example $\DecreasingPatternName$, when~$\DomainMax - \DomainMin \geq 2$.
\item
  Consider the~$\pattern =\DecreasingTerracePatternName$ regular
  expression when $\DomainMax - \DomainMin \geq 3$,
  i.e.~the overlap~$\CharPar{\Overlap}{\pattern}{\DomainMin,\DomainMax}$ is not~$0$.
  For any two words~$\wordv = \reg{>=^{(k)}>}$ and~$\word= \reg{>=^{(l)}>}$
  in~$\Language{\pattern}$ with~$k, l$ being positive integers, the set of their superpositions
  wrt~$\Tuple{\DomainMin, \DomainMax}$ contains at most two words,
  namely~$\reg{>=^{(k)}>>=^{(l)}>}$ and~$\reg{>=^{(k)}>=^{(l)}>}$.
  Then, the value
  of~$\CharAll{\VariationOfMax}{\pattern}{\DomainMin,\DomainMax}{\wordv,
  \word}$ equals~$-1$ and is reached for the superposition $\reg{>=^{(k)}>=^{(l)}>}$.
The same reasoning applies for $\IncreasingTerracePatternName$.
\end{itemize}
\end{sloppypar}

\input{table_characteristic_variation-of-maxima.tex}

\end{document}